\documentclass[rmp,aps,twocolumn,showpacs,preprintnumbers,amsmath,amssymb]{revtex4-1}
\usepackage[utf8x]{inputenc}
\usepackage{amsmath}
\usepackage{amsfonts}
\usepackage{amssymb}
\usepackage{graphicx}
\usepackage{mathtools}
\usepackage{dcolumn}
\usepackage[usenames,dvipsnames]{color}

\newcommand{\br}{\mathbf{r}}

\newcommand{\bs}{\mathbf{s}}
\newcommand{\bmu}{\boldsymbol{\mu}}

\newcommand \comma {\mbox{\makebox[.1 in]{ },}}

\def \be {\begin{equation}}
\def \ee {\end{equation}}
\def \ben {\begin{eqnarray}}
\def \een {\end{eqnarray}}

\begin{document}

\title{Delocalized excitons in natural light harvesting complexes}

\author{Seogjoo J. Jang}
\affiliation{Department of Chemistry and Biochemistry, Queens College, City University of New York, 65-30 Kissena Boulevard, Queens, NY 11367 \& PhD programs in Chemistry and Physics, and Initiative for Theoretical Sciences, Graduate Center, City University of New York, 365 Fifth Avenue, New York, NY 10016}
\author{Benedetta Mennucci}
\affiliation{Department of Chemistry, University of Pisa, via G. Moruzzi 13, 56124 Pisa, Italy}

\begin{abstract}
Natural organisms such as photosynthetic bacteria, algae, and plants employ complex molecular machinery to convert solar energy into biochemical fuel.  An important common feature shared by most of these photosynthetic organisms is that they capture photons in the form of excitons typically delocalized over a few to tens of pigment molecules embedded in protein environments of light harvesting complexes (LHCs).   Delocalized excitons created in such LHCs remain well protected despite being swayed by environmental fluctuations, and are delivered successfully to their destinations over hundred nanometer length scale distances in about hundred picosecond time scales.  Decades of experimental and theoretical investigation have produced a large body of information offering insights into major structural, energetic, and dynamical features contributing to LHCs' extraordinary capability to harness photons using delocalized excitons.   The objective of this review is (i) to provide a comprehensive account of major theoretical, computational,  and spectroscopic advances that have contributed to this body of knowledge, and (ii) to clarify  the issues concerning the role of delocalized excitons in achieving efficient energy transport mechanisms.  The focus of this review is on three representative systems, Fenna-Matthews-Olson complex of green sulfur bacteria, light harvesting 2 complex of purple bacteria, and phycobiliproteins of cryptophyte algae.  Although we offer more in-depth and detailed description of theoretical and computational aspects,  major experimental results and their implications are also assessed in the context of achieving excellent light harvesting functionality.  Future theoretical and experimental challenges to be addressed in gaining better understanding and utilization of delocalized excitons are also discussed.

\end{abstract}

\date{Accepted for publication in the Reviews of Modern Physics on March 26, 2018; Revised on June 8, 2018}
\maketitle

\noindent

\section{Introduction}

The energy for life on earth begins with harvesting of photons from sunlight. To perform this quantum mechanical process, photosynthetic organisms have developed a wide range of highly tuned forms of light harvesting complexes (LHCs) \cite{amerongen-pe,blankenship-2,hu-qrb35,cogdell-qrb39}, while utilizing surprisingly few kinds of pigment molecules \cite{blankenship-2,croce-nchembio10} as primary units that absorb photons and create electronic excitations (excitons) \cite{davydov,agranovich}. Most LHCs consist of about $10-100$ pigment molecules held by protein scaffolds, with typical inter-pigment distances $\gtrsim 1 {\rm nm}$, and
harness excitons that are typically delocalized over a few to tens of such pigment molecules.  In general, the light harvesting unit of a specific photosynthetic organism is formed by aggregates or super-complexes of those LHCs.

\begin{figure}[hptb]
(a)\makebox[3in]{ }\\
\includegraphics[width=3.2in]{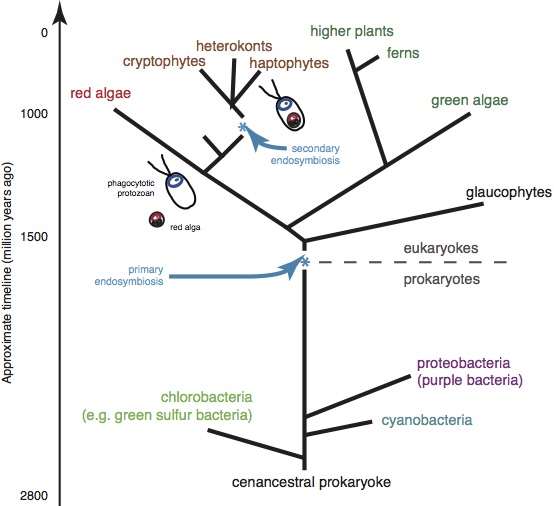}\vspace{.2in}\\
(b)\makebox[3in]{ }\vspace{.1in}\\
\includegraphics[width=3.5in]{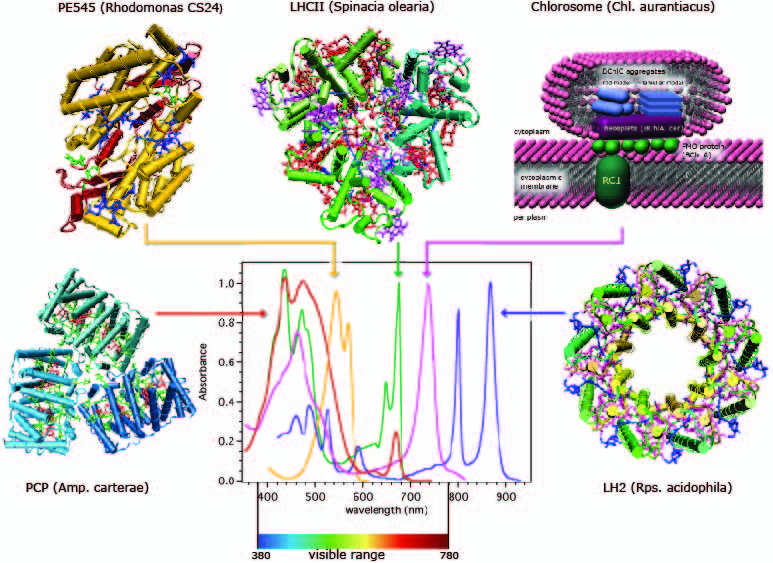}
\caption{(a) A simple evolutionary tree of photosynthetic organisms (provided by Gregory D. Scholes) (b) Major light harvesting antennas and their spectral regions adapted with permission from \cite{scholes-jpcl1}. Copyright 2010 American Chemical Society.}
\label{fig:ps_tree}
\end{figure}

Each LHC exhibits unique structural and energetic features realized through specific arrangement of pigment molecules with finely tuned excitation energies \cite{blankenship-2}, which in turn determine the energetics and the dynamics of delocalized excitons.  These are believed to embody certain design principles that have been developed through long evolutionary adaptation processes, and ensure optimal light harvesting capability for given living environments and specific light conditions.  Figure \ref{fig:ps_tree} provides a simple overview of the evolutionary tree of photosynthetic organisms and major LHCs.      

To what extent do such design principles owe their success to quantum effects associated with the delocalized nature of excitons?   Answering this question might have been the underlying motivation for decades of research on LHCs.  However, only during the past decade, this question has become much more explicit, attracting numerous experimental and theoretical studies.  While there have been excellent review articles on various aspects of LHCs \cite{renger-pr343,hu-qrb35,cogdell-qrb39,renger-pr102,renger-pccp15,jang-wires,cheng-arpc60,jankowiak-cr111,phil-tran-370,olaya-castro-irpc30,pachon-pccp14,schroter-pr567,chenu-arpc66,konig-cpc13,lee-arpc67,lambert-np9,mirkovic-cr117,levi-rpp78,curutchet-cr117,kondo-cr117}, a comprehensive review for the general physics community,  which encompasses all of the energetic, dynamic, and spectroscopic information, seems currently lacking to the best of our knowledge.  Our objective is to offer such a review with particular focus on quantum mechanical aspects of delocalized excitons in LHCs.

Excitons created in most LHCs are generally viewed as Frenkel type \cite{frenkel-pr37,davydov}, for which the Hamiltonian representing a single exciton can be expressed as 
\be
\hat H_e=\sum_{j}E_j |s_j\rangle\langle s_j|+\sum_j\sum_{k\neq j} J_{jk}|s_j\rangle\langle s_{k}| \ ,\label{eq:eh1}
\ee
where $|s_j\rangle$ represents an exciton localized at site $j$, namely, an individual pigment or chromophore in LHCs, with $E_j$ as the corresponding energy.  The term $J_{jk}$ represents the electronic coupling between site local exciton states $|s_j\rangle$ and $|s_k\rangle$. In general, this is a sum of Coulomb and exchange interactions between the two site excitation states.  Another commonly accepted assumption of Frenkel-type exciton, although not essential, is that $|s_j\rangle$'s form an orthogonal basis.  This assumption can be justified for most LHCs, where different pigment molecules are separated well enough to have negligible overlap integrals between respective electronic excited states.   Under these assumptions, the exciton Hamiltonian can be diagonalized easily as 
\be
\hat H_e=\sum_j {\mathcal E}_j|\varphi_j\rangle\langle \varphi_j| \ , \label{eq:eh1-diagonal}
\ee
where ${\mathcal E}_j$ is the eigenvalue and $|\varphi_j\rangle$ is the corresponding eigenstate.  This is related to the site excitation state through a unitary transformation with matrix element,  $U_{jk}=\langle s_k|\varphi_j\rangle$, as follows:
\be
|\varphi_j\rangle =\sum_{k} U_{jk} |s_k\rangle\ . \label{eq:exciton_s}
\ee

For Frenkel excitons, delocalization occurs only through a quantum mechanical superposition of site-local exciton states, $|s_j\rangle$'s.  In other words, the extent of the superposition represented by Eq. (\ref{eq:exciton_s}) is solely dictated by the magnitudes of $J_{jk}$'s relative to those of $\delta E_{jk}$'s, where $\delta E_{jk}=E_j-E_k$.   
However, as in most condensed phase molecular systems, Eq. (\ref{eq:eh1}) or (\ref{eq:eh1-diagonal}) alone is severely limited in representing the full quantum nature of excitons in LHCs.  Other factors such as molecular vibrations and polarization response of protein environments have to be taken into consideration as well.  This requires consideration of additional Hamiltonian terms.  

Let us introduce ${\bf X}$ as a collective variable representing all the degrees of freedom coupled to the exciton states.  These include the nuclear degrees of freedom of pigment molecules as well as protein residues, and the  environmental electronic degrees of freedom that do not participate directly in the electronic excitations of pigment molecules.  Then, the total Hamiltonian representing the entirety of the exciton and its environments (the bath) can be expressed  as follows:
\be
\hat H_{ex}=\hat H_e+\hat H_{eb}({\bf X})+\hat H_b({\bf X})\ ,  \label{eq:hex}
\ee 
where $\hat H_b({\bf X})$ is the bath Hamiltonian and $\hat H_{eb}({\bf X})$ is the exciton-bath Hamiltonian that can be expressed as
\ben
&&\hat H_{eb}({\bf X})=\sum_{j}\sum_{k}\hat B_{jk}({\bf X})|s_j\rangle \langle s_k| \nonumber \\
&&=\sum_{j}\sum_{k}\left (\sum_{j'}\sum_{k'}U_{jj'}^*\hat B_{j'k'}({\bf X})U_{kk'}\right)|\varphi_j\rangle\langle\varphi_k| \ .
\een
All the parameters entering the Hamiltonian $\hat H_{ex}$ of Eq. (\ref{eq:hex}),  $\delta E_{jk}$'s, $J_{jk}$'s, and $\hat B_{jk}({\bf X})$'s, affect the nature of excitons directly.  In addition,  some features of $\hat H_b({\bf X})$ may also have indirect influence on the dynamics of excitons.    In general, information on all of these is necessary for quantitative characterization of excitons. This in turn leads to reliable assessment of the overall exciton migration mechanism and helps elucidate the design principle of each LHC.    

However, establishing a satisfactory level of information on all the components of $\hat H_{ex}$ of Eq. (\ref{eq:hex}) has remained difficult for most LHCs.  This has been true even for well known LHCs such as the Fenna-Matthews-Olson (FMO) complex \cite{fenna-nature258} of green sulfur bacteria and the light harvesting 2 (LH2) complex of purple bacteria \cite{hu-qrb35,cogdell-qrb39}, for which the X-ray crystal structures \cite{fenna-nature258,fenna-bbrc75,mcdermott-nature374,koepke-structure4} were discovered more than two decades ago. Even when main features of the exciton Hamiltonians of these systems were understood, much still needed to be established regarding the details of the full exciton bath Hamiltonian.  As a result, theoretical interpretations of many spectroscopic data have remained ambiguous for a long time. 

The past decade has seen a surge of new efforts in the spectroscopy, computational modeling, and theoretical description of delocalized excitons in LHCs.   This has become possible through advances in nonlinear electronic spectroscopy with femtosecond time resolution, improvement in computational capability, and advances in quantum calculations.   Another important source of motivation that has inspired a large group of recent works was the suggestion \cite{engel-nature466,collini-nature463,scholes-jpcl1} that coherent quantum dynamics of excitons might play much more central role than what had been perceived before.  While definite assessment of this suggestion remains open, especially under natural light condition, it is true that various attempts to understand the role of quantum coherence have led to significant advances in the spectroscopy and theoretical description of excitons in LHCs. 

A fundamental feature being shared by many LHCs, which has become clearer during the past decade, thanks to a large body of newly gathered information, is the intermediate nature of the terms constituting $\hat H_{ex}$. In other words, for LHCs, it is common that  $\delta E_{jk}$'s, $J_{jk}$'s, and $\hat B_{jk}({\bf X})$'s constituting the Hamiltonian are of comparable magnitudes.   Typically, these parameters are on the same order as those of the room temperature thermal energy and also of the major portion of the energy spectrum comprising $\hat H_b({\bf X})$.  The energetic convergence of these multiple terms endows LHCs with a rich repertoire of pathways and mechanisms for exciton dynamics and energy harvesting.  At the same time, the lack of apparent small parameters in these systems renders simple perturbation theories to be rather unreliable quantitatively. Therefore, advanced levels of quantum dynamics theories and computational approaches have become necessary for accurate description of exciton dynamics and relevant spectroscopic observables.   In this sense, delocalized excitons in LHCs have served as prominent testing cases of modern quantum dynamics, electronic structure calculation approaches, and spectroscopic methods.      

The objective of this review is to provide a self-contained exposition of excitons in LHCs, with particular attention to the sources and implications of their quantum delocalization.  For this, we first introduce three major LHCs for which sufficient level of experimental information is available and extensive theoretical and computational studies have also been made. In order to clarify assumptions behind prevailing models of excitons in LHCs, we provide a derivation of the commonly used form of exciton-bath Hamiltonian, and make  critical assessment of underlying assumptions. Then, we offer a comprehensive overview of major computational and quantum dynamical methods to study and model these excitons, and summarize applications of these methods to the three representative systems.  While this review is devoted more to theoretical and  computational aspects as outlined above, we also provide a wide range of experimental results that are crucial for understanding the energetics and the dynamics of excitons.  We also discuss features that are deemed important for their functionality.  Ultimately, this work offers both rigorous theoretical framework and comprehensive information that can stimulate  future endeavor to 
elucidate important quantum mechanical design principles behind efficient and robust harvesting of excitons by LHCs.

\section{Overview of major light harvesting complexes}

\begin{figure}[hptb]
\includegraphics[width=3.6in]{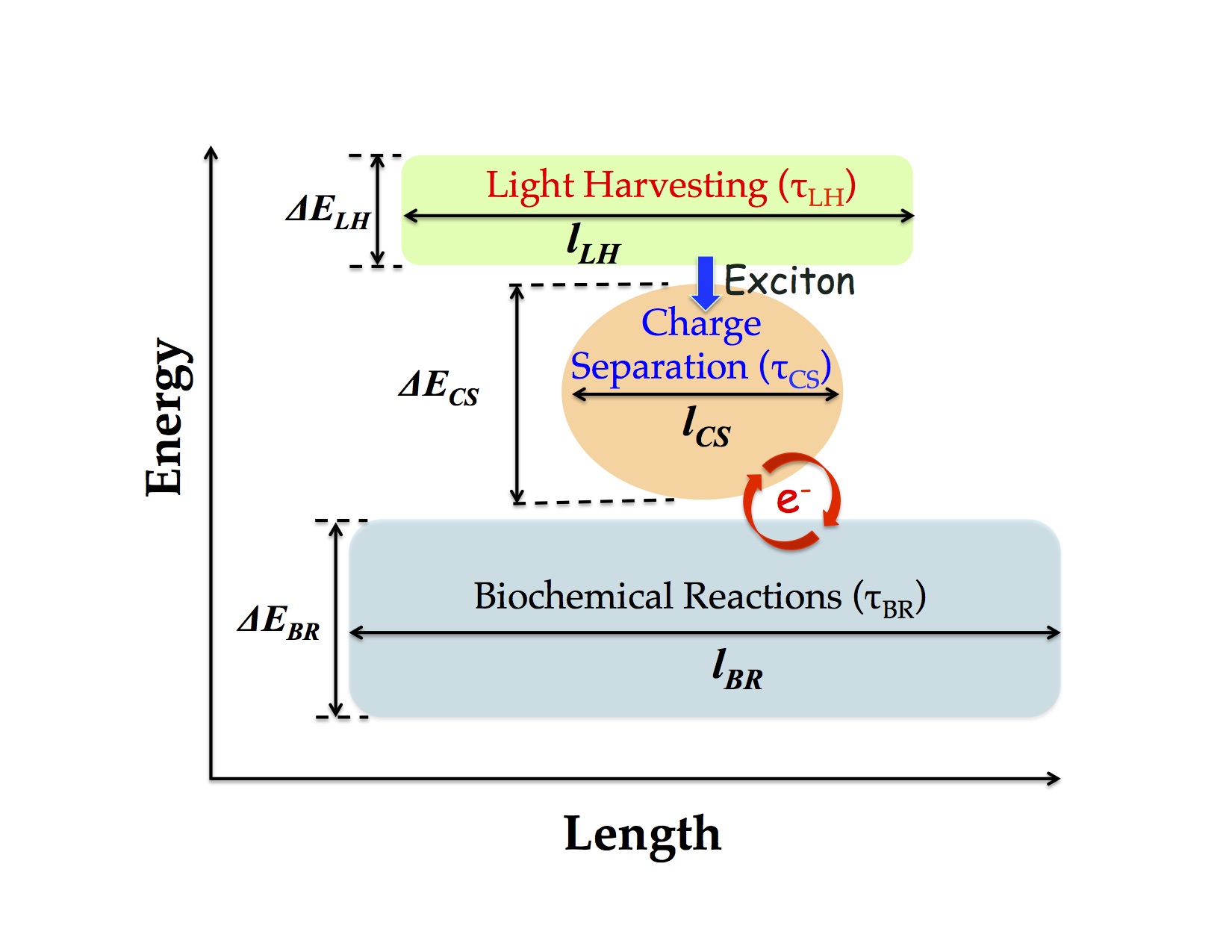}
\caption{A schematic of three domains constituting photosynthesis, performing light harvesting (LH), charge separation (CS), and biochemical reaction (BR).  Their respective time scales are denoted as $\tau_{LH}$, $\tau_{CS}$, and $\tau_{BR}$. The length scales of these domains are denoted as $l_{LH}$, $l_{CS}$, and $l_{BR}$, respectively.  The ranges of energy changes that occur in these domains are respectively denoted as $\Delta E_{LH}$, $\Delta E_{CS}$, and $\Delta E_{BR}$.  }
\label{fig:photosyn_sch}
\end{figure}

\begin{table}
\caption{Time, length, and energy scales of photosynthesis.}
\begin{tabular}{ccc|ccc|ccc}
\hline
\hline
$\tau_{LH}$&$\tau_{CS}$&$\tau_{BR}$& $l_{LH}$&$l_{CS}$&$l_{BR}$&$\Delta E_{LH}$&$\Delta E_{CS}$&$\Delta E_{BR}$\\
\hline
$100{\rm ps}$ & $1{\rm ns}$ & $1{\rm  \mu s }$ & 100nm & 10nm & $1{\rm  \mu m}$& $0.1{\rm eV}$ &$0.5 {\rm eV}$ & $1 {\rm eV}$ \\
\hline
\hline
\end{tabular}
\label{ps-parameters}
\end{table}

Photosynthesis consists of three distinctive processes as shown schematically in Fig. \ref{fig:photosyn_sch}, which we  call here as light harvesting (LH), charge separation (CS), and biochemical reaction (BR).  Although it remains difficult to make definite assessment of the length, time, and energy scales of all the photosynthetic processes, it is possible to offer rough estimates of these parameters based on the observation of many systems studied so far, which are listed in Table \ref{ps-parameters}. There is amazing diversity in how each of these processes is executed and how the three processes are integrated together, as has been reviewed comprehensively down to molecular level  \cite{blankenship-2}. 
Recent efforts to model the chromatophore of purple bacteria also offer unique insights into the synergistic organization of all the photosynthetic processes \cite{cartron-bba1837,strumpfer-jpcl3,sener-elife5}.     In this review, we will focus on only one of the three processes, namely the LH process.  In this section, we present a general introduction of three major LHCs, Fenna-Matthews-Olson (FMO) complex of green sulfur bacteria, light harvesting 2 (LH2) complex of purple bacteria, and phycobiliproteins (PBPs) of cryptophyte algae.

\subsection{FMO complex of green sulfur bacteria}

\begin{figure}[htbp]
  \includegraphics[width=3.5 in]{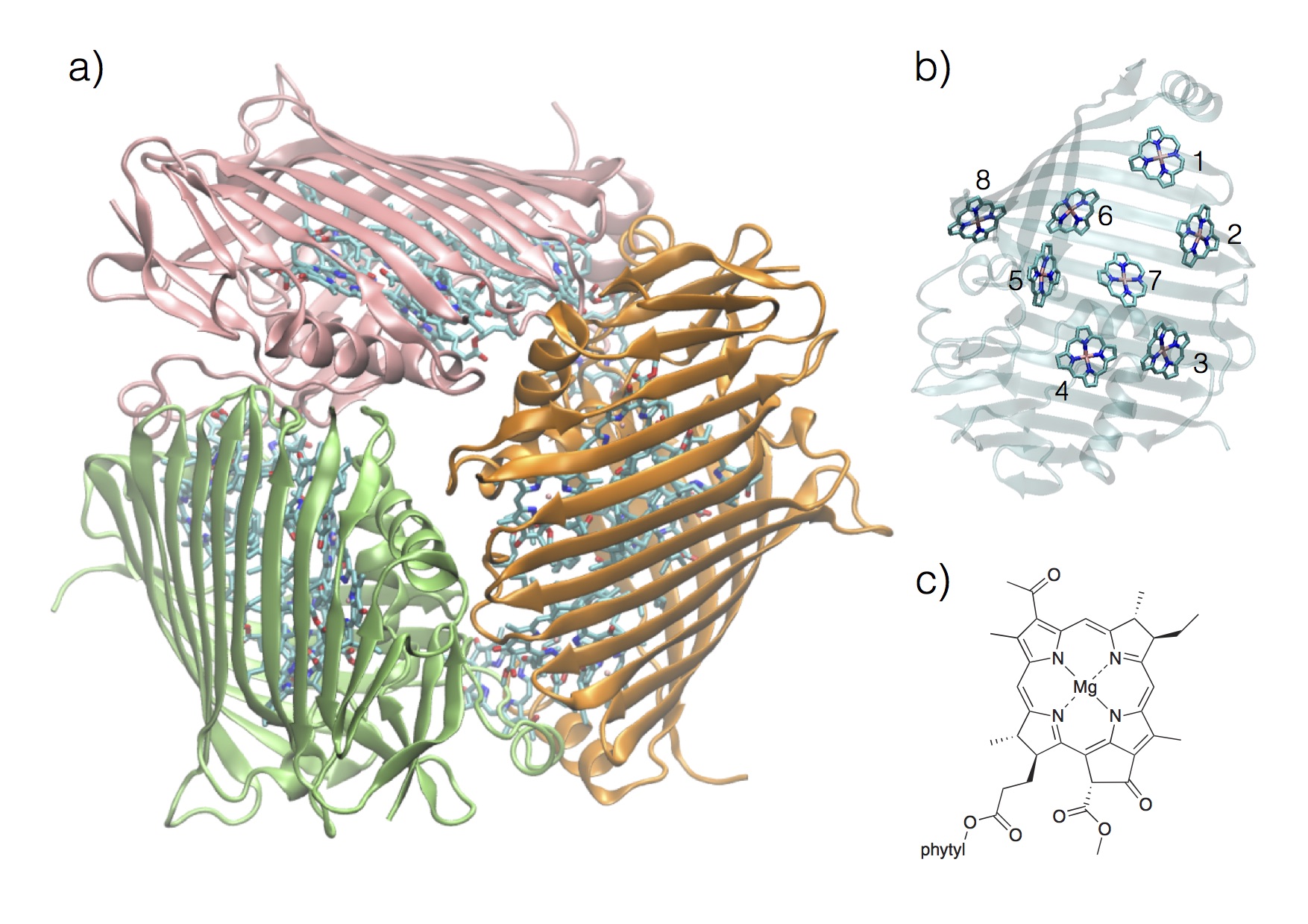}
      \caption{(a) FMO trimer complex. (b) View of the eight BChls of the monomer unit A (only the heavy atoms of the BChls ring are shown). (c) Chemical structure of the BChl pigment.}
  \label{fig:FMO-Fig1}
\end{figure}

FMO complex is an LHC found in green sulfur bacteria \cite{blankenship-2} and serves as the conduit for excitons migrating to reaction centers from a large chlorosome super-complex. Each functional unit of such a conduit is formed by a trimer of FMO complexes assembled in ${\rm C_3}$ point group symmetry, although actual pathway of each exciton appears to be confined to one complex for each transfer.   There are 7 well-established bacteriochlorophyll-$a$ (abbreviated as BChl hereafter) forming each FMO complex, which had long been believed to form a complete set of pigment molecules.  However, recent findings and analyses confirmed \cite{tronrud-pr100,busch-jpcl2} the existence of the eighth BChl as well.  This serves as a linker molecule to the chlorosome, a super-aggregate of BChls serving as the major LHC of the green sulfur bacteria, and mediates initial inception of the exciton into the FMO complex.   
Figure \ref{fig:FMO-Fig1} shows the trimer and the monomer unit of the FMO complex and the constituting BChl.

The FMO complex has historical significance because it is the first LHC for which X-ray crystallography structural data became available \cite{fenna-nature258,fenna-bbrc75} and has been subject to extensive spectroscopic and computational studies since then.  However, its functional significance as an LHC was considered rather minor.  
FMO complex does not have an apparent internal symmetry in its arrangement of pigment molecules.  Thus, it was a nontrivial task to assign complete details of  its exciton states and to determine the extent of their couplings to protein environments.   Earlier spectroscopic and computational investigations \cite{louwe-jpcb101,wendling-pr71,vulto-jpcb102,renger-pr343,cho-jpcb109}  thus focused on quantifying such parameters, as will be described in more detail in the next section.

The zeroth order exciton Hamiltonian of the FMO complex can be expressed through Eq. (\ref{eq:eh1}) where the sum runs over the 8 BChl molecules.
Tables II and III provide representative model parameters for two different species, {\it P. aestuari} and {\it C. tepidum}, with known X-ray crystallography structural data. 
The consensus established through the earlier works was that excitons formed in the FMO complex are modestly delocalized \cite{cho-jpcb109} (up to 2-3 BChl molecules at most) and are weakly coupled to protein environments.    These assessments were also consistent with the functional role of the FMO complex as a ``drain" of highly delocalized excitons of the chlorosome into much more localized ones at the reaction center.  

\begin{table}
\caption{Excitation energies (in ${\rm cm^{-1}}$) of BChls in the FMO complex, for sets A1 \cite{wendling-pr71}, A2 \cite{adolphs-bj91}, A3 \cite{busch-jpcl2}, T1 \cite{vulto-jpcb102}, and T2 \cite{adolphs-bj91}  \label{table-fmo-energies}}

\begin{tabular}{c|c|c|c|c|c}
\hline
\hline
&\multicolumn{3}{c|}{\it P. aestuari}&\multicolumn{2}{c}{\it C. tepidum} \\
\hline
&\makebox[.5in]{A1} & \makebox[.5in]{A2}&\makebox[.5in]{A3}& \makebox[.5in]{T1} & \makebox[.5in]{T2}  \\
\hline
$E_3$&12160&12230 &12195&12140&12210\\
\hline
$E_1-E_3$&190&215 & 200& 260&200\\
\hline
$E_2-E_3$&305&220 & 230 & 460&320\\
\hline
$E_4-E_3$&190&125 & 180 &140 &110\\
\hline
$E_5-E_3$&440&450 & 405 &360&270\\
\hline
$E_6-E_3$&320&330 & 320 &360&420\\
\hline
$E_7-E_3$&300&280 & 270 &290&230\\
\hline
$E_8-E_3$& & &505\\
\hline
\hline
\end{tabular}
\end{table}

\begin{table}
\caption{Electronic coupling constants of FMO complex.  The labels for sets are the same as Table \ref{table-fmo-energies}. \label{table-fmo-couplings}}
\begin{tabular}{c|c|c|c|c|c|c}
\hline
\hline
\makebox[.2in]{$j$}&\makebox[.2in]{$k$}&\multicolumn{5}{c}{$J_{jk}\ ({\rm cm^{-1}})$}\\
\hline
&&\multicolumn{3}{c|}{\it P. aestuari}&\multicolumn{2}{c}{\it C. tepidum} \\
\hline
& &\makebox[.5in]{A1} & \makebox[.5in]{A2}&\makebox[.5in]{A3}& \makebox[.5in]{T1} & \makebox[.5in]{T2}  \\
\hline
1&2&-102&-98.2&-94.8&-106&-87.7  \\
\cline{2-7}
&3&6&5.4&5.5&8&5.5\\
\cline{2-7}
&4&-6&-5.9&-5.9&-5&-5.9\\
\cline{2-7}
&5&7&7.1&7.1&6&6.7\\
\cline{2-7}
&6&-15&-15.2&-15.1&-8&-13.7\\
\cline{2-7}
&7&-14&-13.5&-12.2&-4&9.9\\
\cline{2-7}
&8&&&39.5& &\\
\hline
2&3&32&30.5&29.8&28 &30.8\\
\cline{2-7}
&4&8&7.9&7.6&6 &8.2\\
\cline{2-7}
&5&1&1.4&1.6&2 &0.7\\
\cline{2-7}
&6&14&13.1&13.1&13 &11.8\\
\cline{2-7}
&7&9&8.5&5.7&1 &4.3\\
\cline{2-7}
&8&&&7.9& &\\
\hline
3&4&-56&-55.7&-58.9&-62&-53.5 \\
\cline{2-7}
&5&-2&-1.8&-1.2&-1 &-2.2\\
\cline{2-7}
&6&-10&-9.5&-9.3&-9 &-9.6\\
\cline{2-7}
&7&2&3.1&3.4&17&6.0\\
\cline{2-7}
&8&&&1.4& &\\
\hline
4&5&-69&-65.7&-64.1&-70&-70.7\\
\cline{2-7}
&6&-19&-18.2&-17.4&-19&-17.0\\
\cline{2-7}
&7&-60&-58.2&-62.3&-57&-63.3\\
\cline{2-7}
&8&&&-1.6& &\\
\hline
5&6&89&88.9&89.5&40&81.1\\
\cline{2-7}
&7&-4&-3.4&-4.6&-2 &-1.3\\
\cline{2-7}
&8&&&4.4& &\\
\hline
6&7&37&36.5&35.1& 32 &39.7\\
\cline{2-7}
&8&&&-9.1&&\\
\hline
7&8&&& -11.1&&\\
\hline
\hline

\hline

\hline
\end{tabular}
\end{table}

In 2007, a 2-dimensional electronic spectroscopy (2DES) measurement \cite{engel-nature466} reported direct time domain observation of off-diagonal signals (in the 2DES frequency plot of signals) with beating in time that lasts more than ${\rm 500\  fs}$ at 77 K, which was suggested as being purely due to electronic coherence.   This was viewed by many researchers as surprising because direct observation of electronic coherence surviving more than ${\rm 100\ fs}$ was rare in such disordered environments.   However, considering relatively small magnitudes of electronic couplings (about $30 - 80 \ {\rm cm^{-1}}$) among BChl molecules and weak couplings of their electronic excitations to molecular vibrations or surrounding protein environments, it was not unreasonable to expect the existence of electronic coherence with such time scale.  This also appeared to be consistent with an earlier interpretation of the oscillating anisotropy from a pump-probe spectroscopy as being due to electronic coherence \cite{savikhin-cp223}.  

However, for the electronic coherence to be detected as real time coherent signals,  dephasing due to the inhomogeneity should be eliminated successfully.   Whether the 2DES signal observed \cite{engel-nature466} indeed represents such condition was not clear initially, let alone the question of whether the coherent signals can indeed be confirmed and validated by other independent measurements \cite{thyrhaug-jpcl7,duan-pnas114}.    Another  key suggestion they made \cite{engel-nature466} was that 
 ``wavelike energy motion owing to long-lived coherence terms"  may be  active for the exciton dynamics rather than ``semiclassical hopping mechanism," reviving  
an old debate \cite{knox-pr48,fleming-nature431}.  They also alluded that the FMO complex is implementing a kind of quantum search algorithm.  This suggestion drew particular attention of the quantum information community and motivated exploring LHCs as natural quantum information processing machines.

During the past decade, the FMO complex has been subject to a new level of quantum dynamics studies, electronic structure calculations, all-atomistic simulations, and spectroscopic measurements.  These were particularly focused on (i) describing the exciton dynamics as accurately as possible while accounting for all major atomistic details of the system, (ii) rigorous understanding and modeling of the 2DES signals and their implications, (iii) and deeper understanding of how quantum coherence and other quantum features make positive contribution to the functionality of the FMO complex.

\subsection{LH2 complex of purple bacteria}

LH2 complex is the primary LHC responsible for harvesting and delivering excitons in purple photosynthetic bacteria.  
The first X-ray crystal structure \cite{mcdermott-nature374} was determined for a bacterium called {\it Rhodopseudomonas} ({\it Rps.}) {\it acidophila}, now reclassified as {\it Rhodoblastus acidophilus} \cite{cogdell-qrb39}, revealing a cylindrical form with ${\rm C_9}$ symmetry that contains three BChl molecules, two helical polypeptides, and a carotenoid in each symmetry unit. Soon after, a different LH2 complex with ${\rm C_8}$ symmetry called {\it Rhodospirillum } ({\it Rsp.}) {\it molischianum }, now reclassified as {\it Phaeospirillum molischianum} \cite{cogdell-qrb39},  was also reported \cite{koepke-structure4}.  Another well-known LH2 complex  from {\it Rhodobacter} ({\it Rb.}) {\it sphaeroides}, although its structure has not been determined, is often assumed to have almost the same structure as {\it Rps. acidophila.} All the LH2 complexes firmly confirmed to date are known to have $C_N$ symmetry with $N=8-10$ only \cite{cogdell-qrb39,cleary-pnas110}.   As an LHC with the highest level of symmetry while being finite in its size (cylindrical forms with both the diameter and height about $7 \ {\rm nm}$), LH2 has been subject to a wide range of spectroscopic studies \cite{sundstrom-jpcb103,cogdell-qrb39}. There have also been various theoretical and computational studies \cite{hu-qrb35} modeling/explaining spectroscopic data and also addressing important issues such as coherence length, dynamical time scales, and energetics of excitons.  

\begin{figure}[htbp]
(a)\makebox[3in]{} \\
\includegraphics[width=2.5 in]{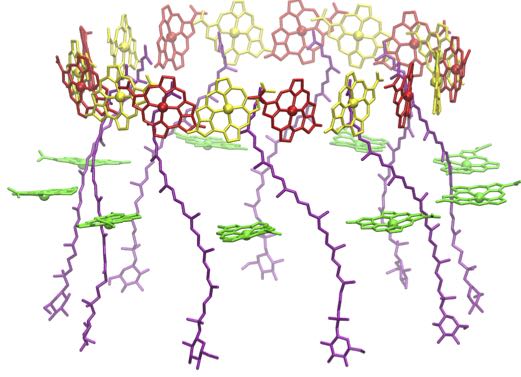} \\
(b)\makebox[3in]{ }\\
\includegraphics[width=2.5 in]{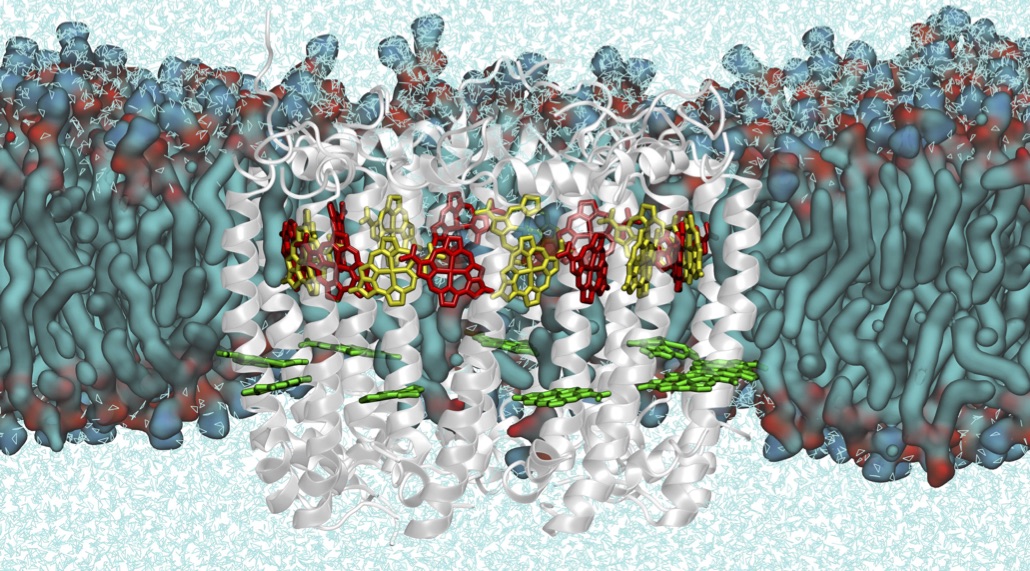} \\
(c)\makebox[3in]{ }\\
\includegraphics[width=2.7 in]{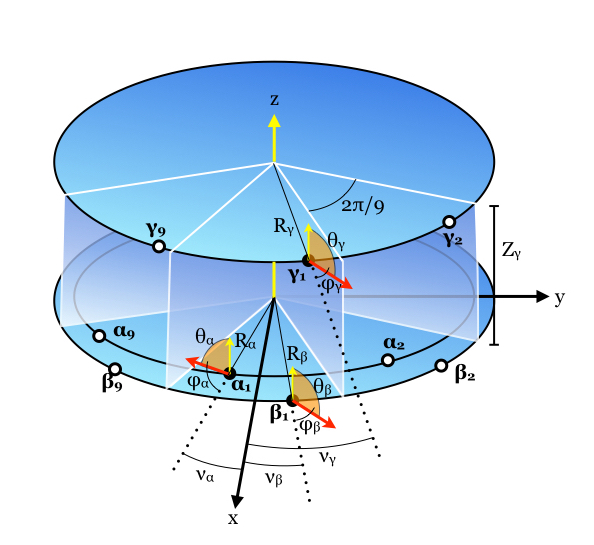}
\caption{(a) Side view of the pigment arrangement in LH2: $\alpha$ and $\beta$ BChl molecules of the B850 ring constitute the upper circular aggregate whereas $\gamma$ BChl molecules of B800 form the lower circular aggregate.  The carotenoids are also reported in purple. (b) Side view of LH2 within the its membrane from a snapshot of an all-atomistic MD simulation (the carotenoids are not shown).
(c) Geometric representation of the arrangement of the BChl molecules in B800 and B850 rings adapted with permission from \cite{montemayor-jpcb122}. Copyright 2018 American Chemical Society.  Here a reflection of 180 degrees has been used for clarity's sake leading the B800 ring above the B850 ring.}
\label{fig:lh2-image}
\end{figure}

Figure \ref{fig:lh2-image}(a) shows the structure of the complex from  {\it Rps. acidophila} without proteins, and Fig. \ref{fig:lh2-image}(b) depicts the whole complex embedded in lipid membrane environments.  Figure \ref{fig:lh2-image}(c) shows a geometric representation of LH2 (with the upside direction switched).   Out of the $3N$ BChl molecules constituting an LH2 with $N$-fold symmetry, $N$ of them in the cytoplasmic side form an exciton band near $800\ {\rm  nm}$ wavelength and the remaining $2N$ in the periplasmic side form an exciton band near $850\ {\rm nm}$ wavelength at room temperature.   Thus, the former is called B800 band and the latter B850 band.    The $N$ BChl molecules constituting the B800 band are arranged circularly, with their nearest neighbor distances about $2\ {\rm nm}$.  The $2N$ BChl molecules constituting the B850 band are also arranged circularly and are formed by two intervening concentric rings ($\alpha$ and $\beta$) of BChl molecules of similar radii.    In this B850 unit, the nearest neighbor (formed by two adjacent $\alpha$- and $\beta$-BChl pairs) distances are about $1\ {\rm nm}$. 

The exciton Hamiltonian of LH2 in the absence of disorder can be compactly expressed as \cite{jang-jpcb105,jang-jpcb111,montemayor-jpcb122} 
\begin{eqnarray}
 \hat H_{LH2}^0&=&\sum_{n=1}^N \left \{ E_\alpha |\alpha_n\rangle\langle \alpha_n|+E_\beta |\beta_n\rangle \langle \beta_n|+E_\gamma|\gamma_n\rangle \langle \gamma_n| \right . \nonumber \\
 &&\hspace{.2in}+J_{\alpha\beta}(0)(|\alpha_n\rangle\langle \beta_n|+|\beta_n\rangle\langle \alpha_n|)  \nonumber \\
 &&\hspace{.2in}+J_{\alpha\gamma}(0)(|\alpha_n\rangle\langle \gamma_n|+|\gamma_n\rangle\langle \alpha_n|) \nonumber \\
 &&\hspace{.2in}\left . +J_{\beta\gamma}(0)(|\beta_n\rangle\langle \gamma_n|+|\gamma_n\rangle\langle \beta_n|) \right\}\nonumber \\
 &&+\sum_{n=1}^N\sum_{m\neq n}^N \sum_{s,s'=\alpha,\beta,\gamma} J_{ss'}(n-m)|s_n\rangle\langle s'_m|\ , \label{eq:h_e0}
 \end{eqnarray}
where $|s_n\rangle$, with  $s=\alpha$, $\beta$, or $\gamma$, represents the excited state where only $s_n$ BChl is excited with its energy $E_s$, and $J_{ss'}(n-m)$ is the electronic coupling between $|s_n\rangle$ and $|s'_m\rangle$.    
The coordinates of BChl molecules and transition dipole moments can also be expressed as follows \cite{jang-jcp118-2,jang-jpcb111,montemayor-jpcb122}:
\begin{equation}
\label{Rxyz}
{\bf R}_{s,n}=\left (
\begin{array}{c}
R_{s}\cos\left [2\pi (n-1)/9+\nu_{s}\right ] \\
R_{s}\sin\left  [2\pi (n-1)/9+\nu_{s}\right ] \\
Z_s  \end{array}\right ) \ ,
\end{equation}
and
\begin{equation}
\label{muxyz}
\boldsymbol{\mu}_{s,n}=\mu_s
\left(
\begin{array}{c}
\sin\theta_{s}\cos[2\pi (n-1)/9+\nu_{s}+\phi_{s}]\\
\sin\theta_{s}\sin[2\pi (n-1)/9+\nu_{s}+\phi_{s}]\\
\cos\theta_{s}
\end{array} \right) \ ,
\end{equation}
where $s=\alpha,\beta,\gamma$ and $n=1,\cdots, 9$.  Table IV provides values of the above parameters for {\it Rps. acidophila}.
\begin{table}[htbp]
   \centering
   \caption{Geometric parameters representing the coordinates and the transition dipole orientations of BChls in {\it Rps. acidophila}.  Both values derived from the crystal structure and molecular dynamics (MD) simulation are shown \cite{montemayor-jpcb122}.} 
   \begin{tabular}{@{} c|c|c|c|c|c|c @{}} 
     \hline\hline 
     &\multicolumn{3}{c|}{Crystal Structure}&\multicolumn{3}{c}{MD Simulation}  \\
     \cline{2-7}
&\makebox[.3in]{$\alpha$} & \makebox[.3in]{$\beta$}& \makebox[.3in]{$\gamma$} &$\alpha$&$\beta$&$\gamma$  \\
 \hline
     $R_s$ (\AA) & 25.8 & 27.0 & 31.0 & 26.0  & 27.5 & 32.1 \\
     $Z_s$ (\AA) & 0 & 0 & 16.8 &0&0 & 16.5 \\
     $\theta_s$ ($^\circ$) & 94.9 & 95.8 & 97.3 & 96.5 & 97.3 & 98.2 \\
     $\nu_s$ ($^\circ$) & -10.2 & 10.2 & 23.6 &-10.2 &10.2 & 23.3 \\
     $\phi_s$ ($^\circ$) & -108.2 & 64.0 & 65.2 &-106.6 & 60.6 & 63.7 \\
      \hline
      \hline
   \end{tabular}
   \label{geometry}
\end{table}

Table V provides data for the electronic coupling constants calculated by various approaches, which will be explained in detail in Sec. IV.  For the B800 band, the magnitudes of the nearest neighbor electronic couplings are $\sim 20\ {\rm cm^{-1}}$.  These electronic couplings are estimated to be smaller, by about a factor of $2 \sim 3$, than the  disorder in the energy and the dynamic coupling to protein environments of the excitation of each BChl.  
For the B850 band, the magnitudes of electronic couplings between nearest neighbor BChls (formed by two adjacent $\alpha$ and $\beta$ BChl pairs) are known to be about $ 200 -300\ {\rm cm^{-1}}$.  These are comparable to the disorder in the excitation energy of each BChl and are somewhat larger than the reorganization energy of the coupling between the excitation of a BChl to its environment.    Therefore, excitons in this B850 unit are more delocalized than those in the B800 unit, although the debate on the coherence lengths of excitons in both units still remain somewhat unsettled.  In particular, the B850 exciton serve as a unique example of intermediate coupling regime, where all major energetic and dynamical parameters are of comparable magnitudes.  This regime defies simple classifications normally applicable in limiting situations.  For example, there have been different estimates \cite{dahlbom-jpcb105,book-jpcb104,jang-jpcb105,trinkunas-prl86} of the delocalization lengths of excitons ranging from about 4 to the entire set of the BChl molecules.

\begin{table}[htbp]
   \centering
   \caption{Major electronic coupling constants among BChls in LH2 of {\it Rps. acidophila} (see Fig. 3 for notations). All values are in ${\rm cm^{-1}}$.  C-1: Transition dipole coupling data based on X-ray crystal structure \cite{sundstrom-jpcb103}.  C-2: Transition density cube method data based on X-ray crystal structure \cite{krueger-jpcb102}.  C-3: TD-DFT/MM data based on X-ray crystal structure \cite{sgatta-jacs139}. C-4: RASSCF/RASPT2 data with MM solvation based on X-ray crystal structure \cite{sgatta-jacs139}. C-5: TD-DFT data with MMPol solvation based on X-ray crystal structure \cite{cupellini-jpcb120}.  MD-1: TD-DFT data with MMPol based on structures from MD simulation \cite{cupellini-jpcb120}. MD-2: TrESP calculation data based on structures from MD simulation \cite{montemayor-jpcb122}.    }
\begin{tabular}{c|c|c|c|c|c|c|c}
     \hline\hline
     Parameter& C-1&C-2&C-3&C-4&C-5& MD-1 & MD-2\\
     \hline 
         $J_{\alpha\beta}(0)$&322&238&336&563&362&339&$245$ \\
     $J_{\alpha \beta}(1)$&288&213&288&474&409&317&$140$\\
      \hline 
     $J_{\alpha\alpha}(1)$&&-46&-91&-83&-87&$-66$&$-59$\\
     $J_{\beta\beta}(1)$&&-37&-63&-69&-59&$-51$&$-29$\\
     \hline 
     $J_{\beta\alpha}(1)$&&11&23&22&24&18&$14$\\
     $J_{\alpha\beta}(2)$&&13&24&26&25&20&$13$\\
     \hline
     $J_{\gamma\gamma}(1)$&-22&-27&-47&-46&-50&-32&$-25$\\
     $J_{\alpha\gamma}(1)$&&27&44&46&59&42&$28$\\
     $J_{\beta\gamma}(1)$&&23&$-11$&$-11$&$-6$&$-3$&$3$\\
     $J_{\gamma\alpha}(0)$&&-13&-19&-22&-20&-16&$-12$\\
     $J_{\gamma \beta}(0)$&&5&-8&-9&-12&-10&$-3$\\
     \hline
      \hline
   \end{tabular}
   \label{MDparams}
\end{table}

Both the B800 and B850 units serve as active absorbing units of photons of LH2, but the excitons formed in the B800 unit transfer quickly to the B850 unit, which serves as main reservoir and relayer of excitons.  Pump-probe and photon echo spectroscopy measurements \cite{jimenez-jpc100,pullerits-jpcb101} suggested that this transfer occurs within about $1\ {\rm ps}$ and is fairly insensitive to temperature.  This is much larger than theoretical estimates based on the F\"{o}rster theory \cite{forster-ap,forster-dfs,scholes-arpc54}, and has been the subject of theoretical and experimental studies \cite{mukai-jpcb-103,scholes-jpcb104,kimura-jpcb107,herek-bj78,jang-prl92,cheng-prl96,jang-jpcb111,novoderezhkin-jpcb107}.

Other important issues concerning the LH2 complex include understanding the role of carotenoids and their dark states, and elucidating the effects of hydrogen bonding on the structure and energetics of BChls.   More recently, the availability of the atomic force microscopy (AFM) images of membranes containing aggregates of LH2 complexes \cite{bahatyrova-nature430,Scheuring-Science309,Scheuring-JMB358,Scheuring-pr102} made it possible to investigate the nature of arrangement of LH2 complexes. Along with  theoretical studies \cite{cleary-pnas110,jang-jpcl6}, this information helps addressing the dynamics of excitons in the aggregates of LH2 and their physical implications.

\subsection{Phycobiliproteins (PBP) of cryptophyte algae}

Phycobiliproteins (PBPs) are primary LHCs of cryptophyte algae, a diverse group of eukaryotic single-cell photosynthetic organisms \cite{blankenship-2} that can be found in both sea and fresh water.  These organisms live under water where spectrum of incident light lacks the chlorophyll-absorbing region but still has significant intensities in blue and green spectral regions.   
The pigment molecules of PBPs are linear tetrapyrroles called bilins, which are covalently attached to the protein scaffold through single or double cysteine bonds. 
In the FMO  and LH2 complexes, tuning of excitation energies of pigment molecules and their electronic couplings are main factors dictating the excitonic features, respectively, as they contain a single type of pigment. In PBPs, however, different types of bilins can be found in the same complex, and the possibility to tune their protonation states or their conformation provides further mechanisms to adapt the spectral range for efficient light-harvesting.  Another distinctive feature is that pigment molecules are bound to proteins by covalent boding.   

\begin{figure}[htbp]
  \includegraphics[width=3.5in]{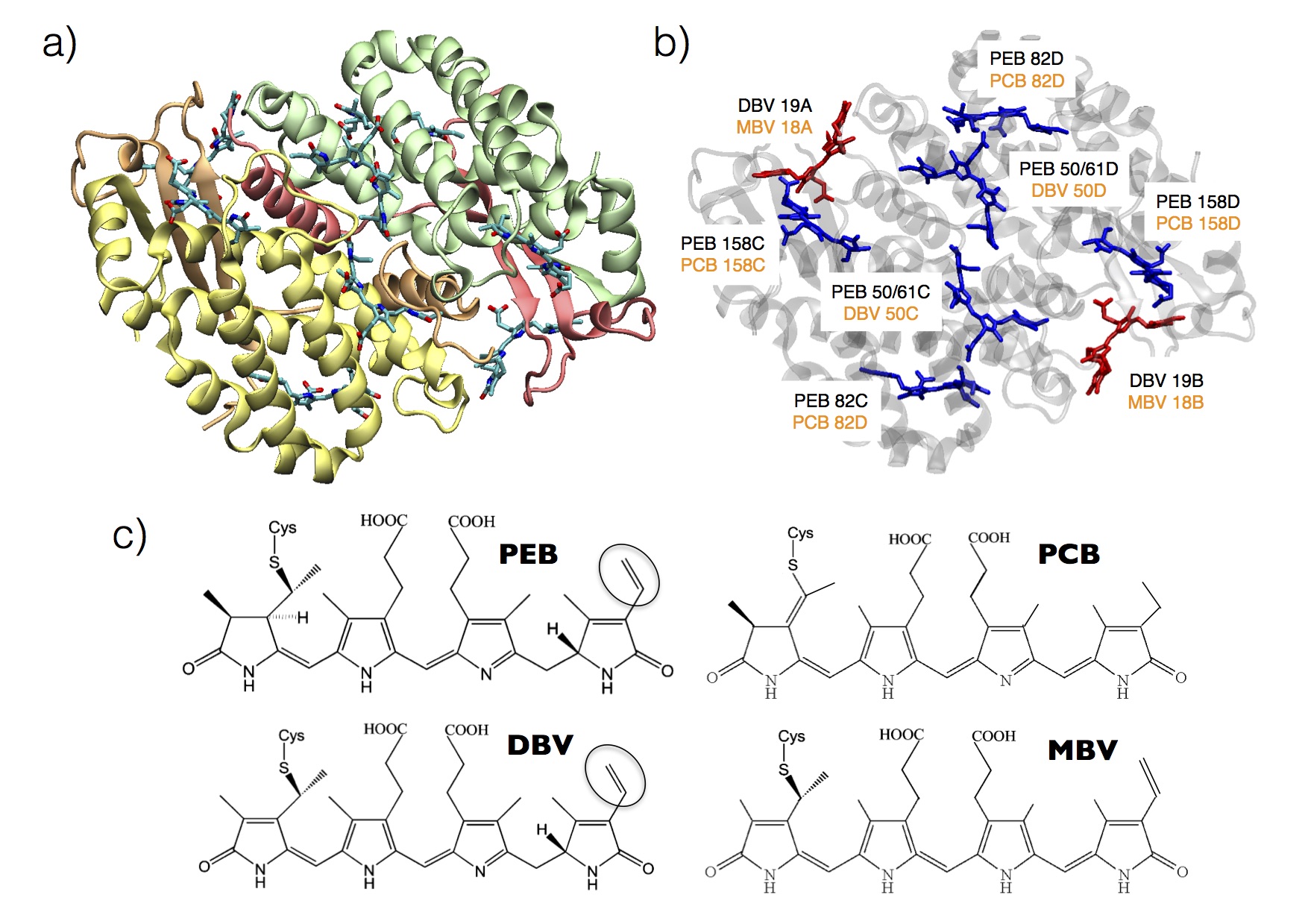}
   \caption{(a) PBP complex. (b) View of the eight bilins (only the heavy atoms are shown) present in PE545 (black) and PC645 (orange or grey in print). (c) Chemical structure of the  different bilins present in PE545 (PEB,DBV) and PC645 (PCB,DBV, MBV). The ellipses indicate the double bond which is used to create the second cysteine bond to the protein in PE545(PEB) and PC645(DBV).}
  \label{fig:PE545-Fig1}
\end{figure}

There are two well-known PBPs.  One is phycoerythrin 545 (PE545), the structure of which was first determined with rather low resolution \cite{wilk-pnas96}  and later with higher resolution \cite{doust-jmb344}.  The other is called phycocyanin 645 (PC645), the detailed structure of which was confirmed quite recently \cite{harrop-pnas111}. Figure \ref{fig:PE545-Fig1} shows the structures of PBPs and bilins. 

The capability of PE545 to photosynthesize under low-light conditions has been an important subject of both experimental \cite{doust-jpcb109,collini-nature463,wong-nchem4,harrop-pnas111}  and computational \cite{hossein-nejad-jpcb115,huo-jpcl2,curutchet-jacs133,curutchet-jpcb117,viani-jpcl4,aghtar-jpcl5,viani-pccp16,aghtar-jpcb121} studies.  A possible explanation of this efficiency is in terms of the flexible structural nature of the pigments.  This allows the optimal modulation of the absorption and energy transfer processes through local pigment-protein interactions. 
The crystal structure of PE545 contains eight bilins \cite{doust-jmb344}.  In particular, each $\alpha$ chain (A and B) contains a 15,16-dihydrobiliverdin (DBV), whereas each $\beta$ polypeptide chain (C or D) is linked to three phycoerythrobilins (PEB). The corresponding pigments are labeled ${\rm DBV_{19A}}$, ${\rm DBV_{19B}}$, ${\rm PEB_{158C}}$, ${\rm PEB_{158D}}$, ${\rm PEB_{50/61C}}$, ${\rm PEB_{50/61D}}$, ${\rm PEB_{82C}}$, and ${\rm PEB_{82D}}$, where the subscripts denote the protein subunit and cysteine residue linked to the chromophore. The central ${\rm PEB_{50/61D}}$ pigments are linked to the protein by two cysteine residues. The overall PE545 structure displays a pseudo-symmetry about the 2-fold axis relating the $\alpha_1\beta$ and $\alpha_2\beta$ monomers \cite{doust-jmb344}.

\begin{table}
\label{table-PE545-couplings}
\caption{Electronic coupling constants of PE545 and PC645. MD1: CIS/MMPol calculations on structures obtained from an MD simulation (couplings are obtained as a sum of Coulomb interactions of transition densities and an environment term) \cite{curutchet-jpcb117}. MD2: CIS/MM calculations on structures obtained from an MD simulation (couplings are obtained as dipole-dipole interactions of transition dipoles positioned at the center-of-mass of the bilin sites. The resulting couplings were then multiplied by a value of 0.72 to take into account screening effects from the environment.) \cite{lee-jacs139}. C1: CIS/PCM calculations on the crystal structure \cite{mirkovic-pps6,collini-nature463}.}
\begin{tabular}{c|c|c|c|c|c}
\hline
\hline
\makebox[.2in]{$s$}&\makebox[.2in]{$s'$}&\multicolumn{4}{c}{$J_{ss'}\ ({\rm cm^{-1}})$}\\
\hline
&&\multicolumn{2}{c|}{PE545} &\multicolumn{2}{c}{PC645} \\
\hline
& &\makebox[.5in]{MD1} & \makebox[.5in]{ MD2 }& \makebox[.5in]{C1 } & \makebox[.5in]{MD2}  \\
\hline
$H_1$&$H_2$& 71.7 & 163.9 & 319.4 & 212.3 \\
\cline{2-6}
&$M_1$& -21.5 & -29.5  &-43.9& -53.4\\
\cline{2-6}
&$M_2$& 24.5 & 20.6 & -9.6& -10.8\\
\cline{2-6}
&$M_3$& 34.0 & 38.8 &25.3& 34.9 \\
\cline{2-6}
&$M_4$& 12.1 & 17.9 &23.8& 31.9\\
\cline{2-6}
&$L_1$& 2.2 & 2.7 &-20.0& -19.9\\
\cline{2-6}
&$L_2$& -46.6 & -45.3 &-46.8& -43.4\\
\hline
$H_2$&$M_1$& -15.2 & -22.1& 7.7& 11.0\\
\cline{2-6}
&$M_2$& 19.1 & 29.8 &43.9& 49.0\\
\cline{2-6}
&$M_3$& -16.0 & -19.3 &29.5& 24.4\\
\cline{2-6}
&$M_4$& -35.6 & -38.1 & 30.5& 33.7\\
\cline{2-6}
&$L_1$& -39.3 & -47.8 &21.5& 19.9\\
\cline{2-6}
&$L_2$& 1.4 & 3.2 &48.0& 48.9\\
\hline
$M_1$&$M_2$& -6.1 & -6.3 &4.3& 3.4\\
\cline{2-6}
&$M_3$& 7.3 & 7.9 & 86.2& 77.6\\
\cline{2-6}
&$M_4$& 6.4 & 6.4 &3.4& 2.3\\
\cline{2-6}
&$L_1$& -27.3 & -37.4& 53.8& 69.3\\
\cline{2-6}
&$L_2$& -3.7 & -3.5 &-14.7& -16.0\\
\hline
$M_2$&$M_3$& 6.8 & 6.2 &-2.9& -2.0\\
\cline{2-6}
&$M_4$& 8.2 & 8.7&-86.7& -78.2\\
\cline{2-6}
&$L_1$& 3.5 & 3.6 &-15.8& -17.2\\
\cline{2-6}
&$L_2$& 26.3 & 38.7 &49.3& 66.8\\
\hline
$M_3$&$M_4$& 4.0 & 6.2 &7.8& 10.8\\
\cline{2-6}
&$L_1$& -11.4 & -13.0 &29.0& 11.8\\
\cline{2-6}
&$L_2$& -36.1 & -48.2 &-10.7& -12.4\\
\hline
$M_4$&$L_1$& 34.3 & 45.0 &11.0& 11.2\\
\cline{2-6}
&$L_2$& 11.6 & 13.8 &10.0& 10.7\\
\hline
$L_1$&$L_2$& -4.3 & -4.8 &48.0& 9.8\\
\hline
\hline

\hline

\hline
\end{tabular}
\end{table}

Though the protein scaffolds of the PE545 and PC645 complexes are nearly identical, the compositions of pigment molecules and their $\pi$-conjugation lengths are different.
In PE545, the lowest energy states are localized on the peripheral ${\rm DVB_{19}}$ molecules. For the PC645 complex, which functions at lower absorption energy, the DBV molecules reside at the dimer interface and play the role of the source states, while the phycocyanobilin (PCB) and the mesobiliverdin (MBV) molecules form lower energy intermediate and sink states, respectively \cite{mirkovic-pps6,collini-nature463,huo-jpcl2,lee-jacs139}.

We can express the exciton Hamiltonian for both PE5454 and PC645 commonly by abbreviating the two pigments with the highest excitation energies $H_1$ and $H_2$, the two lowest ones $L_1$ and $L_2$, and the four intermediate ones as $M_n$, $n=1,\cdots, 4$.  Thus, for PE545, ${\rm PEB_{50/61C(D)}}$ and DBVs are the $H$ and the $L$ pairs, respectively, while for PC645, DBVs are the $H$s and PCBs the $L$s.  
With these notations, the zeroth order exciton Hamiltonian for both can be expressed as  
\begin{eqnarray}
\hat H_{PBP}^0&=&\sum_{n=1}^2 \{E_{Hn} |H_n\rangle\langle H_n|+ E_{Ln} |L_n\rangle\langle L_n|\} \nonumber  \\
&+&\sum_{n=1}^4 E_{Mn}|M_n\rangle\langle M_n|  +\sum_{s}\sum_{s'\neq s} J_{s,s'} |s\rangle \langle s'| \ ,  \label{eq:h_e0_pbp}
\end{eqnarray}
where $s$ and $s'$ run over all possible sites.   
Table VI provides data for these electronic coupling constants calculated by various approaches. 
As LHCs with their structures known most recently among the three being considered here, spectroscopic data as well as the X-ray data played key role early on.  Detailed account of these spectroscopic studies are provided in the next section.  

\section{Spectroscopy of Excitons in Light Harvesting Complexes}
Absorption spectroscopy can identify distinctive signatures of LHCs in terms of the positions and widths of major excitonic peaks (see Fig. \ref{fig:ps_tree}(b)).  Given the structural information of an LHC, it is possible to assign some exciton states corresponding to the peaks of the absorption lineshape directly.  However, unambiguous modeling of the entirety of the absorption line shape was difficult in the beginning, and assignment of  some peaks, especially in large LHCs without apparent symmetry, still remains difficult.  Excitation energies of pigment molecules in LHCs can easily be modulated by pigment-protein interactions, typically in the range of $1,000\ {\rm cm^{-1}}$.    Despite significant advances, present day electronic structure calculation methods cannot yet reliably determine these modulations with sufficient efficiency and accuracy.  In addition, the three important factors influencing the absorption lineshape, namely, electronic couplings between pigment molecules, vibronic couplings, and the disorder/fluctuations are of comparable magnitudes in the range of $100 -500\ {\rm cm^{-1}}$, making it difficult to discern their effects on detailed features of the lineshape.    

Earlier efforts to interrogate the energetics and the dynamics of excitons in LHCs employed conventional sub-ensemble nonlinear spectroscopic techniques \cite{mukamel}.  For example, pump-probe and photon-echo spectroscopies \cite{sundstrom-jpcb103,cho-jpcb109} gave time resolved information on the response of a subensemble of excitonic states.  Hole burning spectroscopy \cite{purchase-pr101,jankowiak-cr111} allowed identification of narrow zero-phonon exciton states and their distributions hidden in the ensemble lineshape.  For LH2 and LH1 complexes of  purple bacteria and for other limited examples of LHCs, single molecule spectroscopy (SMS) \cite{oellerich-pr101,saga-jppc15,schlau-cohen-pnas110,vanoijen-science285} has also played an important role.   These nonlinear spectroscopic and SMS data helped extracting  information on exciton relaxation kinetics, homogeneous broadening, and the disorder/fluctuations.  However, they in general fell short of offering unambiguous interpretation of experimental signals for excitons in LHCs, the complexity of which typically causes multitudes of competing models/scenarios to be viable for interpreting the spectroscopic signals. 

Further advances in laser technology have made it possible to conduct a general 2DES \cite{jonas-arpc54,cho-cr108,fuller-arpc66} and allowed investigating new energetic and temporal details of exciton dynamics in LHCs \cite{ginsberg-acr42,schlau-cohen-cp386}.    2DES is a femtosecond pump-probe technique that resolves both pump and probe frequencies. It thus correlates the visible-light absorption spectrum, which enables related excitonic states to be identified by cross-peaks. 2DES is similar to transient absorption spectroscopy. However, two excitation pulses are used cooperatively to excite the sample followed by a third ``probe-pulse'' which interacts with the sample after the pump-probe time delay.   More detailed account of these spectroscopic studies are outlined below for the FMO complex, LH2 complex, and the PBPs of cryptophyte algae. 

\subsubsection{Spectroscopy on FMO complex}

\begin{figure}[htbp]
\includegraphics[width=3.2in]{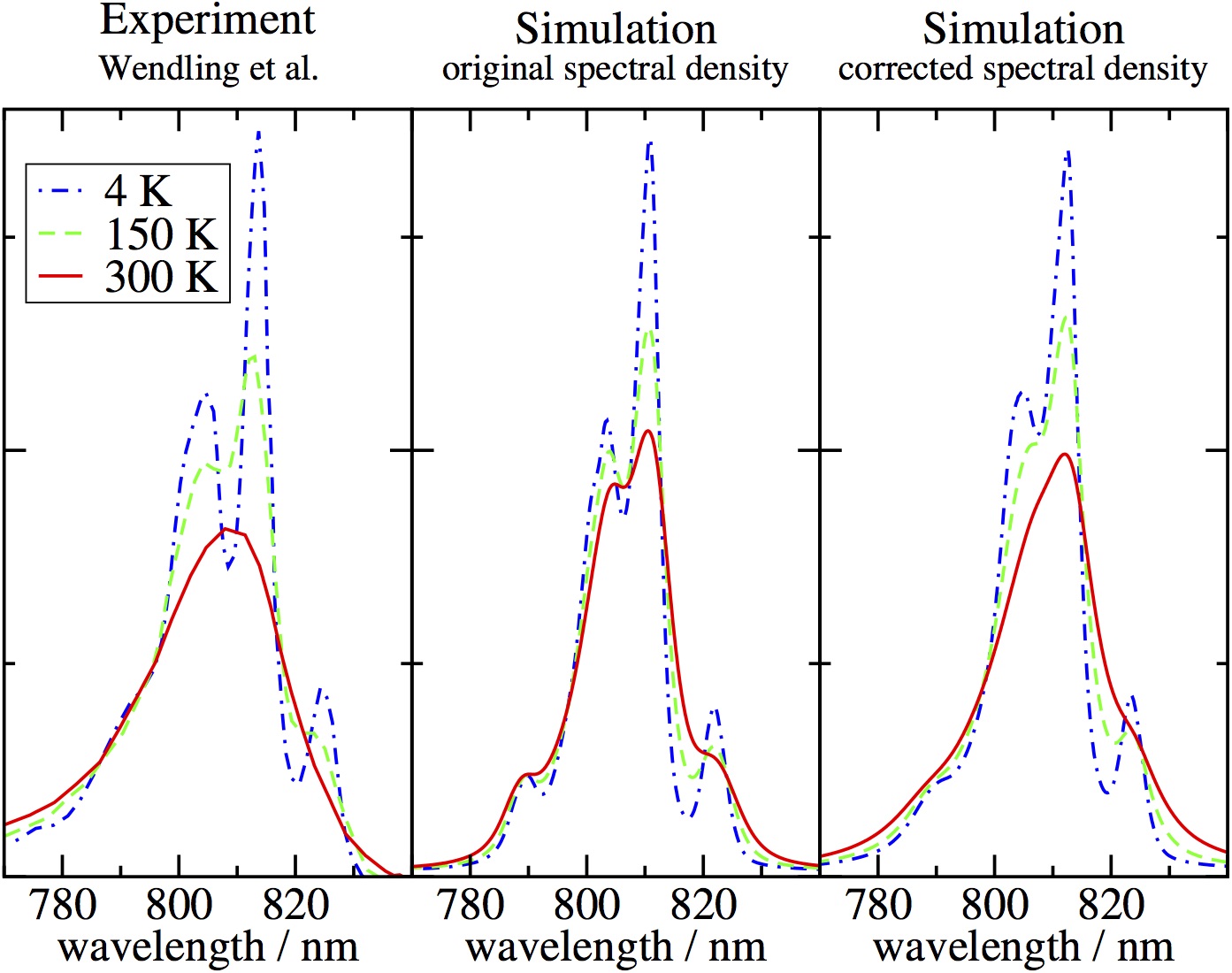}
\caption{Experimental and theoretical absorption spectra of FMO complex at three different temperatures adapted from \cite{renger-jpcb116} with permission. Copyright 2012 American Chemical Society. The two simulation data in the middle and right panel show that improvement in the bath spectral density brings qualitative features of theoretical line shapes closer to those of experimental ones.}
\label{fig:Renger-FMO}
\end{figure}

Already in 1990s, high quality steady state linear spectroscopic data for FMO complexes such as absorption, linear dichroism (LD), and circular dichroism (CD) became available \cite{louwe-jpcb101,wendling-pr71,vulto-jpcb102,miller-pr41,savikhin-pr48}.  Earlier efforts \cite{pearlstein-pr31,gulen-jpc100,louwe-jpcb101,wendling-pr71,vulto-jpcb102} to explain these spectroscopic data were based on simple exciton models of the form of Eq. (\ref{eq:eh1}), without including bath interactions. Thus, line broadening due to environmental relaxation of exciton states and inter-exciton dynamics were not properly taken into consideration in these works.  Nonetheless, the fittings of spectral data and the resulting model parameters turned out to be reasonable.  This indicates that the exciton-bath coupling and inter-exciton couplings in the FMO complex are not dominant factors.   Later, the quality of fitting was improved by including proper lineshape functions and utilizing more systematic fitting algorithms \cite{adolphs-bj91}.   The model parameters obtained from these fittings are compared in Tables II and III.  Figure \ref{fig:Renger-FMO} shows experimental absorption lineshapes of the FMO complex compared with a recent set of theoretical lineshapes \cite{renger-jpcb116}.  More detailed account of computational methods underlying these theoretical calculations will be provided in the later sections of this work. 

Subensemble nonlinear spectroscopies such as pump probe \cite{vanamerongen-jpc95,savikhin-biochemistry33,savikhin-pr48,buck-bj72,freiberg-jpcb101,savikhin-cp223}, hole burning \cite{johnson-jpc95,ratsep-jpcb103}, and photon echo \cite{prokhorenko-jpcb106} were employed to gain information on the dynamics and the energetics of exciton states that are not clearly visible in ensemble linear spectroscopic data.  These experiments revealed signatures of relaxation dynamics ranging from sub-hundred femtoseconds to hundred picoseconds. However,  definite assessment and quantitative quantum dynamical modeling of these data have not been pursued extensively.   

Advances in 2DES technique made it possible to access new information on exciton states of the FMO complex that had not been available otherwise. Earlier success was made in  
determining detailed excitonic pathways \cite{brixner-nature434,cho-jpcb109} that were largely consistent with a model exciton Hamiltonian developed earlier \cite{vulto-jpcb102,vulto-jpcb103}.
This was soon followed by direct observation of beating signals lasting more than 500 fs at 77 K \cite{engel-nature466}.  Engel and coworkers made further progress, and have reported more detailed experimental results \cite{hayes-njp12,hayes-bj100} including the evidence that beating signals can be observed even at room temperature \cite{panitchayangkoon-pnas107}.   While these were exciting results that have motivated a broad community of experimentalists and theoreticians, interpretation of such 2DES signals was not straightforward  because of the fact that the FMO complex has more than two exciton states and the exciton-bath couplings are more complicated than simple models that are typically used for analyzing 2DES signals.   There was strong possibility that the beating signals might have originated from any of electronic, vibrational, and vibronic contributions.  In fact, whether all of these contributions can be discerned by any 2DES was not clear at all \cite{butkus-cpl545}.   Thus, there has been ongoing debate and various analyses on the origins and sources of the beating signal \cite{christensson-jpcb116,tiwari-pnas110,fransted-jcp137,tempelaar-jpcb118,fujihashi-jcp142,liu-cp481,plenio-jcp139}.  Furthermore, such beating signals have not yet been independently confirmed in different 2DES data by other groups \cite{thyrhaug-jpcl7,duan-pnas114}.  
   
Due to the unique arrangement of pigment molecules in the FMO complex, different exciton states have well-defined directions of transition dipole moments. Thus, control and information of the light polarization can play an essential role in characterizing the dynamics directly associated with electronic coherence.  Indeed, utilization of the information on polarizations of light pulses has been shown to be critical in deducing structural data \cite{read-bj95}.     More recent work demonstrated that full polarization control allows accurate description of electronic structure and population dynamics \cite{thyrhaug-jpcl7}.

With further advances in spectroscopic and sample preparation techniques, efforts to clarify detailed characteristics of excitons have continued in different directions.  A new hole burning spectroscopy of FMO trimer and modeling in terms of excitonic calculations were reported \cite{kell-jpcl5}.  Transient absorption and time-resolved fluorescence spectroscopies were conducted for  several mutants of {\it C. tepidum} \cite{magdaong-jpcb121}.  A new 2DES spectroscopy provided evidence that FMO indeed functions as a conduit of exciton energy \cite{dostal-nchem8}.   A single molecule spectroscopy has finally become possible for the FMO complex \cite{lohner-sr6}.

\subsubsection{Spectroscopy on LH2}
Thanks to the high symmetry of the LH2 complex, the number of independent parameters minimally necessary for fitting its lineshape is relatively small compared to the total number of pigment molecules involved. Indeed, the average values of excitation energies of $\alpha$-, $\beta$-, and $\gamma$-BChls, nearest neighbor electronic couplings, and the magnitudes of the disorder for each of the B800 and B850 units seem sufficient for fairly accurate fitting of ensemble lineshape.  However, on the other hand, the easiness of fitting the ensemble lineshape has also been a contributing factor for the lack of clear consensus on some details of the exciton Hamiltonian.   

For example, modeling of both absorption and CD spectra \cite{georgakopoulou-bj82} based on the crystal structure support the assumption that the excitation energy of $\alpha$-BChl is about $300\ {\rm cm^{-1}}$ higher than that of $\beta$-BChl.  However, more recent computational studies \cite{cupellini-jpcb120,montemayor-jpcb122} suggest that the conformations and local environments of $\alpha$- and $\beta$-BChls are virtually the same in membrane environments without indicating significant difference in their excitation energies.    Efforts to refine the exciton-bath model of LH2 by combining more comprehensive spectroscopic data and advanced computational tools have continued, for example, based on temperature dependent absorption spectra \cite{urboniene-bj93,zerlauskiene-jpcbl112} and combination of temperature dependent absorption, fluorescence, and fluorescence-anisotropy \cite{pajusalu-cpc12}.   As yet, there has not been any attempt to explain all existing steady state spectroscopic data over all the temperatures based on a universal exciton model.

The determination of the X-ray crystal structures of LH2 complexes coincided with new advances in nonlinear spectroscopic theories and techniques.  Thus, LH2 soon became an important testing ground for evolving nonlinear spectroscopic techniques.  Along with the standard pump-probe  and hole burning spectroscopies, various versions of four wave mixing optical spectroscopy have been
used to interrogate mechanistic details and time scales of exciton dynamics \cite{sundstrom-jpcb103,lampoura-jpcb104}.  Ultrafast fluorescence up-conversion \cite{jimenez-jpc100}  provided information on time scales of exciton dynamics within B800 and B850 units.  It also showed that the time scale of the exciton transfer from the B800 unit to B850 at room temperature is about $1.5\ {\rm ps}$.  Pump-probe spectroscopies at a few different temperatures \cite{pullerits-jpcb101} also provided similar estimates, and demonstrated that the transfer time decreases with temperature moderately from 1.5 ps at 4.2 K to 0.7 ps at 300 K.  
Time resolved transient absorption spectroscopy of isolated and native membrane-embedded LH2 complexes of {\it Rb. sphaeroides} at 10 K \cite{timpmann-jpcb104} showed that the inter-LH2 exciton transfer time is larger than 1 ps. Three pulse photon echo peak shift suggested that \cite{agarwal-jpca106} the  intra-LH2 exciton relaxation takes about 200 fs, and  the inter-LH2 exciton transfer about 5 ps.   Combination of various spectroscopic techniques were made to interrogate the delocalization length of excitons and its dynamical localization with time scales less than 200 fs \cite{book-jpcb104}. 

As in the case of the FMO complex, real time coherent beating signal was observed in 2DES data of LH2 complex.  Angle resolved coherent four wave mixing spectroscopy showed evidence for coherent dynamics that can be well separated from the relaxation signal due to energy transfer \cite{mercer-prl102}. 2DES has also been used to determine parameters of the Hamiltonian for LH3 complex \cite{zigmantas-pnas103}, which is similar to LH2 and appears under low light condition.  Clear peaks corresponding to exciton transfer from the B800 unit to the B850 unit were shown to emerge even at about 200 fs after excitation \cite{harel-pnas109}. Recent 2DES spectroscopy suggested new aspects on the possible role of dark states of carotenoids \cite{ostroumov-science340} and dark charge transfer states of BChls \cite{ferretti-srep6} in energy transfer dynamics.

\begin{figure}[htbp]
\includegraphics[width=3.2in]{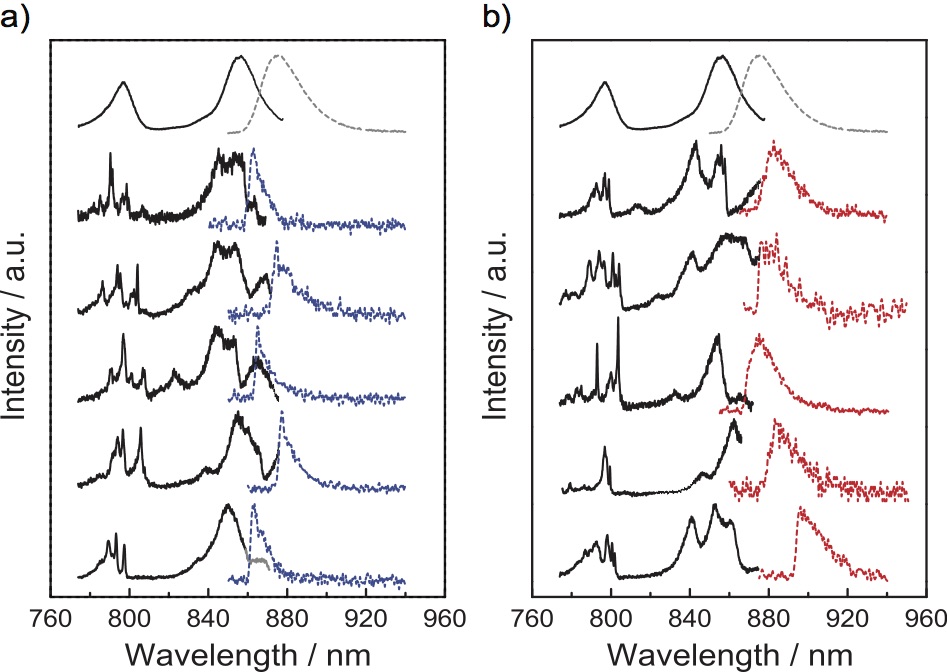}
\caption{Single LH2 fluorescence-excitation spectra (solid black lines)  and emission spectra (dashed blue and red lines, or dashed grey lines in print) provided by J\"{u}rgen K\"{o}hler.   The corresponding ensemble lineshapes are shown in the top.  The left panel is for LH2 complexes with sharp emission line shapes, and the right panel is for LH2 complexes with broad emission line shapes.  Detailed experimental method and more comprehensive set of data are available in \cite{kunz-jpcb116}.  }
\label{fig:Kohler-SMS}
\end{figure}

Hole burning spectroscopy \cite{purchase-pr101,wu-jpcb101-2,jankowiak-cr111} has long played a unique role in gaining quantitative information on the low lying exciton states, and has helped validating the exciton Hamiltonian model for LH2.  However, more direct experimental demonstration of the validity of the Frenkel exciton model for LH2 came from low temperature single molecule spectroscopy based on fluorescence-excitation technique \cite{vanoijen-science285,berlin-plr4,kunz-bj106,lohner-pr123,brotosudarmo-bj97,kunz-jpcb116}.   In particular, the SMS lineshapes for the B850 unit clearly consisted of two major exciton peaks ($k=\pm 1$), which are perturbed due to the disorder, and one or two minor exciton peaks at higher energies.   Figure \ref{fig:Kohler-SMS} shows line shapes reported from a recent SMS experiment \cite{kunz-jpcb116}.

Following the success of low temperature SMS experiments, room temperature SMS experiments were also conducted to investigate the nature and the dynamics of excitons at physiological temperature \cite{bopp-pnas94,bopp-pnas96}.   However, the noisiness of the data made it difficult to obtain any definite microscopic information.   More recently, room temperature single molecule emission spectroscopy of LH2 was shown to be able to identify three emissive states switching at room temperature \cite{schlau-cohen-pnas110}. Single molecule femtosecond pump probe spectroscopy employing ultrafast phase coherent excitation was also reported recently \cite{hildner-science340}.

How excitons migrate in aggregates of LH2 complexes has important implication for understanding the design principle of efficient energy conversion.  To this end, detailed information on the arrangement of LH2 complexes is needed. AFM images have revealed various patterns of  aggregates depending on light conditions \cite{Scheuring-Science309}. The extents of order and disorder also vary with other growth conditions and species \cite{sturgis-biochemistry48,olsen-jbc283}.  AFM images and pump probe spectroscopy of aggregates reconstituted in phospholipids have become available \cite{Sumino-jpcb117}.   Time resolved spectroscopy of exciton migration in arrays of LH2 complexes has also been reported \cite{pflock-jpcb115-2}.

\subsubsection{Spectroscopy on PBPs of cryptophyte algae}
\begin{figure}[htbp]
\includegraphics[width=3.5in]{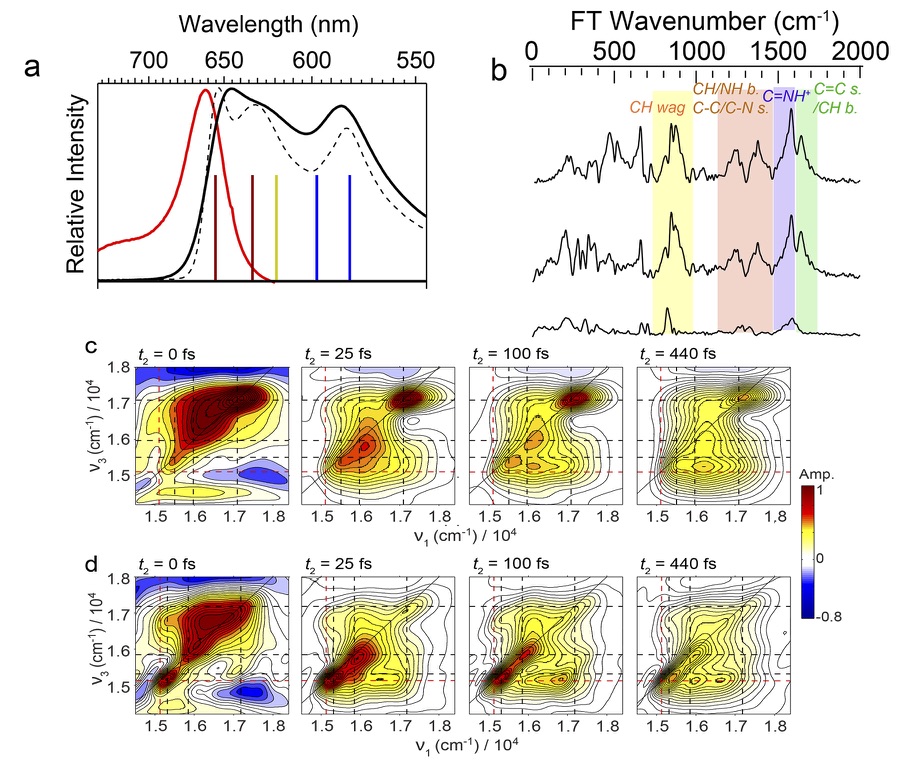}
\caption{Spectroscopic data for PC645 adapted from \cite{dean-chem1} with permission from Elsevier.  (a) Absorption lineshapes at $77\ {\rm K}$ (dashed black line) and ambient temperature (solid black line), and fluorescence lineshape (solid red line, or grey line in print) at ambient temperature. (b) Vibrational frequencies extracted from transient absorption by Fourier transform at three different emission frequencies. (c) 2DES taken at $295\  {\rm K}$.  $t_2$ refers to population time.  (d) 2DES taken at $77\ {\rm K}$.}
\label{fig:pbp-spectra}
\end{figure}
Linear ensemble spectroscopy, polarization anisotropy, and transient grating were employed early on as a tool to augment X-ray crystallography to construct an exciton model for PE545 \cite{doust-jmb344}.  Oscillations lasting up to about 1 ps were observed, which were assigned mostly coming from vibrational coherence \cite{doust-jmb344}. Transient absorption spectroscopy \cite{doust-jpcb109} also revealed broad time scales of energy transfer within the PE545 ranging from 250 fs to picoseconds.   A comprehensive set of absorption, CD, fluorescence, and time resolved transient absorption at both ${\rm 77\ K}$ and ${\rm 300 \ K}$ were reported \cite{novoderezhkin-bj99}.  This work also employed the experimental results to construct the exciton Hamiltonian and model exciton transfer kinetics.

Photon echo spectroscopy  \cite{collini-nature463} of both PE545 and PC645 at ${\rm 294 \ K}$ reported oscillatory cross peaks in the two dimensional representation 
lasting  more than $400\ {\rm fs}$, which were initially interpreted as originating from electronic coherence. Further interrogation of the sources and implications of these signals have continued, and ensuing studies \cite{turner-pccp14,mclure-jpcb118} clarified that most of them have vibrational origin, confirming the earlier suggestion \cite{doust-jmb344}.  In fact, this was not surprising considering the covalent nature of the pigment-protein bonding and rather strong vibronic coupling \cite{kolli-jcp137}.  Recent studies \cite{oreilly-ncomm5,dean-chem1} suggest that vibronic couplings are in fact being utilized positively for efficient and robust light harvesting capability.  Figure \ref{fig:pbp-spectra} provides the absorption and fluorescence lineshapes of PC645, vibrational spectra obtained from different sections of transient absorption, and 2DES profiles at two different temperatures.

\section{Exciton-bath Hamiltonian}
The fact that the Frenkel exciton Hamiltonian of Eq. (\ref{eq:eh1}) provides reasonable description of the electronic spectra of many LHCs was well established at phenomenological level.  However, computational efforts to validate its assumption and to determine the parameters directly through first principles calculation are fairly recent and are relatively in early stage.   For further progress in this direction, it is important to clarify assumptions and approximations involved in the exciton Hamiltonian, or more generally, the  exciton-bath Hamiltonian (EBH).  To this end, we here provide a comprehensive review of quantum mechanical assumptions implicit in the EBH typically used for LHCs and computational methods to calculate key elements of the EBH such as the excitation energies, electronic couplings, and the spectral densities of the bath.    

\subsection{Derivation}
\subsubsection{Aggregates of pigment molecules}
Consider an aggregate of pigment molecules or chromophores (pigment molecules or part of them) more generally. The total molecular Hamiltonian representing the aggregate can in general be expressed as 
\be
\hat H_c=\sum_{j=1}^{N_c} \hat H_{j}+\frac{1}{2}\sum_{j=1}^{N_c}\sum_{k\neq j}^{N_c}\hat H_{jk} \ , \label{eq:hc-def}
\ee
where $N_c$ is the total number of chromophores, $\hat H_j$ is the full molecular Hamiltonian of the $j$th chromophore, and $\hat H_{jk}$ is the interaction Hamiltonian between the $j$th and the $k$th chromophores. 

For the $j$th chromophore, we denote the positions and momenta  of the $i$th electron as ${\bf r}_{j,i}$ and ${\bf p}_{j,i}$, and those of the $l$th nucleus as ${\bf R}_{j,l}$ and ${\bf P}_{j,l}$. Given that the $j$th chromophore has $L_j$ nuclei and $N_j$ electrons, $\hat H_j$ in the above equation can be expressed as
\ben 
\hat H_{j}&=&\sum_{l=1}^{L_j}\frac{\hat {\bf P}_{j,l}^2}{2M_{j,l}}+\frac{1}{2}\sum_{l=1}^{L_j}\sum_{l'\neq j}^{L_j}\frac{Z_{j,l} Z_{j,l'} e^2}{|\hat {\bf R}_{j,l}-\hat {\bf R}_{j,l'}|} \nonumber \\
&+&\sum_{i=1}^{N_j}\frac{\hat {\bf p}_{j,i}^2}{2m_e} -\sum_{l=1}^{L_j}\sum_{i=1}^{N_j}\frac{Z_{j,l} e^2}{|\hat {\bf R}_{j,l}-\hat {\bf r}_{_j,i}|} \nonumber \\
&+&\frac{1}{2}\sum_{i=1}^{N_j}\sum_{i'\neq i}^{N_j}\frac{e^2}{|\hat {\bf r}_{j,i}-\hat {\bf r}_{j,i'}|} \nonumber \\
&=&\hat T_{j,n}+\hat V_{j,nn}+\hat T_{j,e}+\hat V_{j,en}+\hat V_{j,ee}\ ,  \label{eq:hj_def}
\een
where the terms following the second equality are abbreviations of the kinetic ($\hat T$) and potential ($\hat V$) energy operator terms.  The subscripts $n$ and $e$ in these terms respectively refer to nuclear and electron degrees of freedom. 
 
The second term in Eq. (\ref{eq:hc-def}) represents interactions between all chromophores. These are pairwise at the level of explicit description of all Coulomb interactions among electrons and nuclei.  Namely, each component, $\hat H_{jk}$, represents the interaction between the $j$th and the $k$th chromophores, and consists of four potential terms as follows: 
\ben 
\hat H_{jk}&=&\sum_{l=1}^{L_j}\sum_{l'=1}^{L_k}\frac{Z_{j,l} Z_{k,l'} e^2}{|\hat {\bf R}_{j,l}-\hat {\bf R}_{k,l'}|} \nonumber \\
&-&\sum_{l=1}^{L_j}\sum_{i=1}^{N_k}\frac{Z_{j,l} e^2}{|\hat {\bf R}_{j,l}-\hat {\bf r}_{k,i}|} -\sum_{l=1}^{L_k}\sum_{i=1}^{N_j}\frac{Z_{k,l} e^2}{|\hat {\bf R}_{k,l}-\hat {\bf r}_{j,i}|} \nonumber \\
&+&\sum_{i=1}^{N_j}\sum_{i'= 1}^{N_k}\frac{e^2}{|\hat {\bf r}_{j,i}-\hat {\bf r}_{k,i'}|}  \nonumber \\
&=&\hat V_{jk,nn}+\hat V_{jk,ne}+\hat V_{jk,en}+\hat V_{jk,ee} \ , \label{eq:hjk_def}
\een
where each term in the last line is again an abbreviation of the corresponding interaction potential term. 

One can employ the standard quantum mechanical procedure to derive an EBH corresponding to Eq. (\ref{eq:hj_def}).  Appendix A provides a detailed description of such procedure, based on the assumption that the ground electronic state and the site excitation state can be defined in terms of {\it direct products of adiabatic electronic states of independent chromophores defined at reference nuclear coordinates}.    The default choice for these reference nuclear coordinates are those of the optimized ground electronic states of chromophores.    Thus, the ground electronic state and the site excitation states are defined by  Eq. (\ref{eq:g-def}) and Eq. (\ref{eq:sj-def}).  Collecting all the terms in Appendix A, $\hat H_c$ can thus be expressed as 
\ben
\hat H_c&= &E_g^{c,0}|g\rangle\langle g|+\sum_{j=1}^{N_c}(\hat J_{jg}^{c,0}|s_j\rangle \langle g|+\hat J_{gj}^{c,0}|g\rangle \langle s_j|) \nonumber \\
&+& \sum_{j=1}^{N_c} E_j^{c,0} |s_j\rangle\langle s_j|+\sum_{j}\sum_{k\neq j} J_{jk}^{c,0}|s_j\rangle\langle s_k|\nonumber \\
&+&\sum_j \hat B_{j}^{c,0}|s_j\rangle\langle s_j| +\hat H_{b}^{c,0}  \ .\label{eq:h_c_ebh}
\een
In the above expression,  $E_g^{c,0}$ is the energy of the ground electronic state and is defined by Eq. (\ref{eq:def-egc}).  $\hat J_{jg}^{c,0}$   and $\hat J_{gj}^{c,0}$ are  the electronic couplings  between the site excitation state $|s_j\rangle$ and the ground electronic state $|g\rangle$, and are defined by Eqs. (\ref{eq:j_jgc}) and (\ref{eq:j_gjc}).  These terms originate from interactions between electrons in the excited state of the $j$th chromophore  and the nuclear degrees of freedom in the ground electronic state of all others.  Note that these are in general operators with respect to the nuclear degrees of freedom and may be responsible for nonadiabatic transitions, although small.    
 $E_{j}^{c,0}$ is the electronic energy of the site excitation state $|s_j\rangle$ as defined by Eq. (\ref{eq:def-ejc}), $J_{jk}^{c,0}$ is the electronic coupling between site excitation states $|s_j\rangle$ and $|s_k\rangle$, and is defined by Eq. (\ref{eq:j_jk}). $\hat B_{j}^{c,0}$ represents the coupling between $|s_j\rangle$ and all the nuclear degrees of freedom, as described by Eq. (\ref{eq:bcj}), and $\hat H_{b}^{c,0}$ represents the bath Hamiltonian originating from the nuclear degrees of freedom of all the chromophores. The definition of this is given by Eq. (\ref{eq:hcb}).  

It is important to note that we have labeled all the terms defined in Eq. (\ref{eq:h_c_ebh}) with an additional superscript $0$.  This was to make it clear that they are defined with respect reference adiabatic states of independent pigment molecules.  In fact, even for simple aggregates of pigments, there are many-body effects of other pigments, which affect the definition and calculation of adiabatic electronic states of each pigment molecule, making the actual values of parameters and Hamiltonian terms different from the zeroth order ones in Eq. (\ref{eq:h_c_ebh}).   For LHCs, these implicit effects due to other pigment molecules are expected to be less significant than those due to protein environments in general, which can be considered together.    

\subsubsection{Light harvesting complex}
The presence of the protein environment in the LHC affects the molecular Hamiltonian of the aggregate in all of its terms, and adds an additional one referring to its own degrees of freedom.  Expressing all these in the site excitation basis,  one can obtain the following expression for the Hamiltonian of the LHC:
\ben
&&\hat H_{_{LHC}}= (E_g^c+E_g^r) |g\rangle\langle g|\nonumber \\
&&+\sum_{j=1}^{N_c}\left \{(\hat J_{jg}^c+ \hat J_{jg}^r)|s_j\rangle \langle g|+(\hat J_{gj}^c+\hat J_{gj}^r)|g\rangle \langle s_j|\right\} \nonumber \\
&&+\sum_{j=1}^{N_c} (E_j^c+E_j^r )|s_j\rangle\langle s_j|+\sum_{j}\sum_{k\neq j} (J_{jk}^c+ \hat J_{jk}^r)|s_j\rangle\langle s_k|\nonumber \\
&&+\sum_{j=1}^{N_c} \sum_{k=1}^{N_c} (\hat B_{j}^c\delta_{jk}+\hat B_{jk}^r)|s_j\rangle\langle s_k| +\hat H_{b}^c+\hat H_{b}^r \ , \label{eq:h_lh}
\een
where $\delta_{jk}$ in the last line is the Kronecker-delta symbol.  All the terms with superscript $c$ represents contributions from chromophores, which also include all the implicit effects on the adiabatic state of each chromophore by the protein environments and the other chromophores in the ground electronic state. The  terms with superscript $r$ represent explicit contributions of the protein environment.

Equation (\ref{eq:h_lh}) is the final and the most general form of the exciton-bath Hamiltonian for LHC, and includes all possible interactions between single excitons and other degrees of freedom.  The terms that are missing here are intra-pigment non-adiabatic terms, spontaneous emission terms of the site excitation states and interaction terms with the radiation, which will be considered later.   The common assumption implicit in most theoretical models developed so far is that the environment induced nonadiabatic coupling between the ground and the exciton state (cross terms between $|g\rangle$ and $|s_j\rangle$ of Eq. (\ref{eq:h_lh})) are negligible compared to the spontaneous emission term and the interaction terms with the radiation.  Thus, we also assume that they are negligible.

Combining the contributions of the chromophores and protein environments, we define the following parameters: 
 \ben
E_g&=&E_g^c+E_g^r \\
E_j&=&E_j^c+E_j^r \ ,  \\
J_{jk}&=& J_{jk}^c +\langle \hat J_{jk}^r \rangle \ ,  \label{eq:j_jk_tot}\\ 
\hat B_{jk}&=&\hat B_{j}^c\delta_{jk}+\hat B_{jk}^r +\hat J_{jk}^r-\langle \hat J_{jk}^r \rangle\ , \label{eq:bjk} \\
\hat H_b&=&\hat H_{b}^c+\hat H_{b}^r  \ .
\een 
Employing the above definitions and neglecting the environment induced nonadiabatic terms between the ground electronic and the single exciton states, as noted above, Eq. (\ref{eq:h_lh}) can be simplified as 
\ben
\hat H_{_{LHC}}&=&  E_g |g\rangle\langle g|+\hat H_{e}+ \sum_{j=1}^{N_c}\sum_{k=1}^{N_c} \hat B_{jk}|s_j\rangle\langle s_k|+\hat H_b   \nonumber  \\
&=&E_g |g\rangle\langle g|+\hat H_{ex}\ , \label{eq:h_lhc-gen}
\een
where $\hat H_{e}$ has the same form as Eq. (\ref{eq:eh1}) and the second equality serves as the definition of $\hat H_{ex}$.  As can be seen above, this also has the same form as Eq. (\ref{eq:hex}) except that the variables ${\bf X}$ is not explicitly shown.  
The full determination of $\hat H_{_{LHC}}$ given by Eq. (\ref{eq:h_lhc-gen}) is a challenging task in general.  Furthermore, even if the full information on the Hamiltonian of the above type were available, the actual quantum dynamical calculation poses as another theoretical challenge. Thus, in consideration of practicality, further approximations have  been made.

\subsubsection{Site diagonal coupling to bath of harmonic oscillators}

Given the structural and spectroscopic stability of LHCs, it is reasonable to assume that the displacements of the bath degrees of freedom remain small enough to be governed by almost linear restoring forces.  Under this condition, the bath Hamiltonian can be approximated as a set of harmonic oscillators, and the  exciton-bath interaction terms can be assumed to be linear in the bath coordinates.  While these interaction terms may in general involve both diagonal and off-diagonal terms of the exciton Hamiltonian in the site excitation basis, most models of LHCs developed and considered so far have assumed the existence of only diagonal couplings.   At least for the case of the FMO complex, explicit calculation of normal modes confirmed that \cite{renger-jpcb116} off-diagonal exciton-bath couplings are about an order of magnitudes smaller than diagonal ones, offering microscopic justification for such assumption. 

Thus, the typical form of the EBH that has been used for LHC so far assumes that $\hat B_{jk}$ is diagonal and linear in the displacement of bath coordinates.  Within this approximation, Eq. (\ref{eq:h_lhc-gen}) can be expressed as
\ben
\hat H_{_{LHC}} &\approx& E_g|g\rangle\langle g|+\sum_{j=1}^{N_c} E_j |s_j\rangle\langle s_j|+\sum_{j=1}^{N_c}\sum_{k\neq j} J_{jk}|s_j\rangle\langle s_k| \nonumber \\
&&+\sum_{j=1}^{N_c} \sum_n \hbar\omega_n g_{j,n} (\hat b_n+\hat b_n^\dagger)|s_j\rangle\langle s_j|\nonumber \\
&&+\sum_n\hbar \omega_n \left (\hat b_n^\dagger \hat b_n+\frac{1}{2}\right )\ ,  \label{eq:h_lhc_ho}
\een 
where $\hat b_n^\dagger$ and $\hat b_n$ are raising and lowering operators of the $n$th normal mode constituting all the bath degrees of freedom.  
This approximation also ignores Duschinsky rotation, which can be significant if excitation causes nontrivial structural change of chromophores. 

Even though the exciton-bath coupling is assumed to be diagonal in the site excitation basis, certain delocalized vibrational normal modes can be coupled to different site excitation states together, which are called common modes.  Thus, in general, the complete specification of the exciton-bath coupling requires specification of  the following spectral densities of the bath for all pairs of $j$ and $k$:
\be
{\mathcal J}_{jk}(\omega)=\pi\hbar \sum_n \delta (\omega-\omega_n)\omega_n^2 g_{j,n}g_{j,k}\ . \label{eq:spectral-density}
\ee
For practical calculations, it is convenient to decompose the spectral density into the sum of contributions from the intramolecular vibrational modes of the given chromophore itself and those of the environments.  It is reasonable to assume that the former is local to each site excitation state in general.  Thus, Eq. (\ref{eq:spectral-density}) can be decomposed into 
\be
{\mathcal J}_{jk}(\omega)={\mathcal J}_{j}^{c}(\omega)\delta_{jk}+{\mathcal J}_{jk}^r(\omega)\ . \label{eq:spectral-density-decomp}
\ee
The reorganization energy of the bath upon the creation of each site excitation is thus given by 
\be
\lambda_j=\lambda_j^c+\lambda_j^r \ ,
\ee
where
\ben 
\lambda_j^c=\frac{1}{\pi}\int_0^\infty d\omega \frac{{\mathcal J}_{j}^c(\omega)}{\omega} \ , \\
\lambda_j^r=\frac{1}{\pi}\int_0^\infty d\omega \frac{{\mathcal J}_{j}^r(\omega)}{\omega}  \ . 
\een

\subsection{Electronic structure calculation}

\subsubsection{Excitation energies}
The EBH for LHC given by Eq. (\ref{eq:h_lhc-gen}) or (\ref{eq:h_lhc_ho}) serves as an efficient framework for describing excitons in LHCs, and can be used for practical calculations even for large aggregates of many interacting pigments by keeping the computational cost limited \cite{curutchet-cr117}. Moreover, as the cost is mainly due to the calculation of localized excitations, accurate quantum chemical methods can in principle be used within this model. However, even with this simplification, typically large molecular dimensions of the pigments present in natural LHCs (around 50-60 heavy atoms) prevent routine use of highly accurate methods such as {\it ab initio} multi-reference or coupled-cluster methods.    

A good compromise between cost and accuracy is represented by the Density Functional Theory (DFT) and its excited-state extension, commonly known as Time-Dependent (TD) DFT. Beside its computational efficiency, this approach has two main limitations, (i) to be highly sensitive to the selected functional and (ii) to be a single-reference description. The first issue is particularly relevant for excitations that have a (partial) charge-transfer character. TD-DFT is in fact well known to be inaccurate in those cases due to its intrinsic limit, commonly known as ``self-interaction error"; this arises from the spurious interaction of an electron with itself in the Coulomb term of the DFT Hamiltonian, which is not exactly canceled by the exchange contribution, e.g., as in the Hartree-Fock (HF) approach.

More recently, functionals which introduce range separation into the exchange component and replace the long-range portion of the approximate exchange by the HF counterpart, have been proposed to correct this error \cite{WCMS:WCMS1178}. By using these long-range corrected functionals, CT-like transitions can be semiquantitatively described at TD-DFT level. The second issue, namely of being a single-reference approach, is instead particularly relevant for excitations in long conjugated systems such as carotenoids. In those systems, in fact, a multi-reference approach is compulsory if a correct evaluation of the different (``dark" and ``bright") excited states is needed. For such systems, good performances have been shown by the multi-reference extension of DFT, known as DFT/MRCI approach \cite{grimme99,Kleinschmidt:2009fm,Andreussi:2015gv}.

The optimal electronic structure calculation method for LHCs should not only give accurate excitation energies for the different pigments, but should also provide correct description of the variation of the electronic density upon excitation and the transition density. The former is required for accurate evaluation of excited state properties such as the geometrical gradients, which are used for calculating relaxed geometries and the coupling between electronic and nuclear degrees of freedom (see below). The latter (transition density) determines not only the transition dipole corresponding to a selected excitation but also the inter-pigment interactions defining the electronic coupling, $J_{jk}$ (see below). These requirements on the electronic densities largely limit the range of methods that can be used safely. For example, many semi-empirical approaches, which have been largely used in the past, should be used with care. In fact, by construction (e.g. by parameterization), they often give correct excitation energies but this does not mean that the corresponding transition densities are correct as well. Moreover, only by chance, they could give a reliable description of the variation of the electronic density upon excitation, as excited state properties are not used in their parameterization. Once more, at present, TD-DFT represents the only feasible approach available in most cases even if the same two limitations, (i) and (ii), mentioned in regard to the excitation energies also apply for excited state densities and transition densities. 

As a matter of fact, some DFT functionals can give quite accurate transition energies but may completely fail in correctly describing the change in the electronic density upon excitation. In general, transition densities are known to be less sensitive to errors of the DFT functional than excited state densities \cite{MunozLosa:2008ik}.  However, benchmark data should always be used to be fully confident about the selected functional. Another aspect that should be preliminarily checked before selecting an electronic structure calculation method is the robustness with respect to the change of geometry. Some calculation methods, especially those based on semi-empirical formulation, can behave very differently if the molecular system is out of the minimum region of the potential energy surface. For pigment molecules in LHCs, significant deviations from the minimum geometry are possible due to the temperature-dependent effects and/or to geometrical constraints due to the protein matrix. Indeed, this is a very important aspect to consider when selecting a reliable electronic structure calculation method.   In this respect, at present, systematic studies providing satisfactory references are not available to the extent of our knowledge. 

\subsubsection{Electronic couplings}
The electronic coupling $J_{jk}$ in Eq. (\ref{eq:eh1}), (\ref{eq:h_lhc-gen}), or (\ref{eq:h_lhc_ho}) is a key quantity determining both the dynamics and the mechanism of exciton transfer.  The magnitude of the electronic coupling, compared to the exciton-phonon interactions, determines whether states localized on different sites are mixed to generate delocalized exciton states.  An accurate computation of the electronic coupling is thus necessary not only to predict the rates of energy transfer processes, but also to determine their mechanism.   
As shown in Eq. (\ref{eq:j_jk_tot}), $J_{jk}$ consists of two terms, one direct interaction term between chromophores, $J_{jk}^c$ and the other environmental contribution, $\langle \hat J_{jk}^r\rangle$.  Hereafter, we denote this latter contribution simply as $J_{kj}^r$.  Thus, Eq. (\ref{eq:j_jk_tot}) can be expressed as
\be
J_{jk}=J_{jk}^c+J_{jk}^r \ . \label{eq:j_jk_tot-2}
\ee 

Within the derivation of Appendix A, the direct electronic coupling between chromophores $J_{jk}^c$ is defined by generalizations of Eqs. (\ref{eq:j_jk}) and (\ref{eq:vjk_ee}) that include the implicit effect of environments as well.  Following the convention, this can also be expressed as  
\be
 J_{jk}^c=\int d{\bf r} d{\bf r}' \rho_j^{tr*}({\bf r})\rho_k^{tr}({\bf r}')\frac{e^2}{|{\bf r}-{\bf r}'|} \ , \label{eq:j-jk-coulomb}
 \ee
 where $\rho_j^{tr*}({\bf r})=\rho_j^{01}({\bf r})$ and $\rho_k^{tr}({\bf r}')=\rho_k^{10}({\bf r}')$, transition densities of chromophores $j$ and $k$ including the implicit effects of environments.  Thus, this represents the Coulomb coupling between de-excitation of the $j$th chromophore and the excitation of the $k$th chromophore within the given LHC.   More generally, additional terms due to  exchange interactions and overlap between orbitals of different chromophores should be included.   This is missing in Eq. (\ref{eq:j-jk-coulomb}) because it has been assumed that electrons from different chromophores are distinguishable and have zero overlap.   For most LHCs, the inter-chromophore distances are far enough to make the Coulomb term by far the most dominant one.  The approximation of Eq. (\ref{eq:j-jk-coulomb}) is well justified under such condition.  Of course,  
short-range non-Coulomb interactions are not always negligible. 
For example, contributions originating from charge-transfer states were shown to make modest contribution to the nearest-neighbor couplings in the
B850 unit of LH2 \cite{scholes-jpcb103} and to constitute significant portion of the interaction between 
the special pairs of purple bacteria and PS1 \cite{Madjet:2009cz}, where the BChls are closely packed together.

The simplest but still widely used approximation for the Coulomb coupling is to use the expansion of the transition densities up to dipolar terms, which leads to the following point transition dipole approximation:
\begin{equation}\label{eq:PDA}
J_{jk}^{c,dp} ={\dfrac{{\bmu}_j \cdot {\bmu}_k}{r_{jk}^{3}}} - \dfrac{3({\bmu_j} \cdot {\mathbf{r}}_{jk}) ({\bmu_k} \cdot {\mathbf{r}}_{jk})}{r_{jk}^{5}} \ ,
\end{equation}
where ${\mathbf{r}}_{jk}$ is the distance vector between pigments $j$ and $k$, and ${\bmu_j}$ and ${\bmu_k}$ are the corresponding electronic transition dipole moments. 

The dipolar term given by Eq.~\eqref{eq:PDA} yields the well known $r^{-3}_{jk}$ asymptotic dependence of the singlet electronic coupling.
Where applicable, the dipole approximation has the clear advantage of only needing experimental data, namely the transition dipole moments and the distance between the centers of pigment molecules, provided that the orientation of the transition dipole moments is known.
For this reason, the dipole approximation has been widely employed \cite{jang-jpcb105,georgakopoulou-bj82,damjanovic-pre65,adolphs-bj91,Wan2014}. However, particular care should be taken in using the approximation because it can result in significant error in the estimation of the Coulomb coupling between proximate pigment molecules.  For example, this is the case for the electronic couplings between the nearest neighbor BChl molecules in the B850 unit of LH2 complex and for the two central bilins of PBPs of cryptophyte algae.  On the other hand, for FMO complex, the transition dipole approximation serves as reasonable approximation for all the electronic couplings.

Another widely used method to compute $J_{jk}^c$ is based on the projection of the transition densities onto atomic transition charges (TrCh). 
The Coulomb coupling is then computed as the electrostatic interaction between those charges:
\begin{equation}\label{eq:VTresp}
 J_{jk}^{\text{c,TrCh}} = 
 \sum_{\substack{n \in j\\m \in k}} \,\dfrac{q_{n}\,q_{m}}{r_{nm}}\ ,
\end{equation}
where the indices $n$ and $m$ run over the atoms of $j$ and $k$, respectively, $q_n$ and $q_m$ are the transition charges of atoms $n$ and $m$, and $r_{nm}$ is the distance between them.
Atomic transition charges with various definitions have been used for a long time to compute Coulomb couplings \cite{Chang:1977eo,Carbonera1999,Duffy2013}.
Arguably, a definition of atomic charges that is physically accurate and adequate for electrostatic interaction is the one based on electrostatic potential (ESP) fitting. 
The calculation of the Coulomb coupling with transition ESP charges (TrEsp) was developed by Renger and coworkers \cite{madjet-jpcb110}, and represents now a widely used method \cite{Kenny2016,olbrich-jpcb114,vandervegte-jpcb119}. 

The Coulomb coupling can also be obtained by directly evaluating the integral, Eq. (\ref{eq:j-jk-coulomb}).  The first calculation of this kind was developed by using a numerical integration over a three-dimensional grid \cite{krueger-jpcb102}. This numerical approach is called the Transition Density Cube (TDC) method, and in principle gives the exact Coulomb contribution of the electronic coupling with an appropriately chosen grid.  The TDC method has been employed extensively \cite{scholes-jpcb103,scholes-jpcb104,scholes-arpc54,Bricker2014,Bricker2015}.  However, more reliable and efficient approach is to perform the integration of Eq. (\ref{eq:j-jk-coulomb}) analytically.  This also allows the inclusion of explicit screening of the Coulomb interaction due to the environment \cite{Hsu2001,Iozzi2004,curutchet-jctc5}. 

The coupling is also significantly affected by the presence of 
the environment surrounding the pigments. 
While for site energies, environment effects are generally seen as a shift depending on the nature of the transition and the characteristic of the environment, the electronic couplings are affected through two mechanisms, each leading to a specific change.

First of all, the environment can change both the geometrical and the electronic structure of the pigment molecules and modify their transition properties, i.e. transition dipoles and transition densities. These changes will be ``implicitly" reflected in a change of the Coulomb coupling which will be generally enhanced. 
The second effect is due to the polarizable nature of the environment which acts as a mediator of the pigments' excitations. The resulting ``explicit" effect acts to reduce the magnitude of the direct (Coulomb) coupling. For this reason it is common to say that the coupling is ``screened" by the environment.

The simplest way to account for the explicit effect of the environment is to introduce a screening factor $s$ such that $J_{jk}= sJ_{jk}^c$, where $J_{jk}^c$ is the direct electronic coupling between chromophores with the implicit effect taken into consideration.  This definition is in line with the standard definition of dielectric screening by environment in electromagnetism but may not always be easy to determine.  In practice, one may introduce a total screening factor $s_{t}$ such that $J_{jk}= s_{t}J_{jk}^{c,0}$, where $J_{jk}^{c,0}$ is the bare electronic coupling between chromophores without including the implicit effect.  The difference between $s$ and $s_t$ can be significant, and care should be taken in using the proper definition.  More discussion of this issue will be provided later.  

The screening factor $s$ can be clearly related to the inverse of an effective dielectric constant $\epsilon_{\infty}$ which represents the electronic response of the environment approximated as a dielectric. $\epsilon_{\infty}$ is generally approximated with the square of the refractive index of the medium: when the environment is characterized by a refractive index of 1.4 (typical for hydrophobic region of protein environments) the screening factor is $\sim0.5$ and the electronic coupling is reduced by a factor $\sim4$. 
Within this approximation, the screening does not depend on the interacting chromophores, neither on their relative orientation and distance. 
Moreover, at short distances, the dielectric medium can be excluded from the intermolecular region, leading to more complex effects.  In particular cases, this can also enhance the coupling.

To achieve a more accurate modeling of the environmental effects on electronic couplings, a combination of a quantum mechanical description of the pigments with a continuum model can be introduced. Various formulations of continuum solvation models \cite{tomasi-cr105} can be used to this end.  
By applying the polarizable continuum model (PCM) \cite{mennucci-wires2} described in Appendix B, the response of the environment (i.e. the polarization of the dielectric) is described by a set of induced (or apparent) charges spreading on the surface of the molecular cavity embedding the chromophores. Within this framework, $J_{jk}^{r}$ becomes:
\begin{equation}
\label{eq:EET-VPCM}
J_{jk}^{r,\textrm{PCM}}= \sum_t \left[ \int d\br \rho_{j}^{tr*}(\br) \frac{1}{|\br -\br_t |} \right] q_t(\epsilon_{\infty};\rho_{k}^{tr})
\end{equation}
Conceptually, the electronic transition in the chromophore $k$ drives a response in the polarizable medium, which in turn, affects the transition in the chromophore $j$. It is important to note 
that the induced charges $q_t$ are calculated using the optical permittivity of the medium, in order to account for the fact that only the electronic component of the polarization can respond. 
Also in this case, we can define an effective screening factor as the ratio between the total coupling and the direct interaction, namely $s = (J_{jk}^c + J_{jk}^{r})/J_{jk}^c $. 
A QM/PCM study has been conducted to investigate the dependence of the screening factor on the nature of the interacting chromophores and their relative arrangement, using different pairs (chlorophylls, carotenoids and bilins) extracted from LHCs \cite{scholes-jpcb111}. It has been shown that at large inter-chromophore separation ( $> 2 \ {\rm nm}$ ) the screening factor is practically constant. 
At closer distances, instead, the screening shows an exponential behavior approaching to $\sim1$ for distances of the order of few tenths of nanometer. At such close distance, the chromophores share a common cavity in the medium (the environment cannot access the region where the two chromophores are at close contact) and the screening effect is reduced.

As noted above, the scaling factor $s$ accounts for only the explicit screening.  On the other hand, the environment can also affect the transition densities of chromophores,  causing implicit change of electronic couplings, which we call here $s_{i}$.  A detailed analysis of $s_{i}=J_{jk}^c/J_{jk}^{c,0}$ was reported in detail in a follow-up paper \cite{curutchet-jpcb111}, which showed that  
it is not possible to define a similar empirical expression for $s_{i}$ as has 
been done for $s$.  The total empirical screening factor we defined earlier is given by $s_t=s \cdot s_{i}$. In general, it cannot be fitted by a simple function of the distance. Thus, care should be taken in estimating the distance dependence of the total screening factor.

An alternative way to account for environment effect in the coupling, is through the so-called Poisson-TrEsp method developed by Renger an coworkers \cite{adolphs-pr95,Renger:2011kp}. The TrEsp charges of the pigments are placed in molecule-shaped cavities that are surrounded by a homogeneous dielectric with a dielectric constant, which represents the optical permittivity of the protein and solvent environments. A Poisson equation is then solved for the electrostatic potential of the TrEsp of each pigment and the resulting potential is finally used to calculate the coupling.

Despite the success of these continuum approaches, an atomistic description of the protein environment is expected to give a more complete description. In those cases, in fact, the dielectric response varies locally and specific interactions between the chromophores and the protein can be established.
A classical formulation can still be used by introducing a Molecular Mechanics (MM) description in which each atom of the environment is represented by a fixed point charge to mimic the electrostatic effects. 
In order to properly account for all the environmental effects, however, the MM model has also to be polarizable. Various strategies are possible (see Appendix B). In the context of LHCs, the most used approach is represented by the induced dipole (ID) formulation where an atomic polarizability is added to the fixed charge to describe each atom of the environment.
Within this MMPol framework, the $J_{jk}^{r}$ term becomes \cite{curutchet-jctc5}:
\begin{equation}
\label{eq:EET-VMMPol}
J_{jk}^{r,\textrm{MMPol}}= -\sum_p \left[ \int d\br \rho_{j}^{tr*}(\br) \frac{(\br_p-\br)}{\vert \br_p -\br \vert ^3} \right]  \mu_p(\rho_{k}^{tr})
\end{equation}
where the transition densities $\rho_{k}^{tr}$ induce a response in the environment which is represented by the induced dipoles $\mu_p$. As in the case of the QM/PCM, $\rho_{k}^{tr}$ here is also calculated self consistently with the polarization of the environment. 
A mixed continuum/atomistic strategy has also been proposed \cite{Caprasecca:2012dk}: in this case,  $J_{jk}^r$ is the sum of $J_{jk}^{r,\textrm{PCM}}$ and $J_{jk}^{r,\textrm{MMPol}}$, and both terms are obtained in a fully polarizable scheme.

As a cautionary remark, we note that, in the case of any atomistic description, the QM-environment interactions, particularly those at short range, depend critically on the configuration of the environment. Therefore, several configurations of the whole system need to be taken into consideration to get a correct sampling. Commonly, the sampling is obtained by using classical Molecular Dynamics (MD). 
This sampling is not needed when a continuum approach is employed since it implicitly gives a configurationally averaged effect due to the use of macroscopic properties.

A completely alternative way for the calculation of the electronic coupling between two sites is that based on a full quantum calculation on the dimeric unit, which is also known as super-molecule approach. These schemes are based on the diabatization of the electronic Hamiltonian of the whole donor-acceptor system, and in principle yield the ``exact" coupling including exchange and overlap interactions within the electronic structure method used \cite{Hsu2009,You2014}. 

Consider a molecular system composed of two moieties, which can be either two separate molecules or two fragments of the same molecule.
An electronic structure calculation on the entire system necessarily yields the adiabatic states, which are the eigenstates of the electronic Hamiltonian at a specific nuclear geometry. 
However, a diabatic picture better describes the states involved in energy transfer.  Within this picture, the electronic Hamiltonian is written in a basis of localized states, and is not diagonal. For example, considering only two states, the Hamiltonian matrix reads as follows:
\begin{equation}
 \mathbb{H}_{el} = 
 \begin{pmatrix}
  E_{j} & J_{jk} \\
  J_{jk} & E_{k} \\
 \end{pmatrix}\ ,
\end{equation}
where $E_{j}$ and $E_{k}$ are the energies of the diabatic states, and $J_{jk}$ is the electronic coupling between those states. 
At the avoided crossing point, the condition $E_{j}=E_{k}$ means that the energy gap between the adiabatic states is $2J_{jk}$.  Therefore, $J_{jk}$ may be computed as half of the energy gap.  Obviously, this condition holds at all geometries when the two fragments are identical. 
In more general cases, a localization scheme is needed to define the diabatic states, and to find the transformation matrix between adiabatic and diabatic bases. 

There is no unique choice for the diabatic states, the choice of which also can alter the definitions of Coulomb and exchange interactions \cite{vura-weis-jpcc114}.  Several schemes have been proposed, such as  the fragment excitation difference (FED) \cite{Hsu2008} or the more recent fragment transition difference (FTD) scheme \cite{Voityuk2014}.
Within this framework, an additional operator (which we shall call $\hat{Y}$), representing an observable that has its extrema in the diabatic states, is introduced; $\hat{Y}$ can be defined in such a way that it has eigenvalues 0 and $\pm1$. 
The localized states are therefore those states that diagonalize $\hat{Y}$. 
The eigenvectors of the matrix $\mathbb{Y}$ form the unitary matrix transformation $\mathbb{U}$ (i.e., $\mathbb{U}^{\dagger}\mathbb{Y}\mathbb{U}$ is diagonal), which is the adiabatic-to-diabatic (ATD) transformation, within the assumption that the two adiabatic states are a linear combination of the localized states of interest. By applying the same transformation to the diagonal energy matrix, one can obtain the Hamiltonian in the diabatic basis, and the electronic coupling $J_{jk}$ as the off-diagonal elements.  Namely,
\begin{equation}\label{eq:coupling_fxd}
 J_{jk} = (E_{k}-E_{j})\dfrac{Y_{jk}}{\sqrt{(Y_{j}-Y_{k})^{2} + 4Y_{jk}^{2}}}\ .
\end{equation}

Applying the transformation $\mathbb{U}$ to the adiabatic states yields states that are localized as much as possible, and similar to the initial and final states.
However, in many cases, a 2-state adiabatic basis is not sufficient to retrieve completely localized states.  In fact, an adiabatic state could be a combination of many diabatic states of both the donor and the acceptor. 
Moreover, charge-transfer states can mix with excitonic states, and vice-versa \cite{Yang2013,Voityuk2013,You2014}.
Another issue with the 2-state model is its inability to compute the couplings between states that are more weakly coupled. 
To overcome these limits, one can resort to a multi-state formulation. The generalization to a multi-state model is not straightforward, as the additional operators only have three different eigenvalues (0 or $\pm$1). This means that the diagonalization of these operators will only separate the adiabatic basis in three subspaces, and therefore there is no unique choice for the transformation $\mathbb{U}$ from the adiabatic to the diabatic basis.

\subsection{Applications of electronic structure calculations to LHCs}
We present here a summary of three computational studies aimed at understanding the molecular origin of the protein  tuning of the excitonic properties of three LHCs introduced earlier, the  FMO complex of green sulfur bacteria, LH2 complex of purple bacteria, and phycoerythrin PE545 of cryptophyte algae.

\subsubsection{FMO complex}

As mentioned in the overview of Sec. II,  the spectral features in FMO arise from the subtle tuning of the individual site energies due to the surrounding protein environment.
To shed light on this effect, the environment-induced changes in the site energies were analyzed in detail using the combination of TDDFT and different classical models based either on the atomistic or continuum descriptions \cite{Jurinovich:2014fw}.

\begin{figure}[htbp]
  \centering
  \includegraphics[width=3.2in]{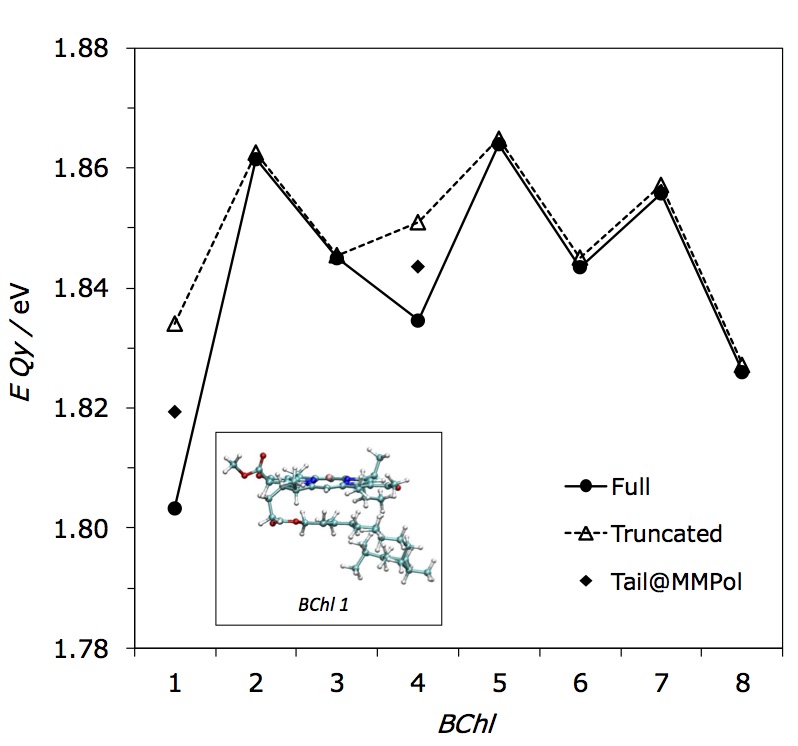}
  \caption{Site energies (eV) of the eight BChls computed on the crystal structure with three different models: the Full model (full line and circles) includes all BChl atoms, the Truncated model (dotted line and triangles) does not include the atoms of the phythyl chain, the Tail@MMPol model (diamonds) describes the atoms of the phythyl chain as MMPol sites. Inset: Structure of BChl 1 for which the chain is folded in a conformation capable of interacting with the porphyrin ring.}
  \label{fig:FMO-Fig3}
\end{figure}

The trimeric crystal structure of \textit{Prosthecochloris aestuarii} (pdb entry 3EOJ, res. 1.30 \AA ), which contains eight BChl per monomer, was used \cite{tronrud-pr100}.
Considering the C$_3$ symmetry of the system, the QM analysis was performed only on the monomer system shown in Figure \ref{fig:FMO-Fig1}.
The corresponding excitation energies of BChl were computed at the TD-B3LYP/6-31G(d) level of theory, the results of which are shown in Fig. \ref{fig:FMO-Fig3}. 

In FMO complex, like in other LHCs, each pigment is confined in a specific binding pocket, surrounded by the protein matrix, which may perturb the pigment's excitation through two distinct effects: 1) a direct effect on the electronic states determined by pigment-protein interactions; 2) an indirect effect due to the modifications induced by the environment on the geometry of the pigment.  The effect of this can be seen clearly in the results of calculations: if the direct environmental effects are switched off, the resulting site energies are still different for the different BChls due to the indirect effects on geometries induced by the local environment. 
In particular, the isolated BChl has  a planar equilibrium structure (excluding the flexible phythyl chain).  However, when embedded in the protein environment, such planarity becomes perturbed differently in different binding pockets. 

The phythyl chain of BChl also plays a role in differentiating the excitation energy through variation in its conformations.  According to the crystal structure, all the chains adopt an outstretched configuration, except for BChl 1 and 4, for which the chain is folded in a conformation capable of interacting with the porphyrin ring (see Figure \ref{fig:FMO-Fig3}).
To investigate these effects, site energy calculations were performed using the crystallographic structure of the pigments and 1) including all the BChl atoms at full quantum mechanical level calculations, 2) replacing the phythyl chain with a methyl group but still treated at the full quantum level, and 3) replacing the entire phythyl chain with classical polarizable MM sites through the MMPol approach (tail@MMPol). In this last model, the boundary between the QM and the classical subsystem was treated by using the link atom scheme. The results shown in Figure \ref{fig:FMO-Fig3} reveal that, when the phythyl chain is close to the porphyrin ring (BChl 1 and 4), the site energy is lowered, as if it introduces an ``additional environmental" effect not present in the other pigments. This effect seems to be largely due to a classical electrostatic and polarization interaction.  In fact, when the chain is treated at MMPol level, a similar lowering of the energy is observed, albeit not as large as in the full QM description.

\begin{figure}[htbp]
  \centering
  \includegraphics[width=3.5in]{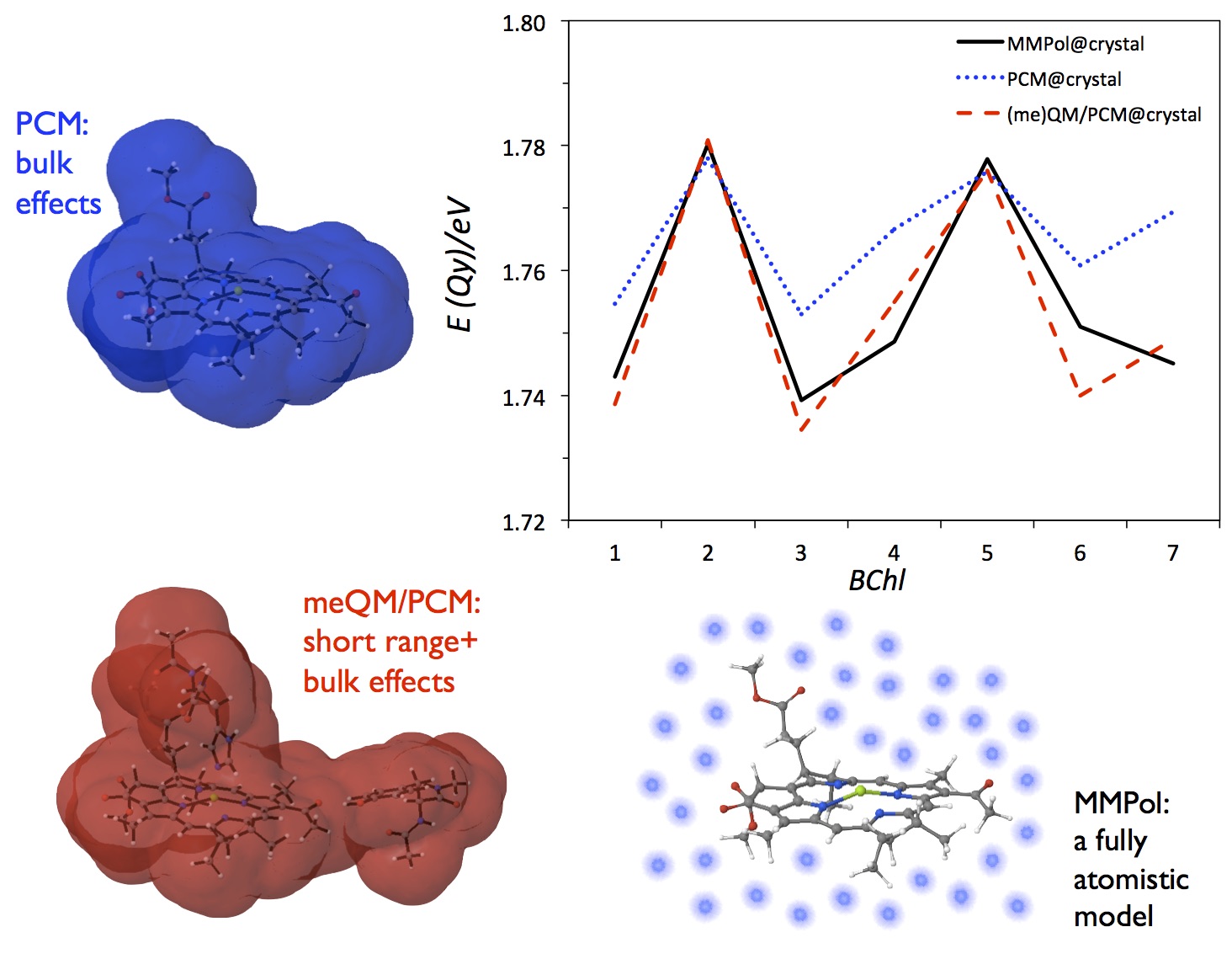}
  \caption{Site energies (eV) for the 1-7 BChls computed on the crystal structure including the environment effects at PCM (dotted line) and MMPol level (full line). A third (me)QM/PCM model combining PCM with a QM description of the interacting residues is also shown (dashed line).}
  \label{fig:FMO-Fig5}
\end{figure}

An important question concerning the role of protein environments in tuning the excitation energies of pigment molecules is the relative importance of specific molecular level interactions versus mean-field (or bulk) effects. This issue can be analyzed through detailed and careful analysis of the two contributions.  As an example, values of excitation energies for  $1$ through $7$ BChl molecules obtained either with the atomistic (MMPol) or the continuum (PCM) model are compared in Figure \ref{fig:FMO-Fig5}. BChl 8 is excluded from this analysis because of the external location of its binding site, being more exposed to the solvent.   This is manifested in drastic difference between PCM and MMPol descriptions of the environment, in particular when the MMPol model is based on the crystal structure and does not contain surrounding water molecules. 

As described in the previous section, the PCM model describes the bulk effect of the environment through an effective dielectric constant, which corresponds to a response averaged over many different configurations of the environment. It is evident that when some specific or directional interactions are present (i.e. H-bond or coordination bonds), their effects cannot be accounted for in such an ``average over the medium" model.  A practical solution for this deficiency is to enlarge the definition of the quantum mechanical subsystem being in the dielectric medium by including not only the pigment but also its proximate protein residues.  Through this minimal environment  (so called ``me") approach, the effects of short-range interactions can be taken into account within the PCM model.
 
The comparison between QM/PCM and (me)QM/PCM shows very similar energies for BChls 2 and 5. These two BChls have weaker axial interactions (a water molecule for BChl 2 and a carbonyl group belonging to the protein backbone for BChl 5) with respect to the other pigments, which are coordinated with the nitrogen of a histidine residue. In the latter cases, the site energies are red-shifted by about $160 \ {\rm cm^{-1}}$ when including the QM residues. In principle, the QM/MMPol description should be able to describe both the bulk effect of the protein and the specific electrostatic interactions of the closer residues. Indeed, the QM/MMPol and the (me)QM/PCM site energies are in good agreement for all pigments except for BChl 6. It has to be also noted that, for such a pigment, a positively charged residue (an arginine) directly interacts with the side group of the porphyrin ring.  In this case, the description of this residue at a QM level allows to include non classical interactions (such as possible charge transfers) between the BChl and the protein, which can significantly affect the electronic excitation.

In summary, the inclusion of the interaction with the protein environment results in a redshift of the excitation energies of BChls in the FMO complex, but in ways depending on different local environments. Thus, a simple continuum model alone is not able to properly account for both short and long-range environmental effects. This point is also supported by a detailed electrostatic analysis of the site energy funnel in FMO \cite{muh-pnas104}, which showed that the electric field from the $\alpha$-helices defines the direction of excitation energy flow in the FMO protein, whereas the effects of amino acid side chains largely compensate each other.
Thus, it is important to have a complete and balanced picture of the fine tuning by protein environments in order to
characterize the FMO complex properly. 
An effective and still practical way for this is to include the residues more strongly interacting with the pigments in the definition of the QM subsystem, while keeping a continuum description for the rest. Nonclassical short-range effects can in fact have a non-negligible contribution to the pattern of BChl excitation energies in the FMO complex. 
In addition, the BChl phythyl chain can play a subtle but significant role, given the small differences found in the energies of the pigments. In particular, the largest effects are found when a folded conformation close to the porphyrin ring is possible, such as for BChl 1 and 4. The effects of the tail can be explained as a combination of overlap interaction at the quantum level and electrostatic plus polarization at purely classical level.  It is worthwhile to note that the latter components can be recovered by using an MMPol description of the chain.

\subsubsection{LH2}
The early atomistic level calculations \cite{hu-jpcb101,hu-qrb35,krueger-jpcb102,hsu-jpcb105} were based on the X-ray crystal structures \cite{mcdermott-nature374,koepke-structure4} and focused on calculating electronic coupling constants between excitations.  The outcomes of these calculations, in combination with empirical correction factors to complement numerical errors, laid the foundation for many exciton models that followed.   In addition, already in 2002,  all atomistic MD simulation of LH2 complex was shown to be feasible \cite{damjanovic-pre65}, providing the basis for more recent advances \cite{olbrich-jpcb114,cupellini-jpcb120,jang-jpcl6,sisto-pccp19,montemayor-jpcb122}.  In this section, some of the outcomes of  these recent computational studies for the LH2 antenna complex of \textit{Rps. acidophila} are summarized.

QM/MM calculations \cite{cupellini-jpcb120} have been performed using (i) the crystal structure resolved at 2.0 {\AA} (PDB code: 1NKZ) \cite{Papiz2003} and (ii) the configurations extracted from a room-temperature sampling, carried out through an MD simulation of the LH2 complex within a solvated lipid membrane.  Environmental effects on the calculations of the excitonic parameters have been introduced in terms of the polarizable (MMPol) embedding combined with a TD-DFT description based on the CAM-B3LYP functional and the 6-31G(d) basis set. The comparison of a ``static" description using a single crystal structure with a ``dynamic" one including structural and electronic fluctuations of the pigments coupled to electrostatic and polarization fluctuations of the environment, allows detailed and molecular-level investigation of the key factors contributing to the characteristics of excitons and their changes in temperature.

For the case of the dynamic model, a different Hamiltonian matrix was calculated for each of the 88 configurations extracted from the MD simulation and the resulting ``instantaneous" excitonic parameters were finally averaged to be compared with those obtained on the single crystal structure.
Indeed, the two sets of data show significant differences.
The B850 exciton splitting decreases from 1517 to 1275 cm$^{-1}$ (-16\%) moving from the static description based on the crystal structure to that based on the MD simulation, which is consistent with experimental variation between low and room temperature Davydov splittings (-13\%) \cite{Pajusalu2011}. 
This behavior is also reflected on the absorption spectrum, where the B850 band blue shifts when the temperature is increased, whereas the B800 peak does not shift \cite{trinkunas-jpcb116}. These results suggest that the B850 blue shift is mainly due to a reduction of the inter-pigment couplings. In addition, the site energy difference between B800 and B850 pigments is also reduced on average by $\sim$50~cm$^{-1}$ in the calculation based on the MD simulation. 

The static and the dynamic models also give different descriptions of the environment effects. 
Using the crystal structure, the inclusion of the MMPol environment causes a redshift of the BChls' excitations of $\sim$880 cm$^{-1}$, with small differences among the different pigment types ($<$ 25 cm$^{-1}$). A red shift is also obtained for the structures coming from the MD trajectory.  However,  in this case, the shift in the excitation energies of $\alpha$ and $\beta$ BChls of the B850 unit is about 830 cm$^{-1}$, whereas the shift for B800 is only 610 cm$^{-1}$.  When considered within the approximation of transition dipole moments, an interesting  environment effect can be observed with respect to their orientations.
In both models, the environment does not affect the out-of-plane tilt, but it has a non-negligible influence on the in-plane tilt. 
The changes are almost the same for $\alpha$ and $\beta$ BChls of the B850 unit.

The different features that can be found between the two static and dynamic models are well reflected in the exciton delocalization.
There are various measures of exciton delocalization \cite{dahlbom-jpcb105,jang-jpcb105}, and a well-known measure is the following inverse participation ratio (IPR):
\begin{equation}\label{eq:IPR}
 L_{D}^{^{\rm IPR}} = \left[ \langle \sum_{k} \left( U_{jk} \right)^{4} \rangle\right]^{-1}
\end{equation}
where $U_{jk}$ is the unitary transformation as defined above Eq. (\ref{eq:exciton_s}), $j$ is the index for each exciton state, $k$ is the index for each site excitation state, and $\langle \cdots\rangle$ represents averaging over all exciton states with proper thermal weights.  By definition, $L_{D}^{^{\rm IPR}}$ ranges from 1, for a fully localized state, to the number of total sites for a completely delocalized state. 

Due to the $C_N$ symmetry (with $N=8-10$) of the LH2 complex, in the absence of disorder, the exciton states are all doubly degenerate with the exception of the lowest energy exciton states for both the B850 and B800 units (plus the highest exciton state for B850 with odd $N$). In such a perfectly symmetric arrangement for $N=9$, for example, the delocalization lengths are maximum 18 (9) for the non degenerate states of the B850 (B800) unit, and 12 (6) for the doubly degenerate states.
The calculations in the static model indeed give delocalization lengths which are close to the maximum values, even in the B800 unit, despite small electronic couplings between BChls. Instead, adding disorder in the BChls' site energies through the dynamic model results in a localization of the excitons with a general reduction of $L_{D}^{^{\rm IPR}}$. 
In the B850 unit, the large couplings allow the exciton to remain delocalized even in the presence of disorder but with a significant reduction of $L_{D}^{^{\rm IPR}}$ to an average value of typically 8$\pm$2. On the contrary, in the B800 ring the average delocalization length reduces to $\sim$ 1.4.  

Although $L_D^{\rm IPR}$ is a well established measure of delocalization, its physical implication in the partially delocalized regime is not always clear.  For this reason, different measures of delocalization length have been tested, which have different degrees of sensitivity \cite{dahlbom-jpcb105,jang-jpcb105}.  For example, a simple alternative measure can be defined as follows: 
\begin{equation}\label{eq:JDS}
 L_{D}^{^{\rm JDS}} = \langle \sum_{k} {\rm min}\{1,N_c U_{jk}^2\} \rangle \ .
\end{equation}
This measure does not give a weight more than one for each site unlike $L_{D}^{^{\rm IPR}}$ while approaching the same values in both full localized and delocalized limits. For the B850 unit \cite{jang-jpcb105}, this measure was shown to be more sensitive than $L_D^{^{\rm IPR}}$ to the nature of the disorder and also results in larger estimate of delocalization length.   

\begin{figure}[htbp]
  \centering
  \includegraphics[width=3.5 in]{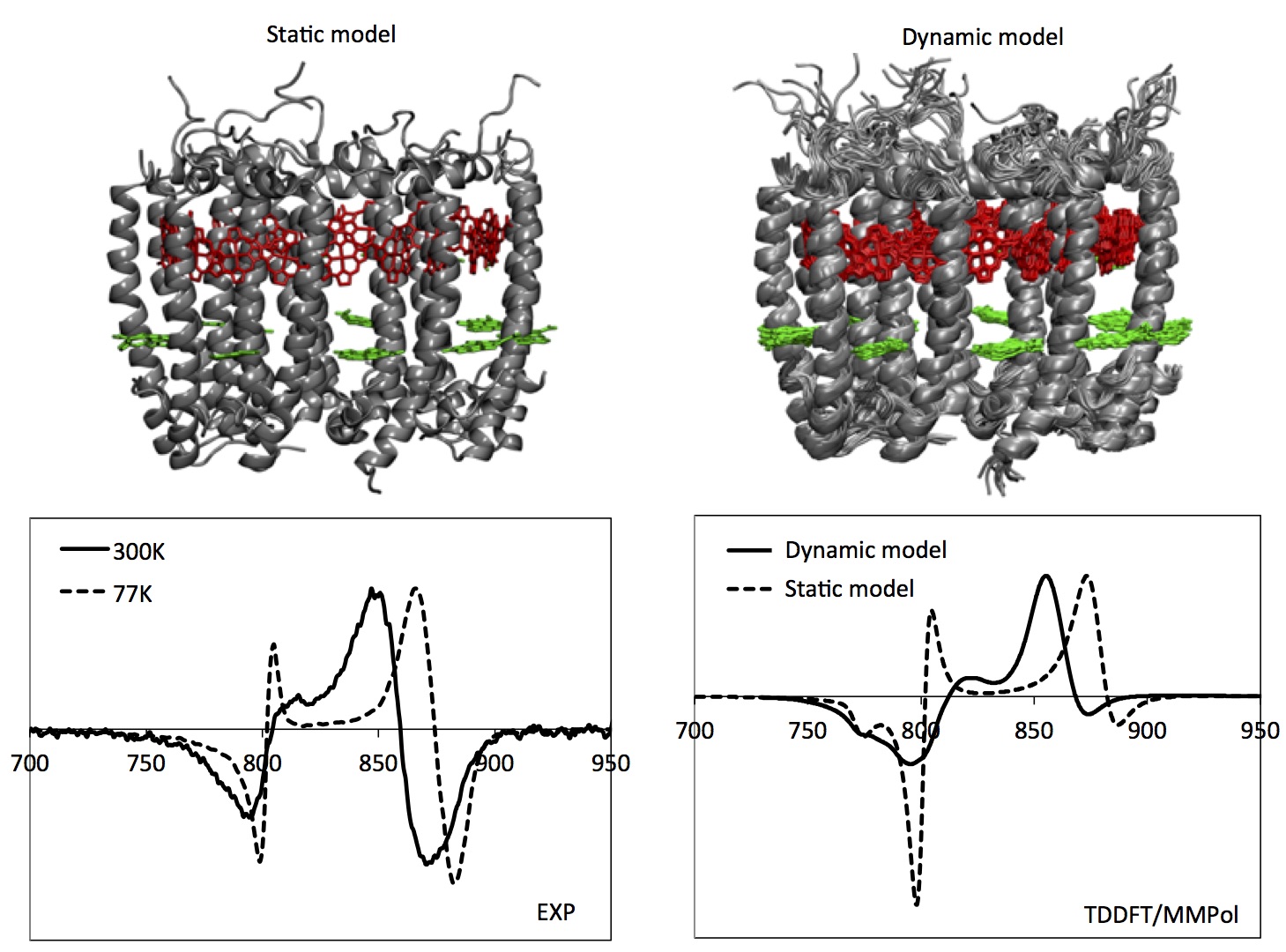}
  \caption{Upper Panel: The crystal structure used in the static model (left) and examples of configurations extracted from MD used in the dynamic model (right) of LH2. Lower panel: (left) measured CD spectra at 77 K (dotted line) and room temperature (full line) and (right) calculated TDDFT/MMPol spectra using the static (dotted line) and the dynamic model (full line).}
    \label{fig:LH2comp}
\end{figure}

The computational results obtained for both the single crystal structure and MD configuration can also be used to simulate the circular dichroism (CD) spectrum of LH2. The CD spectrum represents a fingerprint that is unique for each LHC as it is determined by the nature of the excitons and the geometrical characteristics of the aggregate of pigments. Figure~\ref{fig:LH2comp} compares the two calculated CD spectra with those measured at low (77K) and room temperature. In the case of the dynamic model based on the MD simulation, the spectrum reported in Fig.~\ref{fig:LH2comp} is the average of the instantaneous spectra obtained for each of the 88 configurations.

The experimental CD spectrum is characterized by two ``couplets" corresponding to the excitonic signal due to the B800 and B850 ring, respectively. Due to the increase of the temperature (from 77 K to room temperature) both the broadening and the relative intensities of the couplets change. In particular, the B800 positive band almost vanishes.  The CD spectrum is sensitive to the inter-ring coupling, and the B850 couplet borrows intensity from the B800 band. The small B800--B850 mixing happens despite the site energy differences and the static disorder, and breaks the symmetry of the B800 and B850 couplets. The only way to reproduce the asymmetrical B800 couplet is to consider some B800--B850 mixing. Unlike the spectrum obtained from the crystal structure, here the B800 couplet amplitude is nearly half of the B850 couplet. This can be explained as an effect due to the  disorder in excitation energy of each BChl, combined with the small intra--B800 coupling.

The modeling of the CD spectra suggests the static and the dynamic models can be effectively used to represent 
 LH2 in two different situations, namely, at zero-temperature limit and at room temperature. 
The main features of the CD spectra and their temperature dependence are in fact well reproduced.
Moreover, this analysis indicates that the difference in the exciton structure of LH2 at low and room temperatures are mainly related to fluctuations in the relative orientations of the BChls (and of the corresponding couplings), rather than changes in the ring size as was previously suggested \cite{Pajusalu2011}.

Despite impressive advances as described above and recent demonstration of the capability of {\it on the fly}  calculations \cite{sisto-pccp19}, further improvements in accuracy and efficiency are still needed. Some of key issues to be addressed in this respect include the differences in the excitation energies of $\alpha$-, $\beta$-, and $\gamma$-BChls, solvatochromic shift due to hydrogen bonding, and contribution of carotenoids.   Although some of the early exciton models based on the X-ray crystal structure suggested that \cite{rancova-jpcb116} $\alpha$- and $\beta$-BChls have about $300\ {\rm cm^{-1}}$ difference in their excitation energies, QM/MM calculations for configurations extracted from  MD simulation showed that the two energies are virtually the same in actual protein environments \cite{cupellini-jpcb120}. Combination of MD simulation and DFT calculations also demonstrated that hydrogen bonding can cause red shift of excitation energies by about $500\ {\rm cm^{-1}}$ or larger \cite{jang-jpcl6,montemayor-jpcb122}. Moreover, accounting for the contribution of the possible multi-configurational character of ground and excited states of the pigments can lead to new insights with respect to DFT-based investigations, as it has been shown by recent {\it ab initio} multi-reference calculations on bacteriochlorophylls and carotenoids of LH2 \cite{anda-jctc12,sgatta-jacs139}.

\subsubsection{PE545}

The effect of the protein (and the surrounding solvent) on the excitonic properties of PE545 has been investigated through QM/MMPol calculations for a series of structures extracted from a classical MD simulation of the LHC in water at room temperature \cite{Curutchet:2011hl}.  The QM/MMPol results were complemented with those based on a continuum (PCM) description of the protein and water environment. The comparison between the two sets of calculations can in fact be used to show how the heterogeneous properties of the protein can modify the local screening of the electronic couplings between the bound pigments.
MMPol results were averaged over 140 configurations of PE545 and water molecules extracted from the MD simulation, whereas PCM results were obtained using the cluster of the pigments as found in the ultrahigh resolution crystal structure of the complex. In all cases, the first low-lying $\pi-\pi^*$ excited state of the 8 bilins were computed together with all the corresponding electronic couplings between them. The full transition densities of the pigments were used for these calculations employing  Eqs. (\ref{eq:j-jk-coulomb}),  (\ref{eq:EET-VPCM}), and (\ref{eq:EET-VMMPol}). A comprehensive set of calculation results performed at the CIS/6-31G level are available \cite{Curutchet:2011hl}.

The heterogeneous nature of the protein environment can significantly modulate the coupling through both changes in the transition density of each pigment and in the extent of screening of the corresponding interactions.  In the QM/MMPol approach, both effects are included in the coupling term by considering both interchromophore distances and orientations, as well as the heterogeneous polarization of the protein and solvent environment.  For more quantitative understanding, it is useful to compare the screening factor (and a related effective permittivity) defined differently for each pair $j$ and $k$ as follows: 
\begin{equation}
\label{eq:epsjk}
s_{jk}=\frac{1}{\epsilon_{jk}}= \frac{J_{jk}^c + J_{jk}^{r,\textrm{MMPol}}}{J_{jk}^c}\ ,
\end{equation}
where $J_{jk}^c$ is the Coulomb interaction given by Eq. (\ref{eq:j-jk-coulomb}), which includes the implicit effect of the  environment 
on transition densities, whereas $J_{jk}^{r,\textrm{MMPol}}$ is the explicit MMPol term as described in Eq. (\ref{eq:EET-VMMPol}). 

\begin{figure}[htbp]
  \centering

  \includegraphics[width=3.5in]{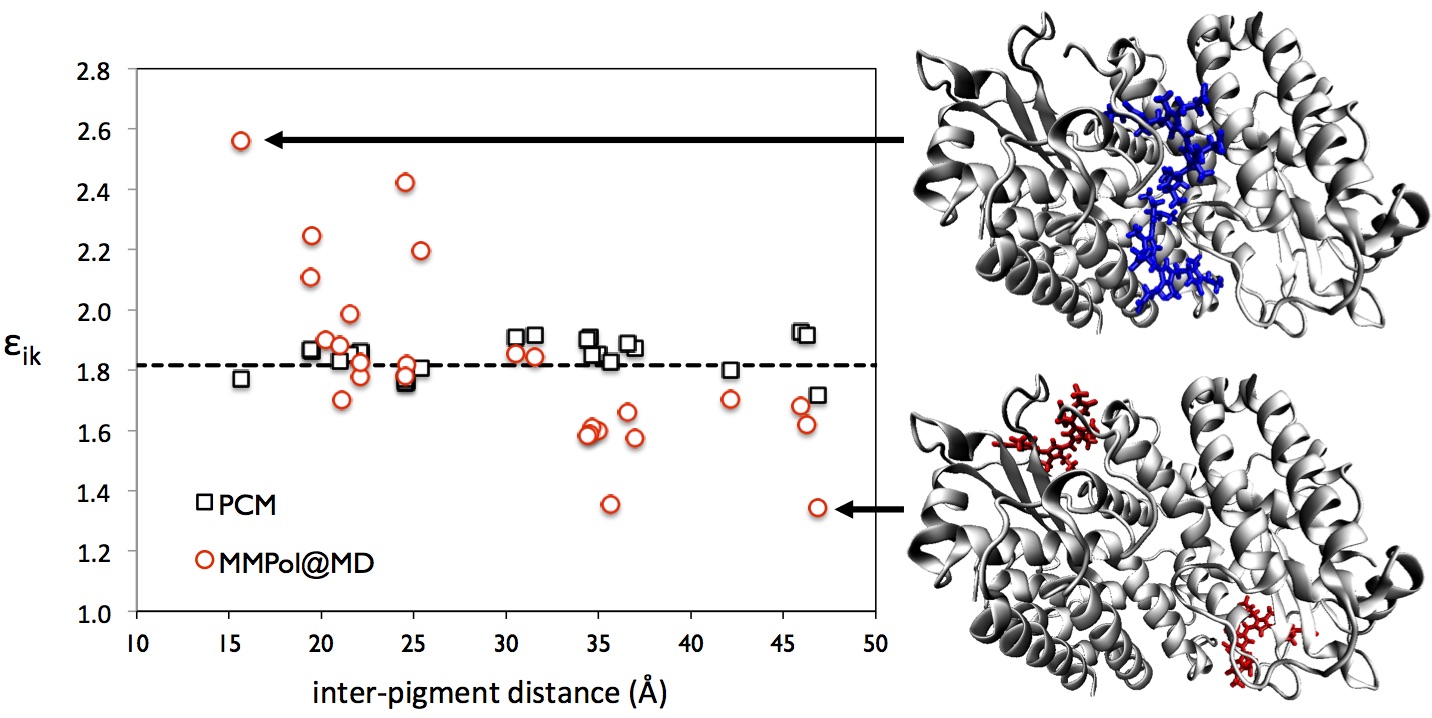}
  \caption{Effective dielectric constant corresponding to each pair ($\epsilon_{ik}$ ) reported as a function of the interchromophore distance: circles refer to the average QM/MMPol@MD data and squares to the QM/PCM calculations on the crystal structure. The dotted line indicates the average value obtained on all MMPol@MD values. The two insets indicate the two pairs showing the smallest (${\rm DBV_{19A}-DBV_{19B}}$) and largest (${\rm PEB_{50/61C}-PEB_{50/61D}}$) $\epsilon_{ik}$ value.}
    \label{fig:PE545-Fig2}
\end{figure}

As illustrated in Figure \ref{fig:PE545-Fig2}, PCM values of $\epsilon_{jk}$ are similar for all pairs while MMPol values present a significant spread  (from 2.6 to 1.2). On the other hand, PCM and MMPol values become very similar when averaged over all pairs. The agreement between averaged results can be explained through a careful analysis of the components of each model. In the continuum method, the screening is described in terms of a set of induced charges spreading on the surfaces of the cavities embedding the pigments. These charges represent the polarization of the environment induced by the electronic transition in the donor.  These are calculated in terms of the optical component of the dielectric permittivity used to represent the mixed protein-water environment (namely 2.0). In the MMPol approach, instead, the screening is calculated in terms of induced dipoles 
originating from the electronic transition in the donor. Within this description, the induced dipoles are determined by the atomic polarizabilities used to mimic the protein (and the water) atoms. The latter, when used to simulate the macroscopic polarization of the whole environment, gives an effective permittivity of 2.3, which is very close to that used in the PCM model.  The similarity between the continuum and the atomistic model, however, disappears when the pigment pairs are analyzed separately.   Each of them in fact is expected to feel a different local environment determined by the variety of residues, respectively presenting a different degree of polarization.

The smallest effective permittivity (1.35) is experienced by the peripheral ${\rm DBV_{19A}}$-${\rm DBV_{19B}}$ pair. On the contrary, for the central ${\rm PEB_{50/61C}-PEB_{50/61D}}$ pair, the coupling is significantly more attenuated by the protein than in other pairs and the resulting effective permittivity is the highest (2.57). 
To understand the origin of these differences, the screening obtained by the MMPol approach can be dissected into contributions arising from the different protein residues and waters. 
Examination of the central pair shows that all the polypeptide chains, together with waters, act to reduce the interaction, resulting in a large screening effect. 
A completely different picture appears for the peripheral pair. In this case, it seems that the protein is organized in such a way as to enhance the electronic coupling while further effect of the surrounding water adds a strong screening contribution.

\subsection{Bath spectral density}
As represented by Eq. (\ref{eq:spectral-density-decomp}), the contribution to the bath modes for a molecule embedded in a flexible environment can be classified into two different sources, one from the internal vibrations of the molecule and the other from the motions of the molecule within the environment.
Conventionally, it is assumed that the internal vibrations constitute only high-frequency underdamped modes, which are reflected in sharp peaks of the spectral density. The low frequency intermolecular modes are often viewed as being induced entirely by the surrounding environment, resulting in a continuous contribution to the spectral density.    However, for the case of pigment molecules in LHCs, intramolecular vibrations are also shown to make significant contributions to the low frequency modes.   

Accurate determination of spectral densities ${\mathcal J}_{jk}(\omega)$ of Eq. (\ref{eq:spectral-density}) is a difficult task in general, let alone confirming the validity of the form of the $\hat H_{_{LHC}}$ as assumed in Eq. (\ref{eq:h_lhc_ho}).  For the case where there is only one pigment molecule, the corresponding spectral density can be determined experimentally using either spectral hole burning (SHB) or fluorescence line narrowing (FLN).  The former removes inhomogeneity by burning specific zero-phonon transition and the latter by obtaining fluorescence lineshape following excitation at the red edge of the ensemble absorption lineshape.   Model spectral densities fitted to these experimental ones 
have been used for the construction of EBHs.   

A popular example of the model bath spectral density is the following Ohmic spectral density with Drude cutoff:
\be
{\mathcal J}_{_{Drude}}(\omega)=2\lambda\frac{\omega/\omega_c}{1+(\omega/\omega_c)^2}  \ .
\ee
For this spectral density, the well-known choice of parameters for the FMO complex is $\lambda=35 \ {\rm cm^{-1}}$ and $\hbar\omega_c =106 {\rm\ cm^{-1}}$.    Although not as widely as in the case of the FMO complex, this spectral density was also used for the LH2 complex \cite{meier-jcp107-2,zhang-jcp108,chen-jcp131} with $\lambda=240\ {\rm cm^{-1}}$ and $\hbar\omega_c = 40.8\ {\rm cm^{-1}}$ based on the fitting \cite{meier-jcp107-2} of photon-echo data \cite{jimenez-jpcb101}.  Another well-known form is the Ohmic or super-Ohmic spectral density with exponential cutoff.  A combination of these forms were constructed for the B850 unit of the LH2 complex based on the fitting of fluorescence-line narrowing data as follows:
\ben 
{\mathcal J}_{_{JNS}}(\omega)&=&\pi\hbar \left (\gamma_1\omega e^{-\omega/\omega_{c,1}}+\gamma_2 \frac{\omega^2}{\omega_{c,2}} e^{-\omega/\omega_{c,2}}\right . \nonumber \\
&&\left . +\gamma_3 \frac{\omega^3}{\omega_{c,3}^2}e^{-\omega/\omega_{c,3}} \right)\ ,
\een
with $\gamma_1=0.22$, $\gamma_2=0.78$, $\gamma_3=0.31$, $\hbar \omega_{c,1} =170\ {\rm cm^{-1}}$, $\hbar \omega_{c,2}=34\ {\rm cm^{-1}}$, and $\hbar \omega_{c,3}= 69\ {\rm cm^{-1}}$.  The above spectral density (or simpler version with only Ohmic term) was used for the modeling of low temperature SMS lineshapes \cite{jang-jcp118-2,jang-jpcb115,kumar-jcp138} and calculation of exciton transfer rates \cite{jang-prl92,jang-jpcb111,jang-prl113}.   

Recently, more refined experimental technique called $\Delta$-FLN, which combines the merits of SHB and FNL, was developed \cite{timpmann-jpcb108,ratsep-jl127} and applied to various LHCs \cite{ratsep-jpcb112,pieper-jpcb115,kell-jpcb117}.  In particular,  a recent analysis \cite{kell-jpcb117} suggested that low frequency region of the model Ohmic spectral density is not consistent with the experimental lineshape, proposing a new form  with log-normal distribution in the low frequency limit. While this exposes a potential deficiency of conventional model spectral densities that typically assume algebraic behavior in the small frequency limit, it is not yet clear whether such log-normal distribution is an apparent effect of
residual degree of inhomogeneity and anharmonic contribution of the bath.   In addition, for general LHCs with more than one pigment molecules with similar excitation energies, 
it is not yet clear how even the $\Delta$-FLN approach can accurately determine site-specific spectral densities represented by  Eq. (\ref{eq:spectral-density}).

Computational determination of the spectral densities are feasible, but are also beset with a few theoretical and practical issues.   A well-known approach is to use a mixed quantum-classical approach assuming that the bath is classical, and to calculate the following bath energy gap correlation function through classical MD simulation \cite{zwier-jcc28,olbrich-jpcl2,shim-bj102,aghtar-jpcb117,viani-pccp16,chandrasekaran-jpcb119}: 
\be
{\mathcal C}_{E,jk}^{cl}(t)=\langle \Delta E_j(t)\Delta E_k(0)\rangle_{cl}\ ,
\ee
where $\Delta E_j(t)$ is the difference of the bath energy for $|s_j\rangle$, which is created at time $t=0$, and that for the ground electronic state.  Assuming that this is related to the real part of the quantum correlation function of $\hat B_{jk}$ defined in Eq. (\ref{eq:h_lhc-gen}), one can calculate the spectral density using the following relation \cite{olbrich-jpcl2}:
\be
{\mathcal J}_{jk}(\omega)=\frac{2}{\hbar}\tanh (\frac{\beta\hbar\omega}{2}) \int_0^\infty dt\ {\mathcal C}_{E,jk}^{cl}(t) \cos (\omega t) \ . \label{eq:standard}
\ee
On the other hand, if we rely on the approximation of Eq. (\ref{eq:h_lhc_ho}) and introduce mass-weighted normal mode $Q_n$ such that $\hbar \omega_j g_{j,n}(b_n+b_n^\dagger)=\sqrt{2\hbar}\omega_n^{3/2}g_{j,n} Q_n$, it is straightforward to show that 
\ben
{\mathcal C}_{E,jk}^{cl}(t)&=&2\hbar \sum_n g_{j,n}g_{k,n}\omega_n^3 \langle Q_n(t)Q_n(0)\rangle_{cl}\nonumber \\
&=&\frac{2}{\pi\beta}\int_0^\infty\ d\omega\ {\mathcal J}_{jk}(\omega) \frac{\cos (\omega t)}{\omega} \ .
\een
Then, through cosine transformation, 
\be
{\mathcal J}_{jk}(\omega)=\beta\omega \int_0^\infty dt\ {\mathcal C}_{E,jk}^{cl}(t)\cos (\omega t)\ . \label{eq:spectral-harmonic}
\ee
While this becomes equivalent to Eq. (\ref{eq:standard}) if the vibrational quanta of all the bath modes are smaller than thermal energy, the discrepancy between the two becomes substantial for high frequency vibrational modes.  Given that harmonic oscillator approximation is more appropriate for high frequency modes, which result mostly from intramolecular vibration, Eq. (\ref{eq:spectral-harmonic}), known as harmonic approximation, is considered more reliable than Eq. (\ref{eq:standard}) in general \cite{valleau-jcp137,chandrasekaran-jpcb121}.  For the bath modes coming from protein environments, this may not be necessarily true because of anharmonic and nonlinear effects.  

Another important issue is that  a long time window is required in order to achieve a complete sampling for low-frequency vibrations. For example, a vibration at $\sim$10~cm$^{-1}$ has a period of $\sim$3~ps, requiring a sampling time window of at least 100~ps. Conversely, to sample intramolecular modes, one needs a rather short time step, i.e. $\le 5$~fs \cite{chandrasekaran-jpcb119,valleau-jcp137,shim-bj102}.
These two opposing factors make the mixed approach computationally very expensive, requiring tenths of thousands of electronic structure calculations for a single spectral density. 

The most concerning issue of the mixed quantum-classical approach is 
the quality of the MD trajectory involved. In fact, an assumption of the spectral density approach is that the relevant dynamics come from the evolution of the energy gap dictated by the ground-state Hamiltonian. However, the dimensions of the systems under study and the time scales involved compel the nuclear trajectory to be computed with a classical force field, {\it i.e.}, using fitted parameters, whereas the energy gap is calculated with a quantum mechanical Hamiltonian. The Hamiltonian used to propagate the nuclear positions may be critically different from the one used in the calculation of the energy gap, leading to an \emph{effective} excited-state PES that is completely different from the ``real'' quantum mechanical one. The ground-state PES could differ in equilibrium position, normal modes, and frequencies.  All of these factors contribute to the shape of the spectral density \cite{zwier-jcc28,lee-arpc67,kim-jpcl6}. Indeed, it has been observed that the resulting spectral density is strongly dependent on the force field parameters, rather than on the quantum mechanical Hamiltonian \cite{aghtar-jpcb117,chandrasekaran-jpcb119,wang-pccp17}. For this reason, in order to effectively use the mixed quantum-classical method, one should carefully assess the quality of the force field parameters by comparing them to the quantum-mechanically calculated PES, and possibly develop \emph{ad-hoc} parameters specifically designed for subsequent excitation energy calculations \cite{prandi-jcc37}.

Another well-established approach to the calculation of spectral densities is the direct calculation of Huang-Rhys factors from the gradient of the excited-state PES at the ground-state equilibrium geometry. The peaks of the spectral density can be broadened by  gaussian or Lorentzian functions in order to take into account the finite vibrational lifetime and the inhomogeneous distribution of vibrational frequencies \cite{lee-arpc67}. Usually, only the discrete part of the spectral density can be included in such a treatment, but the effect of the surrounding environment on the normal mode frequencies and Huang-Rhys factors can be included by using multiscale approaches \cite{lee-arpc67}.

At the ground-state equilibrium, the gradient of the excited state PES is equal to the gradient of the energy gap $U$.  In this position, the excited-state gradient is often called vertical gradient (VG). For multiple modes, the energy gap can be expressed, in mass-weighted coordinates, as follows:
\begin{equation}
\Delta E_{j}({\bf Q})=E_j-E_g-\sum_{n}\left(\omega_{n}^{2}d_{j,n}Q_{n}+\frac{1}{2}\omega_{n}^{2}d_{j,n}^{2}\right)\ ,
\end{equation}
where the displacement $d_{j,n}$ of mode $n$ is related to the derivative $f_{j,n}$ of the excited-state PES by $d_{j,n} =  -f_{j,n}/\omega_{n}^{2}$. The VG in normal coordinates $\tilde{\mathbf{f}_j}$ can be obtained from the cartesian VG, ${\mathbf{f}_{c,j}}$, as follows:
\begin{equation}
\tilde {\mathbf{f}_j}=\mathbb{P}^{\dag}\mathbb{M}^{\frac{1}{2}}\mathbf{f}_{c,j} \ ,
\end{equation}
where $\mathbb{M}$ is a diagonal matrix containing nuclear masses and $\mathbb{P}$ is a rectangular matrix whose columns are the normal modes expressed in mass-weighted Cartesian coordinates. Finally, the Huang-Rhys factor of each mode can be calculated by
\begin{equation}
S_{j,n}=\frac{\lambda_{j,n}}{\hbar\omega_{n}}
=\frac{\omega_{n}d_{j,n}^{2}}{2\hbar}=\frac{f_{j,n}^{2}}{2\hbar\omega_{n}^{3}} \ .
\end{equation}
In this explicit approach, the linear dependence of the energy gap on all normal coordinates can be assessed.  More recently, it was confirmed that the mixed approach of the MD simulation and direct calculation of Huang-Rhys factor results in a spectral density in fairly good agreement with experimental result \cite{lee-jpcl7}.
Notably, the Huang-Rhys factors may be explicitly calculated also for the intermolecular motions. The dimensions of the system, however, do not allow a quantum-mechanical calculation of normal modes and vertical gradients, which can instead be obtained through classical modeling of the PES and the pigment-protein interactions \cite{renger-jpcb116}. 
However, the intermolecular motions may not be well described by harmonic potentials, and the energy gap could be highly nonlinear with the intermolecular coordinates.

Very recently, an extension of the method has been presented by combining VG with force-field based MD \cite{lee-jacs139}: the sampled configurations are used to initiate QM ground state optimization of chromophore geometries in the presence of the instantaneous local fields provided by the MM partial charges of the surrounding protein environment. Ground and excited state properties are then computed at these optimized geometries to parametrize an ensemble of instantaneous local system-bath model Hamiltonians. The method has been successfully applied to two phycobiliprotein structures, namely PE545 and PC645 complexes.

Contrary to the discrete part of the spectral density, the continuous, low-frequency part is much more challenging to be computed at QM level, and it may require extensive calculations along an MD trajectory. Moreover, the intermolecular motions that give rise to the low-frequency part are strongly dependent on the temperature, as some barriers become accessible only when enough thermal energy is present in the system. Therefore, the size and shape of the continuous spectral density may indeed be dependent on temperature \cite{rancova-jpcb118}. 

\begin{figure}[htbp]
\includegraphics[width=3.4in]{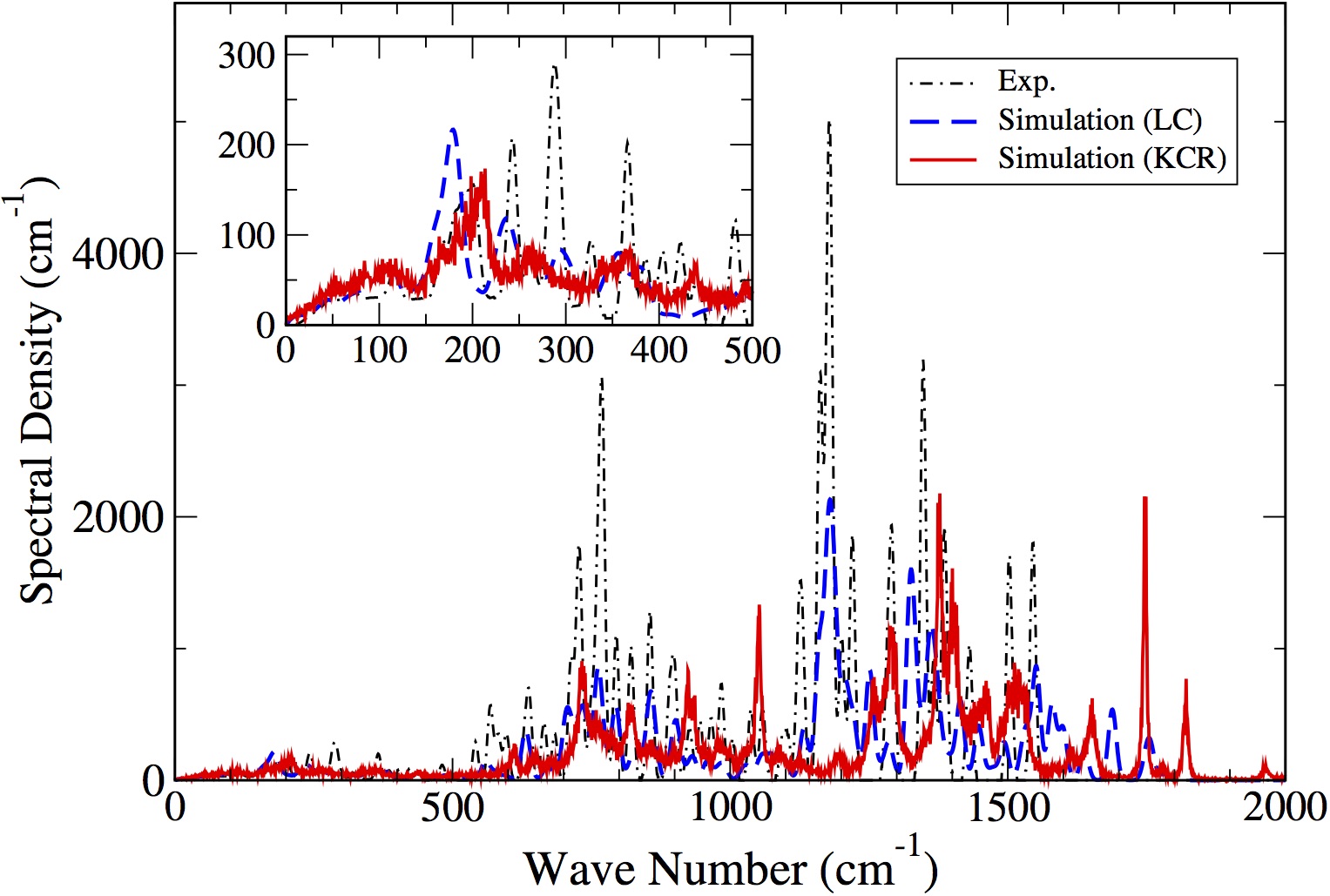}
\caption{Bath spectral densities for BChl-3 of the FMO complex.  ``Exp." represents the experimental  $\Delta$-FNL data in \cite{ratsep-jl127}, ``Simulation (LC)" the theoretical spectral density of \cite{lee-jpcl7}, and ``Simulation (KWC)" the theoretical spectral density of \cite{kim-pccp20}.   The inset shows close-up of the lower frequency region.  All the data have been provided by Young Min Rhee.}
\label{fig:fmo3-spd}
\end{figure}

As an alternative approach, Rhee and coworkers recently developed \cite{park-jpcb116,kim-jctc12} a new scheme that combines interpolated DFT-level calculations of pigment and molecular mechanics calculations of protein environments.   This approach constructs on-the-fly quantum portion of potential energy surface through interpolation from pre-calculated database, and is efficient while maintaining reasonable accuracy of the dynamics.  As a result, dynamics lasting up to about $100\ {\rm ns}$ has been shown to be possible \cite{kim-pccp20}.  Figure  \ref{fig:fmo3-spd} compares the spectral densities calculated in this manner with that based on the mixed MD and HR factor calculation method \cite{lee-jpcl7}.  Although there is discrepancy between the two theoretical approaches, which can be related to motional narrowing effects and different force fields and conditions, both of these are in reasonable agreement with the experimental results.  Although the issues of anharmonic and nonlinear coupling effects of the bath still need to be examined more carefully, the level of agreement as can be seen in Fig. \ref{fig:fmo3-spd} provides a strong support for the validity of the  conventional EBH for the FMO complex.

\section{Exciton-radiation interaction, lineshape functions, and response functions}
The total Hamiltonian for the LHC in the presence of radiation is the sum of Eq. (\ref{eq:h_lhc-gen}) and the matter-radiation interaction Hamiltonian $\hat H_{int}(t)$ as follows:
\be
\hat H_T(t)=E_g|g\rangle\langle g|+\hat H_{ex}+\hat H_{int}(t)\comma \label{eq:h_tot}
\ee
where $\hat H_{ex}$ is the single exciton-bath Hamiltonian defined by Eq. (\ref{eq:h_lhc-gen}).   In general, double exciton states should also be included for a complete description of general four-wave mixing spectroscopy. While this is straightforward, we omit this contribution here for the sake of simplicity. 
We denote the transition dipole vector for the excitation from $|g\rangle$ to $|s_k\rangle$ as $\mbox{\boldmath$\mu$}_k$.  Then, the total electronic polarization operator for the transitions to the single exciton space is given by 
\ben
\hat {\bf P}&=&\sum_{k} \mbox{\boldmath$\mu$}_k \left (|s_k\rangle \langle g|+|g\rangle\langle s_k|\right)  \nonumber \\
&=&\sum_j \left ({\bf D}_j |\varphi_j\rangle\langle g|+|g\rangle\langle \varphi_j| {\bf D}_j  \right)  \ , \label{eq:p_def}
\een
where ${\bf D}_j=\sum_k \mbox{\boldmath$\mu$}_k U_{kj}^*$.    

The theories of absorption and emission lineshapes for Eq. (\ref{eq:h_tot}) are well established.  For example, for the case of radiation with frequency $\omega$ and polarization {\boldmath{${\eta}$}}, the absorption lineshape function is given by 
\ben
&&I (\omega)=\frac{1}{\pi}{\rm Re} \int_{0}^{\infty} dt \ e^{i\omega t } \nonumber \\
&&\times \left \langle e^{\frac{i}{\hbar} t E_g} Tr\left\{e^{-\frac{i}{\hbar}t \hat H_{ex}}|D\rangle \langle D|\hat \rho_b e^{\frac{i}{\hbar} t \hat H_b}\right\}\right\rangle   \ ,  \label{eq:abs-lhc}
\een
where $|D\rangle=\sum_j \mbox{\boldmath{${\eta}$}} \cdot  {\bf D}_j|\varphi_j\rangle$ and $\langle \cdots \rangle$ represents all averaging over the disorder within the ensemble of the sample and polarization direction.  For the sake of completeness, Appendix C provides a derivation of the above expression starting from a more general time dependent Hamiltonian. Alternatively, one can obtain the above expression from the time correlation function of the polarization operator, Eq. (\ref{eq:p_def}).
Similarly, the emission lineshape function is given by 
\ben
&&E(\omega)=\frac{1}{\pi}{\rm Re} \int_{0}^{\infty} d t \ e^{-i\omega t } \nonumber \\ 
&&\times \left \langle e^{-\frac{i}{\hbar} t E_g} Tr\left\{e^{-\frac{i}{\hbar} t \hat H_b} |D\rangle \langle D|\hat \rho_{ex} e^{\frac{i}{\hbar}t \hat H_{ex}} \right \}  \right\rangle \ , \label{eq:ems-lhc}
\een
where $\hat \rho_{ex}=e^{-\beta \hat H_{ex}}/Tr\{e^{-\beta \hat H_{ex}} \}$.  Appendix C provides a derivation of the above expression as well.

Four wave mixing spectroscopies are useful for interrogating the dynamics of excitons because they offer more selective information on the exciton space with appropriate choice of pulse sequences and phase matching conditions.  Theories of four wave mixing and multidimensional electronic spectroscopy are well established \cite{mukamel,cho-cr108,abramavicius-cr109}.  However, in comparison to 2D vibrational spectroscopy for which accurate computational modeling of signals have been shown to be feasible for a broad range of systems, accurate modeling of 2DES remains challenging for most of systems.

\begin{figure}[htbp]
\noindent
(a)\makebox[1.5in]{ }(b)\makebox[1.5in]{ }\\
\hspace{-.5in}\makebox[.2in]{ } \includegraphics[width=2in]{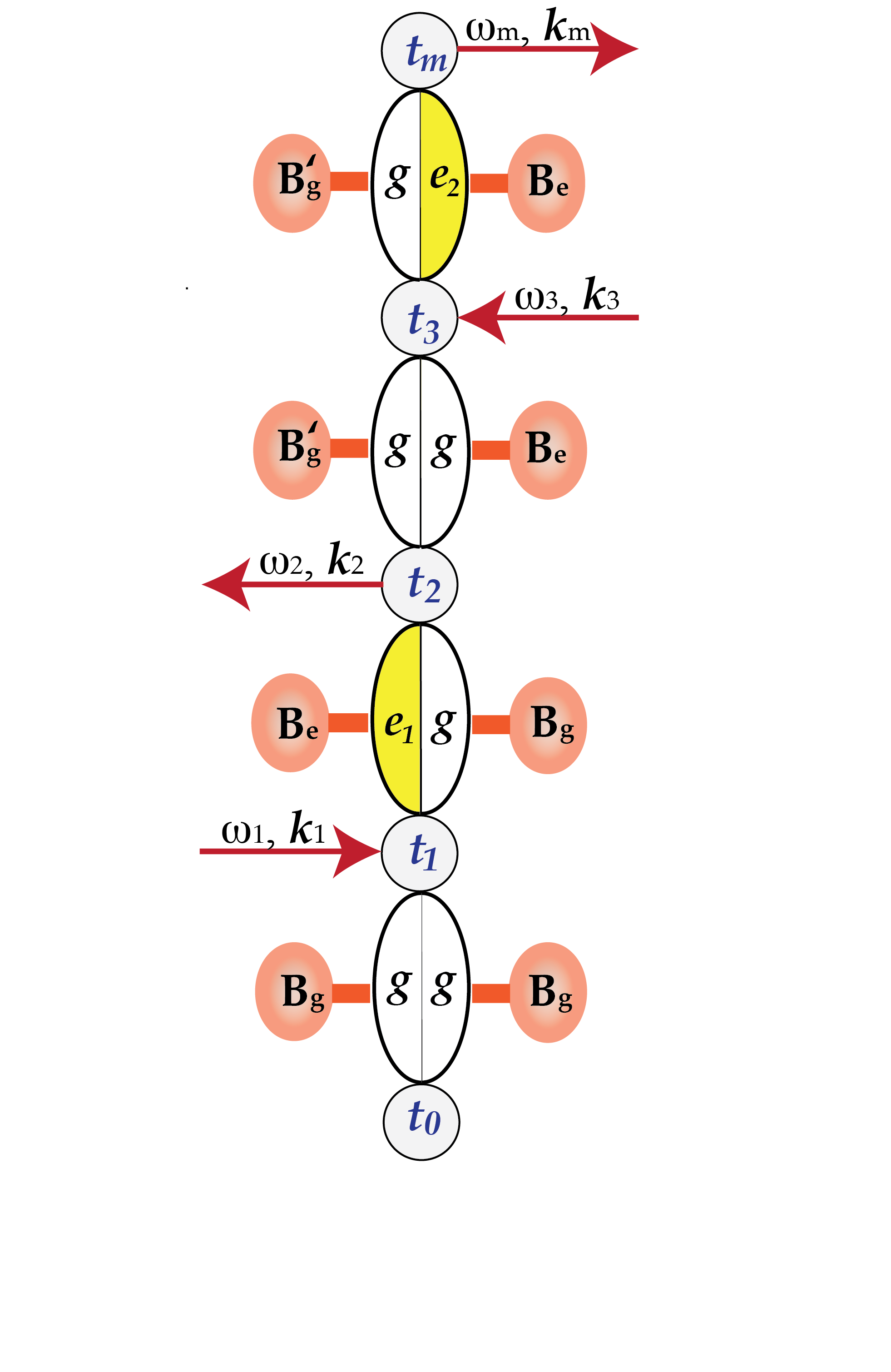}
\hspace{-.4in}\includegraphics[width=2in]{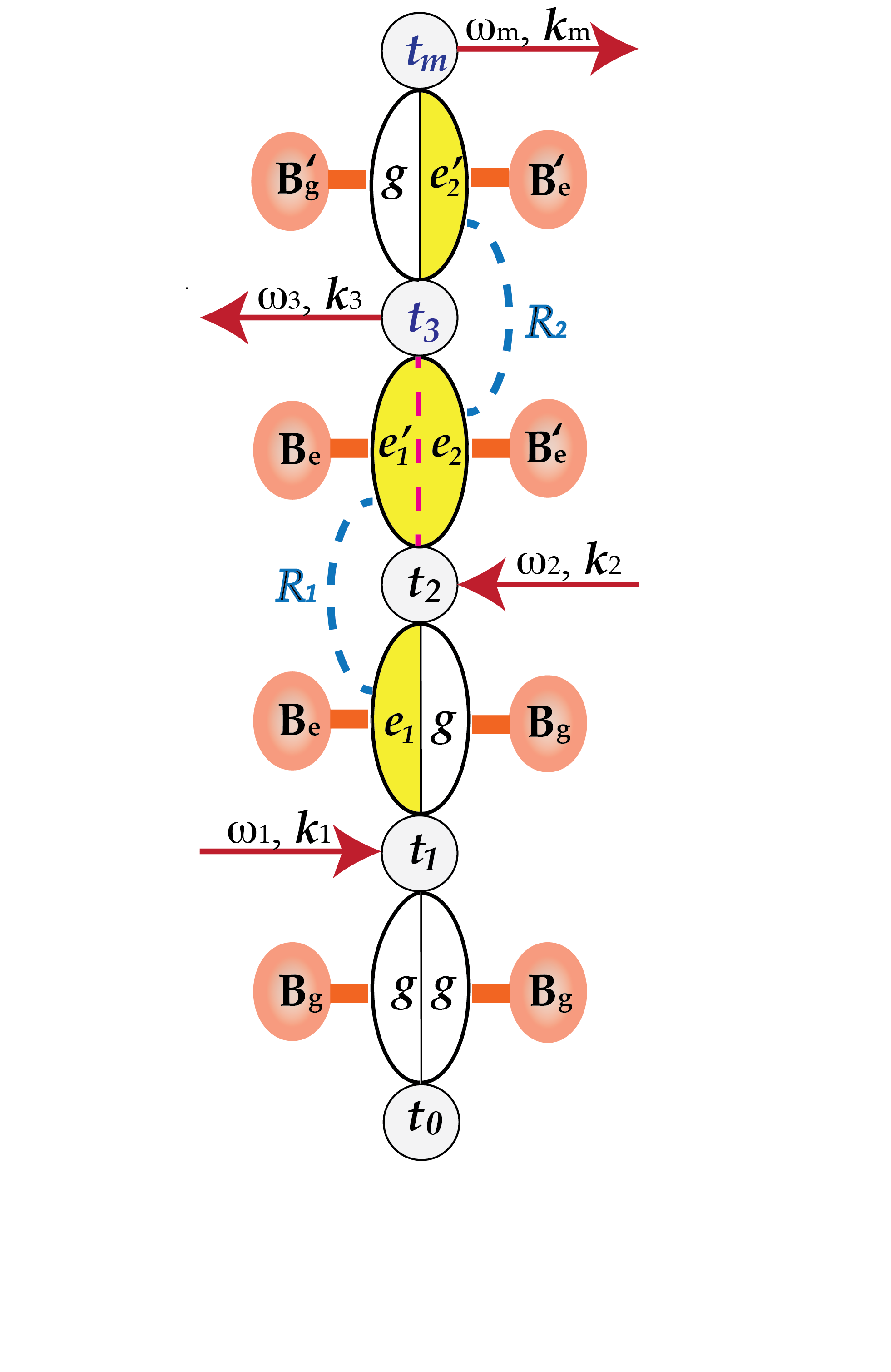}\\
(c)\makebox[1.5in]{ }(d)\makebox[1.5in]{ }\\
\hspace{-.5in}\makebox[.2in]{ }\includegraphics[width=2in]{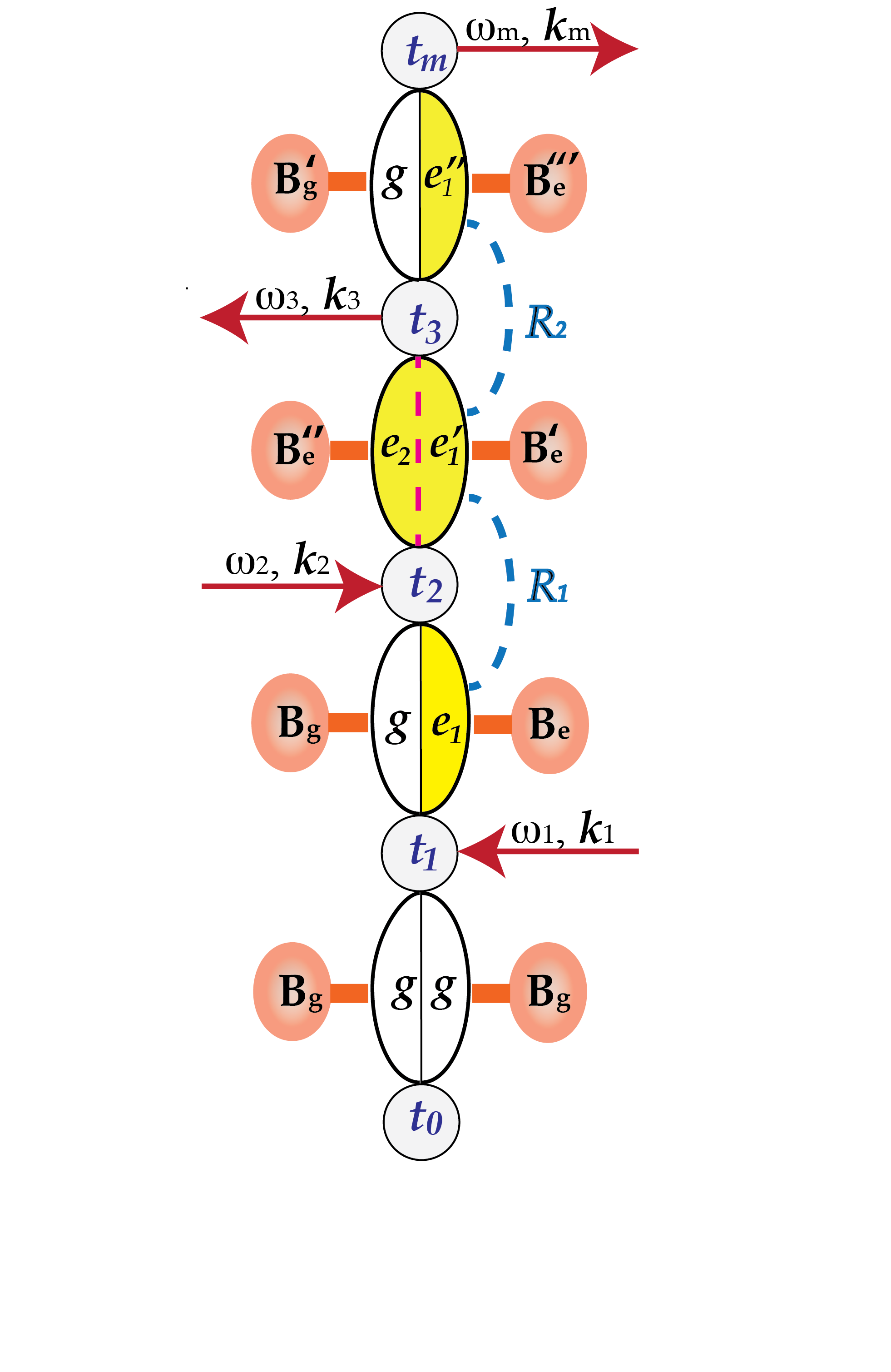}
\hspace{-.45in}\includegraphics[width=2in]{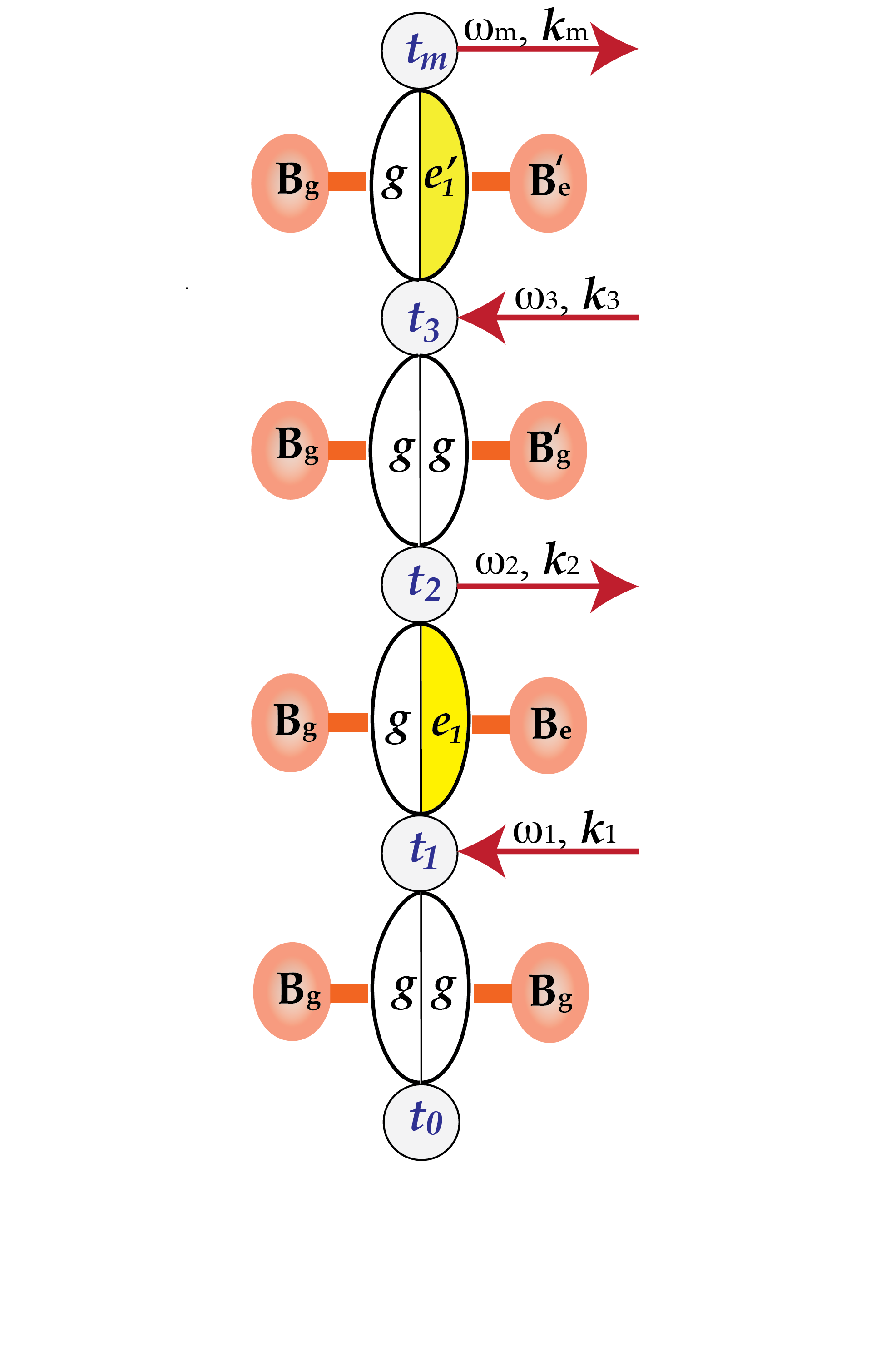}
\caption{Diagrams representing four terms contributing to the third order polarization. The first two terms (a) and (b) correspond to echo signals satisfying the phase matching condition of ${\bf k}_m={\bf k}_3+{\bf k}_2-{\bf k}_1$, and are in general called rephasing terms. The last two terms satisfy a different phase matching condition of ${\bf k}_m={\bf k}_3-{\bf k}_2-{\bf k}_1$ and are in general called non-rephasing terms. Blue dashed curves represent relaxation and decoherence caused by the bath that continue propagating across interaction with radiation pulse, and $R_k$ with different $k$ represents a different bath relaxation operator.  $B_g$ represents the bath of ground electronic state and $B_e$ the bath of the manifold of the single exciton states.  Different bath states created following interaction with the radiation are expressed as different number of primes in the superscript.}
\label{fig:four-wave-diagram}
\end{figure}

Appendix C provides a general expression for the third order polarization, Eq. (\ref{eq:p4t}), the major observable of four-wave mixing spectroscopy, and detailed expressions for its components, Eqs. (\ref{eq:pt1})-(\ref{eq:pt4}).  
The key quantities containing the information on the material system are the third order response functions defined by Eqs. (\ref{eq:pt1})-(\ref{eq:pt4}). 2DES  is a collection of resulting third order polarizations, typically represented with respect to two frequencies of Fourier transform 
for the initial and final coherence times, $t_2-t_1$ and $t_m-t_3$, respectively (see Appendix C for the definition of these times).   In the absence of time dependent fluctuations and considering only transitions between the ground electronic state and single exciton states, the four response functions can be expressed as (see Appendix C for the details of derivation): 
 \ben 
&&\chi^{(1)}(t_m,t,t',t'')=\left \langle e^{\frac{i}{\hbar}(t'-t'')E_g} e^{-\frac{i}{\hbar}(t_m-t)E_g } \right . \nonumber \\
&& \hspace{.3in}\times Tr\left\{e^{-\frac{i}{\hbar}(t_m-t')\hat H_b} |D_m\rangle \langle D_2| e^{-\frac{i}{\hbar}(t'-t'')\hat H_{ex}} \right . \nonumber \\
&&\hspace{.3in}\times \left . \left . \hat \rho_b |D_1\rangle\langle D_3| e^{\frac{i}{\hbar}(t-t'')\hat H_b}e^{\frac{i}{\hbar}(t_m-t)\hat H_{ex}} \right\} \right \rangle\ , \label{eq:chi1-nt}\\
&&\chi^{(2)}(t_m,t,t',t'')=\left \langle e^{-\frac{i}{\hbar}(t_m-t)E_g} e^{\frac{i}{\hbar}(t'-t'')E_g } \right .\nonumber \\
&&\hspace{.3in}\times Tr\left\{e^{-\frac{i}{\hbar}(t_m-t)\hat H_b} |D_m\rangle \langle D_3|e^{-\frac{i}{\hbar}(t-t'')\hat H_{ex}} \right . \nonumber \\
&& \hspace{.3in}\times\left . \left . \hat \rho_b |D_1\rangle\langle D_2|  e^{\frac{i}{\hbar}(t'-t'')\hat H_b}e^{\frac{i}{\hbar}(t_m-t')\hat H_{ex}} \right\} \right \rangle\ ,  \label{eq:chi2-nt}\\
&&\chi^{(3)}(t_m,t,t',t'')=\left \langle e^{-\frac{i}{\hbar}(t_m-t)E_g} e^{-\frac{i}{\hbar}(t'-t'')E_g } \right . \nonumber \\
&&\hspace{.3in}\times Tr\left\{ e^{-\frac{i}{\hbar}(t_m-t)\hat H_b} |D_m\rangle \langle D_3| e^{-\frac{i}{\hbar}(t-t')\hat H_{ex}} \right . \nonumber \\
&& \hspace{.3in}\left . \left . \times  \hat \rho_b |D_2\rangle\langle D_1| e^{\frac{i}{\hbar}(t'-t'')\hat H_b}e^{\frac{i}{\hbar}(t_m-t'')\hat H_{ex}}  \right\} \right \rangle \ , \label{eq:chi3-nt}\\
&&\chi^{(4)}(t_m,t,t',t'')=\left \langle e^{-\frac{i}{\hbar}(t_m-t)E_g} e^{-\frac{i}{\hbar}(t'-t'')E_g } \right . \nonumber \\
&&\hspace{.3in}\times Tr\left\{e^{-\frac{i}{\hbar}(t_m-t'')\hat H_b}|D_m\rangle \langle D_1|\hat \rho_b e^{\frac{i}{\hbar}(t'-t'')\hat H_{ex}} \right . \nonumber \\
&&\hspace{.3in}\left . \left .\times |D_2\rangle\langle D_3| e^{\frac{i}{\hbar}(t-t')\hat H_b}e^{\frac{i}{\hbar}(t_m-t)\hat H_{ex}}  \right\} \right \rangle\ . \label{eq:chi4-nt}
\een 
Figure \ref{fig:four-wave-diagram} provides diagrammatic representations of the above four response functions, which are rephasing and non-rephasing terms of the ground state bleaching and stimulate emission terms.  
Excited state absorption terms are not shown because we limited the consideration to only single exciton states here.
 
Despite recent advances, theoretical understanding of 2DES signals for the exciton states of LHCs remains challenging.  Vibronic couplings of pigment molecules in protein environments are more pronounced than those of isolated pigment molecules.  Their contribution to spectroscopic signals can be significant even for BChls that are known to have very small HR factors.  Due to the fact that the exciton Hamiltonian does not commute with the exciton-bath coupling in general, a specific exciton state $|\varphi_j\rangle$ created by an incoming photon starts decohering and relaxing to other exciton states almost immediately.   These are depicted explicitly in Fig. \ref{fig:four-wave-diagram} by dashed (and blue, on line) lines.  In addition, the states of bath can be complicated in the exciton manifold because of non-adiabatic effects associated with multiple electronic states.  In LHCs, the rates of these decoherence and relaxation processes, which 
are due to vibronic couplings, are comparable to those due to purely electronic couplings that can happen if a specific site excitation state $|s_j\rangle$ can be created.  Furthermore, additional complications can arise due to the fact that excitations created spectroscopically in LHCs are far from localized site excitation states in general.   

Even for highly tuned excitation wavelength, what is being created is most likely to be  a subensemble of exciton states mixed together, which are degenerate in energy but have different extent in linear combinations of exciton states and in vibronic contributions.    Selection of specific polarizations can reduce the size of the subensemble, but the qualitative nature of the subensemble is expected to remain the same.  As a result, any attempt to detect the electronic coherence directly from 2DES in time is expected to be significantly hampered by the effect of subensemble dephasing.  

There are additional complications that can arise in the four-wave mixing spectroscopy of excitons in LHCs.  For the simple case with a well isolated single excited state, dephasing due to the disorder in the excitation energy is cancelled in the rephasing signal when $t_2-t_1=t_m-t_3$.  However,  as can be seen from Eqs. (\ref{eq:chi1-nt}) and (\ref{eq:chi2-nt}) (and also Figs. 12(a) and (b)), the rephasing signal contains multiple contributions involving closely spaced exciton states.  Therefore, full recovery of phase relation is not possible in this case because the disorder affects different exciton states within the subensemble in a different manner.  In addition, the contributions of vibrational modes in both the ground electronic state and single exciton states  are non-negligible.  
Unlike the electronic coherence, vibrational coherence is less susceptible to dephasing and can survive longer.   

The initial  interpretation that 2DES beating signals \cite{engel-nature466,collini-nature463} might reflect purely electronic coherence, is based on the assumption that vibronic couplings are negligible.  Even if this is the case, such coherence may not have significant implication in the context of exciton dynamics unless it links exciton states that are spatially well separated.   Let alone the question of how much the 2DES signal reflects the creation and evolution of excitons under natural irradiation of sun light, unambiguous interpretation of beating signals in general require clear resolution of major features of underlying Hamiltonian governing the dynamics of excitons and details of matter-radiation interaction \cite{ginsberg-acr42}. 
Unambiguous interpretation of spectroscopic signals and detailed computational study of spatio-temporal evolution of excitons are important in this respect.   Significant advances have been made in 
dealing with these issues in the past decade, but much still need to be accomplished.

\section{Theoretical description of exciton dynamics and spectroscopic observables}
\subsection{Overview of theoretical and computational approaches} 
Theoretical research on the dynamics of excitons has a long history dating back to early days of quantum mechanics.  In particular, for excitons formed in molecular aggregates and solids, there have been extensive theoretical and computational efforts to understand exciton migration kinetics and absorption/emission spectroscopic data \cite{kenkre-reineker,agranovich-hoch-pr,agranovich,silbey-arpc27}.  While these have laid fundamental basis for modern research on the dynamics of molecular excitons, they have relied heavily on phenomenological treatment of exciton relaxation and dephasing dynamics, and have focused mostly on steady state properties.  In addition, because systems that were studied consisted of either identical molecules or binary mixtures of host and guest molecules at most, underlying exciton Hamiltonians have had rather simple features. 

With the exception of chlorosome of green sulfur bacteria, which is a superaggregate of BChls, all LHCs can be viewed as host-guest systems of $10-100$ nm length scale sizes.  Proteins serve as host environments, and pigment molecules are guest molecules, serving as site basis for excitations.   The distinctive nature of LHCs compared to simple molecular host-guest systems is that  protein hosts play an active role in tuning excitation energies and controls spatial arrangement of pigment molecules to a great extent.  Therefore, accurate specification of the exciton Hamiltonian, exciton-bath coupling, and bath Hamiltonian terms themselves are important for reliable characterization of LHCs.  Because the magnitudes of these parameters are in the intermediate range, well established theoretical approaches developed for simple or limiting situations are either inappropriate or in need of verification by more advanced approaches.  In this regard, LHCs have motivated new advances in theories and computational methods. 

Theoretical approaches to describe and simulate exciton dynamics can be classified into three classes, depending on the level of information and the degree of accuracy sought after.  The minimal approach is to consider the time evolution of excitation/exciton populations.  Broadly, this can be called master equation (ME) approach.   The next level of description is to consider the time evolution of a reduced exciton density operator (RDO).  We here call this RDO approach.   Finally, the most complete but expensive one is to describe the time evolution of the full density operator (FDO) representing both exciton and the bath.  In practice, this FDO approach relies on being able to effectively represent the bath of infinite size in terms of finite ones.  Below we provide a brief overview of the three major approaches, and then explain how the outcomes of these theoretical calculations enable new understanding of the nature of excitons in LHCs.    

\subsubsection{Master equation approach and rate description}
In ME approach, the physical observables of interest are exciton populations, which are sufficient for calculating exciton mobility and the extent of delocalization.  An obvious choice of the unit of exciton here is each pigment.  In principle, it is possible to formally construct exact and generalized ME \cite{kenkre-prb9} as follows:
\be
\frac{d}{dt}p_j(t)=\sum_{k\neq j} \left\{{\mathcal W}_{k\rightarrow j}(t) p_k(t)-{\mathcal W}_{j\rightarrow k}(t) p_j(t)\right\}\ , \label{eq:me-pigment}
\ee
where $p_j(t)$ is the exciton population at the $j$th pigment molecule and ${\mathcal W}_{j\rightarrow k}(t)$ is the transfer rate of exciton population from the $j$th to the $k$th pigment that can be defined to be formally exact by accounting for all possible pathways.  This approach can serve as a general methodology for all LHCs if accurate rate expressions are available.  However, in practice, calculating the exact rate kernel entails accounting for all the many body quantum interactions and thus remains a challenging theoretical task.  

If the electronic couplings between site excitations of pigment molecules are small compared to other parameters and 
the steady state limit is assumed, one can approximate ${\mathcal W}_{j\rightarrow k}(t)$ with the following F\"{o}rster resonance energy transfer (FRET)  rate expression \cite{forster-ap,forster-dfs}:   
\ben
k_{j\rightarrow k}^F=\frac{J_{jk}^2}{2\pi \hbar^2} \int_{-\infty}^\infty d\omega L_j(\omega) I_k(\omega) \ . \label{eq:kfg_omg}
\een 
where we have adopted a more general definition of FRET by assuming that $J_{jk}$ is not limited to transition dipole interactions.  In the above expression, $L_j(\omega)$ and $I_k(\omega)$ are lineshape functions for the emission of the $j$th pigment and absorption of the $k$th pigment.  For example, let us assume that Eq. (\ref{eq:h_lhc-gen}) can be simplified as 
\ben
\hat H_{_{LHC}} &=& E_g|g\rangle\langle g|+\sum_{j=1}^{N_c} E_j |s_j\rangle\langle s_j|+\sum_{j=1}^{N_c}\sum_{k\neq j} J_{jk}|s_j\rangle\langle s_k| \nonumber \\
&&+\sum_{j=1}^{N_c} \left (\hat B_j |s_j\rangle\langle s_j| +\hat H_{b,j}\right)  \ ,\label{eq:h_lhc-loc}
\een 
where $\hat B_{j}$ and $\hat H_{b,j}$ are exciton-bath coupling and bath Hamiltonian localized to the site exciton state $|s_j\rangle$.  Then, the lineshape functions introduced in Eq. (\ref{eq:kfg_omg}) are defined as  
\ben
&&L_j(\omega)=\int_{-\infty}^\infty dt e^{-i\omega t+i(E_j -E_g)t/\hbar} \nonumber \\
&&\times \frac{1}{Z_{b,j}'}Tr_{b,j}\left\{e^{i(\hat H_{b,j}+\hat B_j)t/\hbar}e^{-i\hat H_{b,j}t/\hbar}e^{-\beta (\hat H_{b,j}+\hat B_j)}\right\} \ , \nonumber \\ &&\label{eq:ld_def}\\
&&I_{k}(\omega)=\int_{-\infty}^{\infty} dt\ e^{i\omega t -i(E_k-E_g) t/\hbar}\nonumber \\
&&\times \frac{1}{Z_{b,k}} Tr_{b,k}\left\{e^{i\hat H_{b,k}t/\hbar} e^{-i(\hat H_{b,k}+\hat B_k)t/\hbar} e^{-\beta \hat H_{b,k}}\right\} \ , \label{eq:iaw_def}
\een
where $Z_{b,j}'=Tr_{b,j}\{e^{-\beta (\hat H_{b,j}+\hat B_j)}\}$ and $Z_{b,k}=Tr_{b,k} \{e^{-\beta \hat H_{b,k}}\}$. 

The normalized emission lineshape (in the unit of $\tilde \nu=\omega/(2\pi c)$) for the $j$th pigment is expressed as
\be
f_j(\tilde\nu)=\tau_j\frac{2^5\pi^3 n_{j,r} \mu_j^2 c}{3\hbar} \tilde \nu^3 L_j(2\pi c\tilde \nu) \ ,\label{eq:fd-ed} 
\ee
where $n_{j,r}$ is the refractive index around the $j$th chromophore.   The molar extinction coefficient for the $k$th pigment is related to the lineshape function by the following relation:
\be
I_k(2\pi c\tilde \nu)=\frac{3000 (\ln 10)n_{k,r} \hbar}{(2\pi)^2 N_A\mu_k^2 \tilde \nu} \epsilon_k (\tilde \nu) \ , \label{eq:ia-ep}
\ee
where $n_{k,r}$ is the refractive index around the $k$th pigment.  Then, Eq. (\ref{eq:kfg_omg}) can be expressed as
\be
k_{j\rightarrow k}^{F}=\frac{9000 (\ln 10)}{128 \pi^5 N_A \tau_j} \frac{n_{j,r}}{n_{k,r}}\frac{J_{jk}^2}{\mu_j^2\mu_k^2} \int_{-\infty}^\infty d\tilde \nu \frac{f_j(\tilde \nu) \epsilon_k (\tilde \nu)}{\tilde \nu^4} \ .\label{eq:kfg-fd-ea}
\ee
This expression becomes the well-known F\"{o}rster's spectral overlap expression \cite{forster-ap} in the limit where $J_{jk}$ is due to the transition dipole-dipole interaction and the dielectric constants around $j$th and $k$th chromophores are the same.

More general expressions than Eq. (\ref{eq:kfg_omg}), which include nonequilibrium \cite{jang-cp275,jang-wires} and inelastic effects \cite{jang-jcp127,jang-wires}, are also available. However, the FRET rate expression or its generalizations, which are based on the second order approximation with respect to the electronic coupling between a pair of pigment molecules, are inappropriate for many LHCs because not many excitons are localized at single chromophores. 

Alternatively, one can extend the unit of exciton population as that residing in a group of strongly coupled pigment molecules.  If groups of chromophores can be identified such that electronic couplings between pigment molecules in different groups are weak enough, the rates of exciton transfer between them can be calculated, again employing a second order approximation with respect to the inter-group electronic couplings.  
This is the idea behind the multichromophoric FRET (MC-FRET) \cite{sumi-jpcb103,jang-prl92}, which was recently re-derived and tested more extensively \cite{ma-jcp142}.   

The use of MC-FRET rates as kernels of ME was formulated more rigorously by introducing the concept of modular excitons \cite{jang-prl113}.   Under the assumption that all the pigment molecules constituting an LHC can be divided into disjoint modules and that only the coarse-grained overall exciton population of each module, $\tilde p_n(t)$, (modular exciton density) is of interest,  one can consider the following generalized ME:
 \be
\frac{d}{dt}\tilde p_n(t)=\sum_{m\neq n} \left\{\tilde {\mathcal W}_{m\rightarrow n}(t) \tilde p_m(t)-\tilde {\mathcal W}_{n\rightarrow m}(t) \tilde p_n(t)\right\}\ . \label{eq:me-module}
\ee
As in the case of Eq. (\ref{eq:me-pigment}), exact formal expression for $\tilde {\mathcal W}_{n\rightarrow m}(t)$ can be found easily \cite{jang-prl113}.   In practice, approximations are needed. Under the assumption that the inter-module electronic couplings are small compared to other parameters and that the exciton density within each module reaches the stationary limit quickly, it is possible to show that \cite{jang-prl113}      
\be
\tilde {\mathcal W}_{ n\rightarrow m}(t)=\sum_{j'j'' }\sum_{k'k''}\frac{J_{j'k'}J_{j''k''}}{2\pi\hbar^2}\int_{-\infty}^{\infty} d\omega L_{j''j'}^n(t,\omega)I_{k'k''}^m (\omega) \comma \label{eq:gme-med-rate}
\ee
where $L_{j''j'}^n(t;\omega)$  and $I_{k'k''}^m(\omega)$ are lineshape matrix elements representing the time dependent emission of module $n$ and the absorption of module $m$.  These are respectively expressed as
\ben
&&L_{j''j'}^n(t;\omega)\equiv 2 {\rm Re} \left [\int_0^t dt' e^{-i\omega t'-iE_gt'/\hbar}  \right . \nonumber \\
&&\times \left . Tr_{bn}\left \{\langle s_{n,j''}|e^{-i\hat H_{b,n}t'/\hbar} \rho_{ns} e^{i\hat H_n t'/\hbar}|s_{n,j'}\rangle \right\} \right ] \label{eq:ed-def} \ , \\
&&I_{k'k''}^m(\omega)\equiv\int_{-\infty}^{\infty} dt e^{i\omega t+iE_gt/\hbar} \nonumber \\
&&\times Tr_{bm}\left\{\langle s_{m,k'}|e^{i\hat H_{b,m}t'/\hbar}e^{-i\hat H_mt'/\hbar}\rho_{bm}^g|s_{m,k''}\rangle\right\} \ , \label{eq:ia-def}
\een  
where $\hat H_n$ ($\hat H_m$) and $\hat H_{b,n}$ ($\hat H_{b,m}$) are respectively the EBH in the single exciton space and the bath Hamiltonian of the $n$th ($m$th) module. 
In the limit of $t\rightarrow \infty$, Eq. (\ref{eq:gme-med-rate}) approaches the MC-FRET rate \cite{jang-prl92,jang-prl113}.

\subsubsection{Reduced density operator approach}
For pigment molecules with moderate or strong electronic couplings, description of the dynamics at the level of a reduced density operator (RDO) in the exciton manifold serves as a better platform than restricting the focus on the exciton population only.  Let us denote the RDO at time as $\hat \sigma(t)$.   Then, the quantum master equation (QME) governing the time evolution equation of $\hat \sigma (t)$ can in general be expressed as 
\be
\frac{d}{dt}\hat \sigma (t)=-\frac{i}{\hbar}[\hat H_e, \hat \sigma (t)]-\hat {\mathcal R}[t;\hat \sigma] +\hat {\mathcal I}[t]\ ,
\ee
where the first term represents the dynamics due to the exciton Hamiltonian, $\hat {\mathcal R}[t;\hat \sigma]$ accounts for the relaxation and dephasing due to environments, and $\hat {\mathcal I}[t]$ is the inhomogeneous term that reflects the effect of the initial condition.

There is a large body of literature available  on the derivation of  formally exact QME and approximations.  For example, for the case of the interaction picture RDO defined as $\hat \sigma_I(t)=Tr_b\{\hat \rho_I(t)\}$, where $\hat \rho_I(t)$ is the total density operator in the interaction picture, the following formally exact QME is well-known:
\ben
&&\frac{d}{dt} \hat \sigma_I(t) =   \nonumber \\
&&-iTr_b\left\{\hat {\mathcal L}_{eb,I}(t) \exp_{(+)} \left[-i\int_0^t d\tau {\mathcal Q}\hat  {\mathcal L}_{eb,I}(\tau) \right] {\mathcal Q} \hat \rho_I(0)\right\} \nonumber \\
&&-\int_0^t d\tau Tr_b\left\{\hat {\mathcal L}_{eb,I}(t) \exp_{(+)} \left[-i\int_\tau^t d\tau' {\mathcal Q} \hat {\mathcal L}_{eb,I}(\tau') \right] \right . \nonumber \\
&&\hspace{.6in}\times  {\mathcal Q} \hat {\mathcal L}_{eb,I} (\tau) \rho_b\bigg \} \hat \sigma_I(\tau) \ .\label{eq:qme-formal}
\een 
In the above equation, $\hat {\mathcal L}_{eb,I}(\cdot)=[\hat H_{eb,I}(t), (\cdot)]/\hbar$, ${\mathcal Q}(\cdot)= 1-Tr_b\{(\cdot)\}$. 

Equation (\ref{eq:qme-formal}), although exact, is not useful in practice because $\exp_{(+)}[-i\int_\tau^t d\tau' {\mathcal Q}\hat {\mathcal L}_{eb,I}(\tau')]$ cannot be determined without full information on the system and bath degrees of freedom.  For practical calculations, approximations are made in general, although recent theoretical advances \cite{shi-jcp120,zhang-jcp125,cohen-njp15,kelly-jcp144} suggest that exact numerical evaluation of this term is feasible.  

One of the simplest and most popular approximation is the 2nd order approximation with respect to $\hat {\mathcal L}_{eb,I}(t)$, which results in
\ben
&&\frac{d}{dt} \hat \sigma_I(t) = -iTr_b\left\{\hat { \mathcal L}_{sb,I}(t) \hat  \rho_I(0)\right\} \nonumber \\
&&-\int_0^t d\tau Tr_b\left\{\hat {\mathcal L}_{sb,I}(t) \hat {\mathcal L}_{sb,I}(\tau) {\mathcal Q}\hat  \rho_I(0)\right\} \nonumber \\
&&-\int_0^t d\tau Tr_b\left\{{\mathcal L}_{sb,I}(t) {\mathcal L}_{sb,I} (\tau) \rho_b\right\} \sigma_I(t^*) \ ,\label{eq:2tnl-2}
\een 
where $t^*$ can be either $\tau$ or $t$.  In the former case, the above equation becomes time-nonlocal (TN).  In the latter case, it becomes time-local (TL).  Which choice works better depends on the nature of the system-bath couplings and the bath dynamics, but most numerical examples so far suggest that the TL equation performs better than the TN equation, 
at least at room temperature where the bath dynamics become most likely Gaussian \cite{palenberg-jcp114,chen-jcp131}. Analyses of the fourth order QME \cite{jang-jcp116} and steady state limits \cite{fleming-pre83} suggest that this is because the 2nd order TL equation can account for some of the effects of the 4th order terms of TN equations.

In Eq. (\ref{eq:2tnl-2}), the super-operator involving the second order correlations of ${\mathcal L}_{sb}(t)$, which appears commonly in both inhomogeneous and homogeneous terms, can be expressed as
\ben
&& \int_0^t d\tau Tr_b\left\{\hat {\mathcal L}_{sb,I}(t) \hat {\mathcal L}_{sb,I} (\tau) \rho_b\right\} (\cdot) \nonumber \\
&&=\sum_j\sum_{k} \int_0^t d\tau {\mathcal C}_{jk}(t-\tau)[\hat s_j(t), \hat s_{k}(\tau) (\cdot)]  +{\rm H.c.} \ , \nonumber \\
\een  
where  ${\rm H.c.}$ denotes Hermitian conjugates of all the previous terms,  $\hat s_j(t)=e^{iH_e t/\hbar}|s_j\rangle\langle s_j|e^{-iH_e t/\hbar}$, and ${\mathcal C}_{jk}(t)$ is the bath correlation function for sites $j$ and $k$ defined as  
\ben 
&&{\mathcal C}_{jk}(t)=\sum_r \omega_n^2 g_{j,r}g_{k,r} \nonumber \\
&&\hspace{.6in}\times Tr_b \left\{ (b_r e^{-i\omega t}+b_r^\dagger e^{i\omega t}) (b_r+b_r^\dagger)\rho_b \right\}  \nonumber \\
&& =\frac{1}{\pi\hbar}\int_0^\infty  d\omega {\mathcal J}_{jk}(\omega) \left (\coth(\frac{\beta\hbar\omega}{2}) \cos(\omega t)-i\sin(\omega t) \right ) \ ,\nonumber \\
 \label{eq:cjkt}
\een
where ${\mathcal J}_{jk}(\omega)$ is the bath spectral densities defined by Eq. (\ref{eq:spectral-density}).  
 The 2nd order QMEs in various forms and approximations have served as major theoretical tools for describing exciton dynamics and calculating spectroscopic data in early pioneering quantum dynamical studies on  LHCs, and still remain as key theoretical methods offering qualitative and/or semiquantitative information. 

As is well-known, the 2nd order approximation with respect to the system-bath interaction causes the QME to be non-positive definite, without rotating wave approximation \cite{pechukas-prl73,kohen-jcp107}, and to be unreliable as the system-bath interaction becomes larger.   An alternative approach addressing this issue but still based on the assumption of weak system-bath coupling is the Lindblad equation \cite{lindblad-cmp48}.  Having been constructed axiomatically, the Lindblad equation guarantees complete positivity and thus can be used to model the effects of environments on any quantum system.   For this reason, it has been used extensively in the quantum information community and was brought to the study of the FMO complex following the 2DES spectroscopy \cite{engel-nature466}.  However, due to the assumption of Markovian-bath intrinsic in the theory and the  phenomenological nature of relaxation and dephasing super-operators used, it has primarily served as a useful analysis tool rather than a quantitative modeling method.  Of course, it is possible to obtain some microscopic expressions for the terms of Lindblad equation, for example, from the 2nd order QME.   However, this involves additional approximations the physical implications of which are not always clear.

Considering typically moderate nature of system-bath interactions found in most LHCs, it is not likely that the errors caused by QMEs based on perturbative approximations are substantial.  However, in general, assessment of the accuracy of the 2nd order QMEs is possible only when benchmarked against more accurate approaches.   For this reason, different approaches employing higher order approximations or nonperturbative methods have been applied to LHCs.

Among various higher order methods developed for decades, the hierarchical equations of motion (HEOM) approach \cite{tanimura-pre47,tanimura-jpsj75,tanimura-jcp142} has gained popularity recently and its results are considered as benchmark data by many researchers.  The HEOM  approach was originally formulated and developed within the path integral influence functional formalism for open system quantum dynamics, and is based on the idea that higher order terms of the system-bath interactions can be accounted for by introducing a hierarchy of coupled auxiliary operators, an assumption valid as long as the bath correlation functions have exponential functional form.  With further advances in practical schemes \cite{ishizaki-jcp130-2,zheng-pc24,tanimura-jpsj75,schroter-pr567} to bring closure to the hierarchy and computational power, application of the HEOM approach to LHCs  became feasible recently.  

There are also other approaches that have been developed more recently and can account for higher order system-bath interactions in different manner.  These include polaron-transformed QME approach \cite{jang-jcp135,jang-njp15,nazir-prl103,mccutcheon-jcp35}, generalized QME approach \cite{shi-jcp120,zhang-jcp125,cohen-njp15,kelly-jcp144}, and transfer tensor approach \cite{rosenbach-njp18,kananeka-jpcl7}.  Well-known variations of QME approaches with modified definitions of system \cite{iles-smith-pra90,iles-smith-jcp144} or system-bath couplings \cite{hwang-fu-cp447,novoderezhkin-jpb50} have also been demonstrated to be accurate enough with appropriate corrections.   

While QME approaches are in general appropriate for describing the effects of quantum mechanical bath, they tend to neglect the effects of classical and stochastic fluctuations, which cannot be captured well in the form of Hamiltonian but may still be substantial in actual environments.  For example, for LHCs embedded in membrane, there can be various noises being propagated from distant sources but having effects on dynamics around pigment molecules.  In addition, some of anharmonic and nonlinear effects of the bath, which are
neglected in constructing the model Hamiltonian with harmonic oscillator bath, may still affect the dynamics by appearing in the form of uncontrollable noises due to the chaotic/irregular nature of the trajectories they incur in general multidimensional space.  The effects of these noises can be significant at ambient condition and may have to be included along with the quantum bath.  While the contributions of these noises to the exciton dynamics can be treated at a simple phenomenological level such as  Haken-Strobl equation \cite{haken-zp5}, more satisfactory QME level description incorporating them into a consistent quantum description of bath is not available to the best of our knowledge.  Functional integration \cite{ritschel-njp13} and Heisenberg picture time evolution \cite{ghosh-jcp131} approaches, which invoke somewhat different approximations but are general otherwise, may be better suited to this end.

\subsubsection{Full density operator approaches}        
The QMEs and the HEOM approaches, despite having been successful, are limited mostly to the harmonic oscillator bath model.  Little is understood regarding the effects of anharmonic contribution of the bath Hamiltonian on the nature and dynamics of excitons in LHCs.  For satisfactory account of these effects, it is often necessary to consider the dynamics of the full density operator (FDO) representing the system and the bath.  In addition, even for the bath of harmonic oscillators, accurate calculation of higher order response functions in general require time evolution of FDO unless a new kinds of coupled QMEs are developed. Well-known FDO approaches are semiclassical dynamics methods \cite{miller-jpca105,cotton-jctc12,huo-jcp133}, mixed quantum-classical approaches \cite{tully-jcp137,zheng-jpcb121} such as surface hopping and mean field approximations.   While these offer more realistic treatment of the bath degrees of freedom and their interactions with excitons, they entail different levels of approximations in the quantum dynamics.  In addition, the bath has to be finite for practical calculations.  Given that these approximations and issues are well addressed, the results of these FDO approaches can offer important information for LHCs that are not accessible through QME and HEOM approaches.

\subsubsection{Computational methods for lineshape and response functions} 
All the QME approaches can be used to calculate the absorption and emission lineshapes, employing the product of the transition dipole and the density operator in either ground or excited state as an initial condition.  Alternatively, a specific QME method can be modified to derive an appropriate lineshape theory using the coherent term of the density operator as the initial condition and applying mixed time evolutions, one on the ground electronic state and the other on the manifold of excited electronic states, respectively, to the two different sides of the density operator.   One popular approach for calculating the lineshape is the so called modified Redfield equation \cite{zhang-jcp108,schroder-jcp124}, which includes the diagonal components of the exciton-bath coupling as part of the effective system part and treat the off-diagonal components in a perturbative manner.   Most recently, this approach was extended further to account for the finite relaxation time of nuclei \cite{dinh-jcp145}.  Modified Redfield equation can also be extended for the simulation of exciton dynamics \cite{yang-cp275,hwang-fu-cp447,novoderezhkin-jpb50}, although care should be taken in identifying the unit of population in this case.

An important issue is that response functions of four-wave mixing spectroscopy cannot be calculated exactly even at a formal level using a conventional QME approach without modification.  The diagrams in Fig. \ref{fig:four-wave-diagram} clarify the reason for this.  Each interaction with pulse corresponds to the creation of a non-equilibrium state, which also depends on previous time evolution of both excitons and the bath that are again dependent on the nature of previous pulses.  For this reason, the relaxation and dephasing terms due to the bath are expected to be functionals of whole sequence of pulses, which are in general different from those of equilibrium bath.   Therefore an FDO approach is necessary in general \cite{mcrobbie-jpca113}. A perturbative approach addressing this issue has also been formulated \cite{jang-bkcs33}.  
 
 \subsection{Applications}
\subsubsection{FMO complex}
Early theoretical works on the FMO complex employed the 2nd order ME and QME approaches.  Application of the Redfield equation \cite{redfield-ijrd1}, the 2nd order TL-QME in the Markovian bath limit, with the secular approximation \cite{renger-jpca102} showed that the approach can reproduce the major features of temperature dependence in the absorption lineshape and the time dependence of pump-probe spectroscopic data.  They  also provided estimate for the relaxation of a higher exciton state (4th state) to be about $2\ {\rm ps}$.    This approach was later combined with an optimal control theory \cite{bruggemann-jpcb108-1} to explore the possibility of controlling the exciton dynamics in the FMO complex through pulse shaping.  On the other hand, the ME approach in the exciton basis was employed \cite{vulto-jpcb103} to describe  time resolved absorption difference spectra and the exciton population dynamics. They have also identified various relaxation time scales of exciton dynamics ranging from sub $100\ {\rm fs}$ up to about $2.3\ {\rm ps}$.    The phenomenological assumption used in this model Hamiltonian was later justified microscopically \cite{adolphs-pr95}. 

The suggestion of quantum information processing \cite{engel-nature466} motivated new theoretical works \cite{plenio-njp10,mohseni-jcp129,rebentrost-njp11,rebentrost-jpcb113,palmieri-jcp130,caruso-jcp131,caruso-pra81,chin-njp12,sarovar-np6,skochdopole-jpcl2,pelzer-jcp136} aimed at understanding the role of entanglement, quantum coherence, and noise.    Earlier versions of these \cite{plenio-njp10,mohseni-jcp129,rebentrost-njp11,caruso-jcp131} employed the Lindblad equation \cite{lindblad-cmp48} or its stochastic implementation, which are based on the assumption of weak system-bath coupling and Markovian bath.  Later works improved these by including the effects of non-Markovian bath \cite{rebentrost-jcp131,caruso-pra81,chin-njp12}, correlated bath fluctuations \cite{rebentrost-jpcb113}, and full consideration of system-bath coupling at the level of HEOM \cite{sarovar-np6}.  According to these studies, the effect of  entanglement, which is limited to mode entanglement, is not significant for single excitons in the FMO complex. On the other hand, dephasing and noise were identified as significant factors for the overall efficiency of energy transport, as espoused by new terms such as environment-assisted quantum transport \cite{rebentrost-njp11,lambert-np9} and dephasing-assisted transport \cite{chin-njp12}.  Studies based on  the Haken-Strobl equation \cite{hoyer-njp12,vlaming-jcp136} have also offered new insights into the potential complications and effects of noise.  It was also suggested that the redundancy \cite{skochdopole-jpcl2} of energy transfer pathways plays an important role.  

Due to the intermediate nature of the exciton-bath coupling in the FMO complex, the Lindblad equation and the 2nd order ME/QME approaches were not perceived as quantitatively reliable.  HEOM calculations were performed to investigate the population dynamics \cite{ishizaki-pnas106}, which served as the first set of key benchmark data in this respect.  This work showed that coherent population dynamics persists up to several hundred femto seconds even at room temperature.  Calculations \cite{chin-np9} based on density matrix renormalization group technique \cite{prior-prl105} also support this result and suggest the contribution of significant nonequilibrium effects. Later,  other approaches have been employed to calculate the excitation dynamics.  For example, non-Markovian quantum state diffusion \cite{ritschel-njp13}, mixed quantum-classical Poisson bracket mapping equation approach \cite{kelly-jpcl2}, and quasiclassical dynamics \cite{cotton-jctc12} have been shown to produce results in reasonable agreement with the HEOM approach.  These approaches are applicable to more general exciton-bath couplings and bath Hamiltonians, and can be used to understand the effects of anharmonic effects of the bath modes and nonlinear exciton-bath couplings.  

Theoretical observation of coherent real time dynamics for the FMO complex was an important step for elucidating the quantum dynamical details of exciton dynamics. However, such oscillatory population dynamics were mostly between proximate BChls with significant electronic couplings, and did not yet imply long range ``wave-like" motion of excitons.
 On the other hand, for the overall description of exciton dynamics and calculation of ensemble-averaged spectroscopic observables, simpler perturbative ME and QME approaches seemed to offer reasonable answers \cite{moix-jpcl2,wu-jcp137,busch-jpcl2}.  Most recently, a full HEOM calculation for trimers \cite{wilkins-jctc11} demonstrated that the time scale of exciton transport is similar to that based on a simple ME approach using FRET rate as its kernel.  This result suggests that the interpretation based on earlier theoretical studies based on approximate theories is valid in general.

 There have been further progress in theoretical modeling of spectroscopic data. While theoretical calculation of linear spectroscopic data was well established and could be used for the refinement of model parameters \cite{adolphs-bj91,busch-jpcl2,jansen-acc42}, accurate modeling and unambiguous interpretation of 2DES signals has remained challenging. One of the earliest simulations of 2DES was based on equation of motion approach, while assuming simple Redfield tensor and secular approximation for the bath relaxation \cite{sharp-jcp132}. This suggested that it is difficult to explain the off-diagonal beating signal observed from 2DES \cite{engel-nature466}.   A different theoretical modeling based on the secular Redfield equation led to a similar conclusion, while leaving the possibility of contribution of correlated bath \cite{abramavicius-jcp134}.  Another theoretical modeling based on the Redfield equation and including the vibronic states explicitly suggested that the beating signal may have originated from vibrational motion \cite{christensson-jpcb116}.  HEOM calculations of absorption spectra \cite{hein-njp14} and 2DES echo spectra \cite{hein-njp14,kreisbeck-jpcl3} on the other hand provided support for coherence signal of electronic nature lasting up to about 500 fs, but in the absence of static disorder and other sources of fluctuations.  Although based on rather simple model Hamiltonian, careful analysis \cite{tiwari-pnas110} demonstrated that electronic coherence can easily dephase when the effect of reasonable magnitude of disorder is taken into consideration, whereas beating that originates from vibrational motion can last much longer.   Th contribution of vibronic terms can be another possibility \cite{plenio-jcp139,mourokh-pre92}.  As yet, because of the sensitivity of 2DES signals to various factors and the lack of fully reliable model and quantum dynamics methods, a definite assessment of the implications of 2DES signals remains open.

From the functionality point of view, to what extent the quantum effects contribute to the overall efficiency of the exciton migration remains a central question even to date \cite{wu-jcp137}. 
Accurate determination of the spectral density of the bath has a significant implication in answering this question.  High frequency modes of the spectral density in general do not affect the mechanistic details of exciton dynamics \cite{abramavicius-jcp140}.  However, low frequency modes with energies comparable to the exciton band width and their coupling strengths can be detrimental to the exciton dynamics mechanism and the interpretation of 2DES spectroscopic data.  There have been various efforts to address these issues through all-atomistic simulations \cite{shim-bj102,olbrich-jpcb115-1} and direct calculation of Huang-Rhys factors \cite{adolphs-bj91}.  The contributions of correlated fluctuations \cite{olbrich-jpcb115-1} and non-Gaussian bath fluctuations \cite{jansen-acc42} were also suggested as potentially significant factors that can complicate the exciton dynamics.  As yet, depending on the approximations and assumptions involved, the assessment of the bath spectral density can be different.  It was once suggested that the exciton-bath coupling might be too strong to allow any coherent dynamics unless the dynamics starts from a coherent initial state \cite{muhlbacher-jpcb116}. However, more recent calculations based on different methods indicate weaker bath spectral densities that are more in tune with older models.   Ultimately, on-the-fly {\it ab initio} nonadiabatic dynamics but with sufficient accuracy may be necessary to settle this issue. New development in first principles linear scaling {\it ab initio} calculation can be a promising avenue to explore in achieving this goal \cite{cole-jpcl4}.

\subsubsection{LH2 complex} 
Soon after the structural information of the LH2 complex became available, advanced quantum dynamical theories, for example, addressing the exciton coherence length \cite{leegwater-jpc100,meier-jcp107}, super-radiance \cite{meier-jcp107}, and four-wave mixing spectroscopy signals \cite{zhang-jcp108,meier-jpcb101} have been developed.  Although based on exciton-bath Hamiltonians of appropriate features but yet with quite simplified forms, these works have not produced quantitative modeling of linear spectroscopic data.  Instead, earlier attempts to fit experimental lineshapes used simple exciton Hamiltonians with disorder terms  \cite{hu-jpcb101},  while dressing each exciton peak with phenomenologically chosen lineshape functions \cite{alden-jpcb101,wu-jpcb102,georgakopoulou-bj82}.  These produced reasonable fitting of linear ensemble lineshapes, which indicates that 
the inhomogeneous broadening due to disorder is large enough to screen the line broadening due to the exciton-bath coupling.  As yet, the disorder is still in the moderate regime that makes it difficult to tell what types of disorder, diagonal or off-diagonal in site excitation basis, are dominant \cite{jang-jpcb105}.  In addition, detailed information on the disorder was shown to be important for
proper interpretation of nonlinear spectroscopic data \cite{yang-jppa142}.   

The SMS data \cite{vanoijen-science285} for the LH2 complex offered the first direct evidence for the delocalized nature of excitons in its B850 unit, and offered new opportunities and challenges 
for theoretical and computational modeling.  On one hand, large gaps between two major excitonic peaks of the B850 unit were inexplicable based on existing exciton models.  This initially led to the suggestion of elliptic distortion of LH2 \cite{vanoijen-science285,dempster-jcp114,mostovoy-jpcb104}.  Later, it was explained by models with modulation in site energies \cite{ketelaars-bj80,hofmann-cpl395} and correlated disorder \cite{jang-jpcb115}.  Another important theoretical issue was elucidating details of SMS lineshapes.   
The first step to understand the physical implication of such lineshapes is to calculate the effects of exciton-bath coupling at each 
single LH2 level.  To this end, an exciton-bath model \cite{jang-jcp118-2} incorporating the bath spectral density \cite{renger-jcp116} extracted from spectroscopic data was developed. 
Lineshapes were then calculated by using the 2nd order TN-QME approximation \cite{jang-jcp118-2,jang-jcp118-1}. The features of the theoretical line shapes \cite{jang-jcp118-2} were in qualitative agreement with  experimental ones \cite{vanoijen-science285}.  However, the widths of the former were much narrower than the latter.  
This was attributed to fluctuations during the long SMS measurement time scales, which are possible even in the very low temperature limit because of numerous excitation and de-excitation processes needed to collect the photons.   Another source for the discrepancy is the approximate nature of the 2nd order TN-QME.  Indeed, later calculations of absorption lineshapes by alternative or higher order methods \cite{schroder-jcp124,chen-jcp131},  while based on somewhat different bath spectral densities and calculated at higher temperatures, 
showed that the 2nd order TN-QME approach underestimates the broadening due to exciton-phonon couplings.   Application of these approaches with improved spectral densities may allow better explanation of SMS lineshapes.

In parallel, there have been advances in the modeling of 2DES spectroscopy data of the LH2 complex, for example, providing intriguing insights into 
anharmonic effects of the bath \cite{rancova-jpcb118}.  More recently, all-atomistic modeling has offered detailed information on physical implications of 2DES signals \cite{vandervegte-jpcb119,sgatta-jacs139}. 
The dark states of carotenoids of LH2, which had long been speculated to be present, was investigated recently based on quantum chemical calculations and simulation of 2DES spectroscopic data \cite{sgatta-jacs139,feng-nc8}.  However, the results obtained so far are not yet sufficient to lead to a general consensus. Thus, further investigations are needed.

In a given LH2 complex, the exciton dynamics within B800, between B800 and B850, and within B850 have all different characteristics. It is generally accepted that the dynamics within B800 can be well described by a hopping model of excitons localized at each BChl.   The exciton transfer from B800 to B850 can also be described by hopping dynamics, but the coherent delocalization of the exciton in the B850 unit has to be taken into consideration.   Two early theoretical studies \cite{mukai-jpcb-103,scholes-jpcb104} accounted for this by employing approximate versions of the MC-FRET theory \cite{sumi-jpcb103,jang-prl92}, which included only the contribution of population terms in the exciton basis and used phenomenological lineshape functions.  Later works \cite{jang-prl92,jang-jpcb111} employing a more satisfactory exciton-bath Hamiltonian \cite{jang-jcp118-2} and full MC effect within the 2nd order TN-QME approach \cite{jang-jcp118-1,jang-jcp118-2} confirmed that neglecting coherence terms in the exciton basis of the MC-FRET theory does not have significant effect on the distribution of transfer rates.  In addition, these works \cite{jang-prl92,jang-jpcb111}  provided more solid evidence that the theoretical results based on the exciton-bath model \cite{jang-jcp118-2} and the MC-FRET theory \cite{sumi-jpcb103,jang-prl92} are indeed consistent with experimental results \cite{jimenez-jpc100,pullerits-jpcb101}.      

 The exciton dynamics within the B850 unit requires full quantum dynamical approach. Although hoping dynamics \cite{abramavicius-pccp6} and Haken-Strobl approach \cite{liuolia-jpcb101} 
 provided some insights,  accurate theoretical description has remained challenging.  Recent applications of the HEOM approach \cite{chen-jcp131,strumpfer-jcp131,strumpfer-jcp137,yeh-jcp137} are important advances in this respect, and  confirm fast decoherence and relaxation of excitons in the range of $100-200\ {\rm fs}$ time scales as seen experimentally \cite{agarwal-jpca106,book-jpcb104}.  However, these calculations are still based on rather simplified spectral densities.  In addition, these have made no or limited consideration of the effects of the disorder. 
 On the other hand,  it was demonstrated that the ensemble dephasing due to disorder, while using relatively simple dynamics theory, is sufficient to explain fast anisotropy decay in LH2 \cite{stross-jpcb120}.  Thus, to clarify which of the two factors determines major time resolved experimental data on the B850 unit, large scale HEOM calculations sampled over sufficient number of realizations of the disorder is necessary.  Some advances have already been made in this direction for the calculation of absorption and emission lineshapes \cite{jing-jcp138}. 
 
\begin{figure}[hptb]
\includegraphics[width=3.2in]{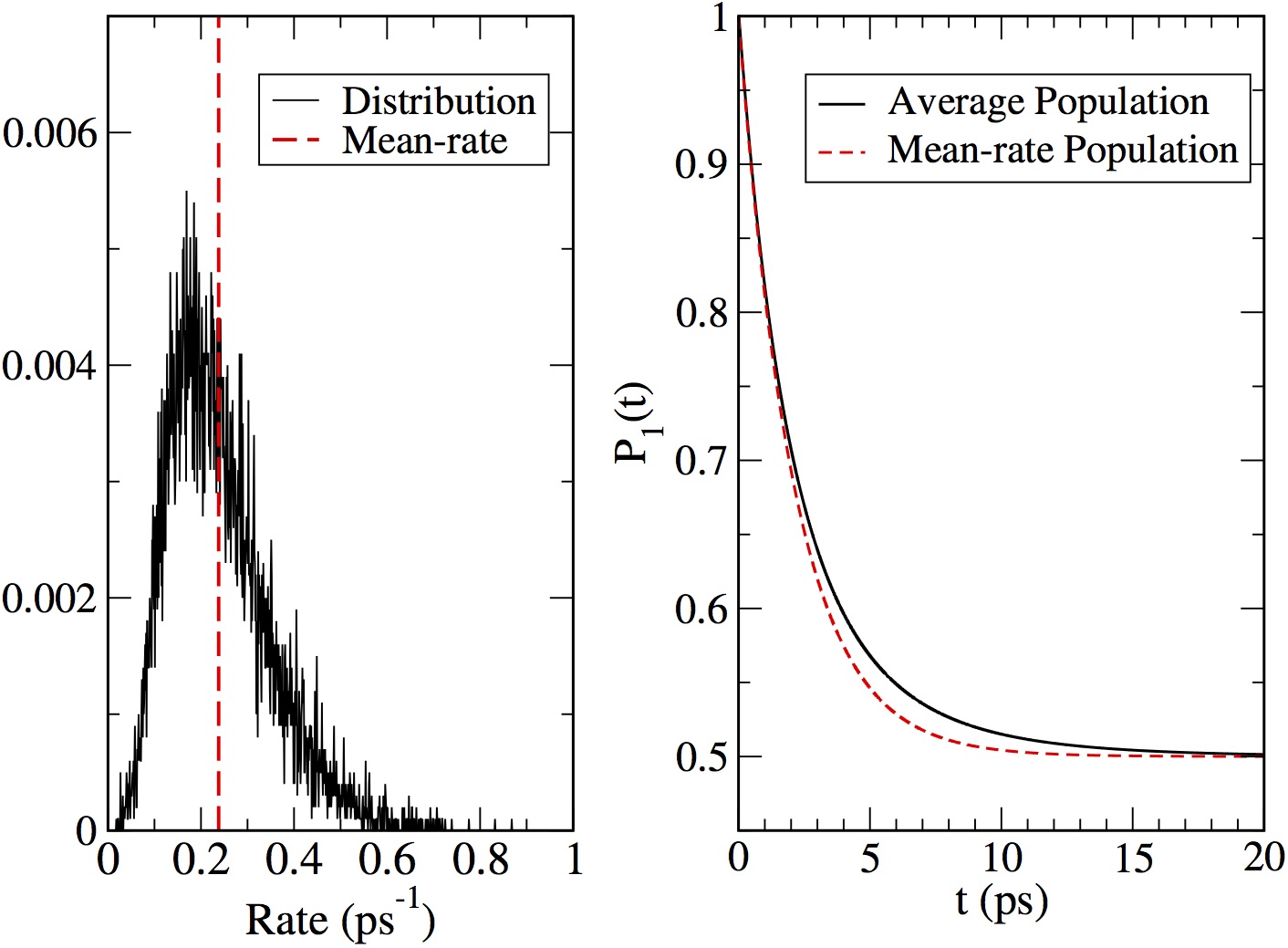}
\caption{Distribution of LH2-LH2 rates (left panel) and average population decay (right panel) of exciton on the initial LH2. More details can be found in \cite{jang-jpcl6}.}
\label{fig:lh2-lh2-rate}
\end{figure}

The dynamics of excitons between LH2s and aggregates of LH2s have important implications for the overall energy collection efficiency, 
and has been  subject to various computational studies \cite{ritz-jpcb105,caycedo-soler-prl104,strumpfer-jcp131,strumpfer-jcp137,xiong-jcp137,yang-jpcc116,jang-prl113,jang-jpcl6}.  The earliest in this endeavor was the kinetic Monte Carlo simulation of exciton dynamics in simple aggregates of LH2, LH3, and LH1 \cite{ritz-jpcb105}.   
In this work, a fixed value of LH2-LH2 exciton transfer time of $10\ {\rm ps}$ was used, which was calculated based on an approximate version of the MC-FRET theory \cite{sumi-jpcb103,jang-prl92} for a single LH2-LH2 distance \cite{ritz-jpcb105}.  This transfer time was also adopted in later simulation of exciton dynamics in the entire PSU of purple bacteria \cite{caycedo-soler-prl104}. While the value has been validated to be in reasonable agreement with an accurate calculation based on the HEOM approach \cite{strumpfer-jcp131}, it is not yet representative of all the inter-LH2 exciton transfer dynamics that are plausible.  An HEOM calculation demonstrated significant potential effect of the correlation of the bath \cite{strumpfer-jcp137}. Application of the GME-MED approach \cite{jang-prl113,jang-jpcl6} showed that the rate can change significantly depending on the realization of the disorder and the inter-LH2 distances that are consistent with AFM images.   Figure \ref{fig:lh2-lh2-rate} shows the distribution of rates at $300\ {\rm K}$ between two LH2 complexes separated by a center-to-center distance of $7.5 \ {\rm nm}$.  In this result, the distribution of rates is purely due to the Gaussian disorder in the excitation energy of each BChl with standard deviation of $200\ {\rm cm^{-1}}$.   Consideration of the distribution of inter-LH2 distances will result in wider distribution of rates \cite{jang-jpcl6}. 

\subsubsection{PBPs}
Modified Redfield equation was employed \cite{novoderezhkin-bj99} to model linear spectroscopic data, excitation anisotropy measurement, and transient absorption spectra, and then to deduce excitation energy transfer kinetics based on these modelings.  Following the observation of the beating signal from the 2DES result \cite{collini-nature463}, an FDO quantum dynamics method called iterative density matrix propagation approach was used for the simulation of the exciton population dynamics \cite{huo-jpcl2}.  The results of this work supported the observation from the 2DES \cite{collini-nature463} that the electronic coherence between central dimers lasts up to about $400\ {\rm fs}$, but also showed that such coherence does not have significant effect on the exciton population dynamics of peripheral pigment molecules.  On the other hand, it was also suggested that quantum vibrational modes 
have significant effect on the exciton transfer \cite{oreilly-ncomm5}.

QM/MM approaches \cite{curutchet-jpcb117,aghtar-jpcl5,lee-jacs139,aghtar-jpcb121} were used to model spectroscopic data and to simulate the exciton transfer dynamics.  The earliest effort was focused on investigating the nature of the bath spectral density, and employed a quantum-classical dynamics simulation by conducting wave-packet dynamics on fluctuating potential energy surfaces \cite{curutchet-jpcb117}.  It was also confirmed that significant portion of the bath spectral density results from the intramolecular vibrations \cite{aghtar-jpcl5}.  The population dynamics were also shown to be non-oscillatory, even much less than that of the FMO complex.   

The methodology of QM/MM simulation has progressed to the level of constructing comprehensive EBH models for PBPs of both PE545 and PC645 with reasonable accuracy \cite{lee-jacs139}. These models were then combined with lineshape theories employing the 2nd order TL-QME approach so as to model the absorption and CD spectra \cite{lee-jacs139}, demonstrating satisfactory agreement between theory and experiment.  Analyses based on these calculations confirmed significant effects of vibronic couplings, and also suggested potential importance of correlated fluctuations of the site excitation energies.  A different QM/MM study, also including the effects of the polarizable medium, showed that fluctuations in couplings and site energies do not have significant effects on the  overall exciton dynamics \cite{aghtar-jpcb121}. Theoretical analysis of a recent 2DES spectroscopy for PC645 \cite{dean-chem1} also suggests that vibronic redistribution enhances the exciton transfer rate by about a factor of $3.5$.

\section{Discussion}
LHCs have characteristics and challenges that are distinctively different from any other biological complexes in relation to their specific function.  While the importance of structural information for LHCs is inarguable, it is necessary to recognize that the functionality of an LHC is determined by how and where excitons are formed and how fast and efficiently the excitons move from one region to another, while conserving their energies as much as possible even in the presence of disorder and fluctuations.  Elucidating these details require reliable information on the energetics and the dynamics of excitons in the excited potential energy surfaces.   In addition to structural data, a comprehensive collection of spectroscopic and computational data are needed even for establishing a Hamiltonian that can represent the behavior of excitons with reasonable accuracy.  Calculation and simulation of quantum dynamical evolution of excitons in these systems are equally challenging due to their complexity and sizes.   Furthermore, the disorder and environmental fluctuations are important factors to consider because they can have significant effects on how excitons are formed and evolve in time. They also make significant contributions to spectroscopic observables.     

Ideally, a spectroscopic measurement with both nanoscale spatial resolution and femtosecond time resolution, or an on-the-fly {\it ab initio} quantum dynamical simulation with sufficient accuracy (with errors less than $100\ {\rm cm^{-1}}$ in energy) and efficiency (to run up to tens of picosecond and to be repeated over many realizations of disorder) is needed to probe excitons in LHCs directly.   However, in the absence of feasible spectroscopic and computational tools meeting these needs at present, the best strategy is to develop a well-designed theory-experiment collaboration focused on addressing prominent issues and testing important assumptions.   Advances made for the FMO complex, the LH2 complex, and the PBPs of cryptophyte algae as reviewed here, provide lessons and future guidelines that can be applied to other LHCs with more complexity and larger scales.

The results of various investigations on these three LHCs serve as concrete examples for addressing many important issues concerning the excitons in LHCs. Among them, we 
discuss here three topics that are in fact intimately connected together and have generated heated discussion and investigation in the past decade, namely, {\it optimality} \cite{cao-jpca113,jang-jpcb111,jang-wires,jang-jpcl6,mohseni-jcp140}, {\it quantum coherence} \cite{phil-tran-370,kassal-jpcl4} and {\it robustness} \cite{wu-jcp137,cleary-pnas110,mohseni-jcp140,huh-jacs136,chen-pre88}.

In a sense, the fact that delocalized Frenkel excitons can be identified and used to understand spectroscopic data is a good evidence that the electronic quantum coherence, defined here simply as having the quality of coherent superposition of site excitation states, plays a significant role in the functionality of LHCs.  However, this does not necessarily mean that a specific phase relationship defining a certain exciton state should be preserved, or change coherently, during a time interval comparable to that of exciton migration.  Rather, the behavior of electronic quantum coherence can be dynamic or inhomogeneous \cite{ishizaki-jpcb115}.  These two features make it extremely difficult to directly detect quantum coherences that are active 
in real time through a spectroscopic measurement.  On the other hand, such features may in fact play crucial role in positive and cost-effective utilization of quantum coherence for robust and optimal exciton dynamics.   In other words, the vast parameter space of quantum superposition makes it possible for excitons to migrate through different regions while conserving energy.  In such dynamics, quantum coherence offers (i) {\it redundancy} to seek for alternatives when one path is blocked and (ii) {\it buffering mechanism} to help counteract negative effects of disorder and fluctuations. 
Thus, genuine understanding of efficient and robust exciton migration in LHCs seems to require quantitative elucidation of such redundancy and buffering mechanisms, which in turn require a holistic approach that considers structure, energetics, dynamics, disorder, and fluctuations all together.

The fact that the rate-based ME approach provides a good description of exciton transfer dynamics in some LHCs (or some of their parts) does not necessarily mean that quantum coherence does not play a role as it has been often assumed.  To the contrary, the rate behavior can be a manifestation of cumulation or averaging of all possible effects of quantum coherences.  In other words,  the extent to which certain signatures of quantum coherences become explicit can be different depending on the starting point or level of averaging.  However, even when an apparent signature cannot be observed, underneath the apparent classical-like phenomenology, quantum dynamical processes taking advantage of the coherently delocalized excitons can play important roles.  

There are two issues that have significant implications but were not addressed carefully in this work.  One is the consideration of multiple exciton states and the other is the dynamics of excitons under natural light condition.  
For a complete description of four-wave mixing spectroscopy and 2DES in particular, it is necessary to include the double exciton space in order to account for excited state absorption.  Three kinds of double excitons can be created in general:  (i) double excitation of a pigment molecule, (ii) double exciton in the same LHC; (iii) single excitons in almost two independent LHCs within the ensemble of spectroscopic measurements.  Except for the trivial case of (iii), calculating energies and couplings that are relevant to (i) and (ii) are challenging, and can make the determination of the Hamiltonian much more complicated.  Nonetheless, investigation of the characteristics of the double excitons  and their dynamics have been shown to be useful even for understanding the behavior of single excitons \cite{bruggemann-jpcb105,trinkunas-prl86}.       

Another important issue is the understanding of the nature and the dynamics of excitons under natural incoherent light sources from the sun and also how/whether lessons learned from femtosecond laser spectroscopy can be used to understand this issue \cite{brumer-pnas109,chenu-cp439,chenu-prl114}.  To this end, formulations involving excitations by steady incoherent light sources have been developed \cite{grinev-jcp143,chenu-jcp144}.  Understanding this issue involves describing photons and their quantum mechanical interactions with nanoscale objects,  and can best be studied 
through quantum optics experiments of clean and more well-controlled nanoscale experimental objects.

\section{Conclusion and Outlook}

Experimental and theoretical evidences cumulated so far support that the majority of excitons residing in pigment molecules of LHCs can be well described by the Frenkel exciton theory.   This is in contrast to the excitons found in most man-made solar energy conversion systems.  For example, excitons created in the first generation semiconductor-based materials are dominantly Wannier excitons.  Even for organic photovoltaic devices consisting of conjugated organic molecules, excitons tend to have intermediate character and cannot be represented by simple Frenkel exciton Hamiltonians.   Why nature takes special advantage of Frenkel excitons, which have been difficult to utilize in man-made systems so far, is an interesting issue to investigate.   A possible explanation based on current experimental and theoretical evidences is that it is related to the properties of proteins as insulating media and to their superb capability to fine tune excitation energies and spatial arrangement of pigment molecules.  In other words, despite the 
short coherence length and fragility of Frenkel excitons compared to Wannier excitons, the versatility of proteins as hosting  
environments allows connecting excitons in energetically and dynamically favorable ways, thereby making  it possible to create efficient and robust exciton-relaying mechanisms.   

The three LHCs reviewed here exemplify different ways proteins can tune the properties of Frenkel excitons.   For the case of the FMO complex, the main mechanism appears to be the downhill structure of site excitation energies through subtle interaction of BChl-protein interactions, whereas weaker  electronic couplings and exciton-bath couplings, also realized through protein's capability to control arrangements of BChls, play supporting roles.  For the case of the LH2 complex, circular arrangement of tens of BChls with nearest electronic couplings on the order of $200-300\ {\rm cm^{-1}}$ in the B850 unit, maintained through stable coordination and hydrogen bonds to protein residues, seems to be the key feature.   Because the excitation-bath couplings are of moderate magnitudes (on the order of $100\ {\rm cm^{-1}}$), the band of discrete Frenkel exciton states spanning about $1,000\ {\rm cm^{-1}}$ are dynamically well connected yet without losing their identity.  Because of alternating in-plane directions of transition dipoles of $\alpha$- and $\beta$-BChls, the lowest exciton state remains almost dark state, preventing radiative loss through super-radiance.  For the case of PBPs, downhill structure of excitation energies by using different pigment molecules and large enough excitation-bath coupling \cite{dean-chem1,killoran-jcp143} realized through covalent bonds, provide efficient pathways  for the migration of Frenkel excitons.        

Another important point to consider and, potentially, a good lesson for developing new generation of photovoltaic systems, comes from the observation of how LH and CS domains are put together in natural photosynthetic systems.   In all the cases known so far, the two domains are always separate, and back exciton transfer pathway from the CS domain to the LH domain is blocked as much as possible through ingenious mechanisms. There is also a ``moderating" mechanism of exciton transfer from LH to RC so as to prevent too fast exciton transfer dynamics that can damage the CS capability.  For purple bacteria, the moderating function is performed by the size of the LH1 complex;  for green sulfur bacteria, the FMO complex serves such a role by acting as a moderating ``valve" of excitons between chlorosome and reaction centers.    

In summary, LHCs are valuable natural systems that offer great insights into the quantum dynamics of coherently delocalized excitons in disordered and insulating host media.   
Experimental and theoretical studies of LHCs involve challenging issues of large scale excited state quantum dynamics in condensed and complex environments, and provide strong motivations and resources for further advances in these areas of research.   Clear molecular level assessment of the natural design principles of LHCs should be based on outcomes of these studies, which in turn can guide the development of genuinely biomimetic solar light harvesting systems.

\acknowledgments
SJJ acknowledges the support of the Office of Basic Energy Sciences, Department of Energy (Grant No. DE-SC0001393) as major sponsor of this research, and the National Science Foundation (Grant. No. CHE-1362926) as the sponsor of research on quantum dynamics methods.  SJJ also thanks Daniel Montemayor and Eva Rivera for their assistance.  BM acknowledges the support of the European Research Council (ERC) through the Starting Grant, proposal No. 277755 (EnLight).  We thank Greg Scholes, Thomas Renger, J\"{u}rgen K\"{o}hler, Young Min Rhee, and David Coker for offering data/images and for valuable comments.  We also thank T\"{o}nu Pullerits and Donatas  Zigmantas for discussion and comments. 

\appendix 
\section{From molecular Hamiltonian to exciton-bath Hamiltonian: A review}
For the $j$th pigment molecule,  $|\Psi_{j}\rangle$ is defined as the full and general molecular quantum state.  
We also abbreviate the collective position state of all the nuclear coordinates of the $j$th chromophore as ${\bf R}_{c_j}$ and define the corresponding position state as follows: 
\be
|{\bf R}_{c_j}\rangle \equiv |{\bf R}_{j,1}\rangle \cdots |{\bf R}_{j,L_j}\rangle \ ,
\ee
where the righthand side represents direct product of states representing $L_j$ (3-dimensional) nuclear coordinates constituting the $j$th pigment, as has been indicated in the main text. 
Then, 
\ben
&&\langle {\bf R}_{c_j}|\hat H_{j}|\Psi_{j}\rangle=\left (T_{j,n} (\nabla_{c_j}) +V_{j,nn}({\bf R}_{c_j})\right . \nonumber \\
&&\hspace{1.2in}\left . +\hat H_{j,e}({\bf R}_{c_j})\right ) \langle {\bf R}_{c_j}|\Psi_{j}\rangle \ , \nonumber \\ \label{eq:rd_hd}
\een 
where 
\ben 
&&T_{j,n}(\nabla_{c_j}) =-\sum_{l=1}^{L_j}\frac{\hbar^2}{2M_{j,l}} \nabla_{j,l}^2 \ , \\
&&V_{j,nn}({\bf R}_{c_j})=\frac{1}{2}\sum_{l=1}^{L_j}\sum_{l'\neq l}^{L_j}\frac{Z_{j,l} Z_{j,l'} e^2}{|{\bf R}_{j,l}- {\bf R}_{j,l'}|} \ ,\\ 
&&\hat H_{j,e} ({\bf R}_{c_j})=\hat T_{j,e}+\hat V_{j,en} ({\bf R}_{c_j}) +\hat V_{j,ee} \ .
\een
In the above expression, the definitions of $\hat T_{j,e}$ and $\hat V_{j,ee}$ can be found from Eq. (\ref{eq:hj_def}). In addition, $\hat V_{j,en}({\bf R}_{c_j})$ is a quantum operator with respect to the electronic degree of freedom, while depending parametrically on nuclear coordinates, as follows:
\be
\hat V_{j,en} ({\bf R}_{c_j})=-\sum_{l=1}^{L_j}\sum_{i=1}^{N_j}\frac{Z_{j,l} e^2}{|{\bf R}_{j,l}-\hat {\bf r}_{j,i}|} \ .
\ee

\subsubsection{Adiabatic approximation for each chromophore}
We denote  the $\alpha$th adiabatic electronic state of the $j$th chromophore at ${\bf R}_{c_j}$ with eigenvalue $E_{j,\alpha}({\bf R}_{c_j})$ as $|E_{j,\alpha}({\bf R}_{c_j})\rangle$. Thus,
\be 
\hat H_{j,e}({\bf R}_{c_j})|E_{j,\alpha}({\bf R}_{c_j})\rangle=E_{j,\alpha}({\bf R}_{c_j})|E_{j,\alpha}({\bf R}_{c_j})\rangle \ .
\ee
Given that  electrons are in one of the adiabatic electronic states, $|E_{j,\alpha}({\bf R}_{c_j})\rangle$, for nuclei at ${\bf R}_{c_j}$, 
\be
\langle {\bf R}_{c_j}|\Psi_{j}\rangle=\Phi_{j,\alpha} ({\bf R}_{c_j})|E_{j,\alpha}({\bf R}_{c_j})\rangle \label{eq:rd_psi} \ ,
\ee
where $\Phi_{j,\alpha}({\bf R}_{c_j})$ is the nuclear wave function for the $\alpha$th adiabatic state of the $j$th pigment.  Alternatively, one can introduce a nuclear state $|\Phi_{j,\alpha}\rangle$ such that 
\be
\Phi_{j,\alpha} ({\bf R}_{c_j})=\langle {\bf R}_{c_j}|\Phi_{j,\alpha}\rangle \ . \label{eq:var_phi_def}
\ee
Inserting Eqs. (\ref{eq:rd_psi}) and (\ref{eq:var_phi_def}) into the right hand side of Eq. (\ref{eq:rd_hd}), we obtain 
\ben 
&&\left (T_{j,n}(\nabla_{c_j}) +V_{j,nn}({\bf R}_{c_j}) +\hat H_{j,e}({\bf R}_{c_j}) \right )\nonumber \\
&&\hspace{1in}\times \Phi_{j,\alpha} ({\bf R}_{c_j}) |E_{j,\alpha} ({\bf R}_{c_j}) \rangle \nonumber \\
&&=|E_{j,\alpha} ({\bf R}_{c_j}) \rangle \left  ( T_{j,n} (\nabla_{c_j}) + U_{j,\alpha} ({\bf R}_{c_j})\right )\Phi_{j,\alpha} ({\bf R}_{c_j}) \nonumber \\
&&-\sum_{l=1}^{L_j} \frac{\hbar^2}{M_{j,l}} \left (\nabla_{j,l} \Phi_{j,\alpha} ({\bf R}_{c_j}) \right) \cdot \left (\nabla_{j,l} |E_{j,\alpha} ({\bf R}_{c_j})\rangle \right) \nonumber \\
&&-\sum_{l=1}^{L_j} \frac{\hbar^2}{2M_{j,l}} \Phi_{j,\alpha} ({\bf R}_{c_j}) \nabla_{j,l}^2 |E_{j,\alpha} ({\bf R}_{c_j})\rangle \ , \label{eq:nonad_sch}
\een
where $U_{j,\alpha} ({\bf R}_{c_j})$ is the effective nuclear potential energy for the $\alpha$th adiabatic state of the $j$th pigment. It is the sum of the nuclear potential energy and the electronic energy as follows:
\be 
U_{j,\alpha} ({\bf R}_{c_j})=V_{j,nn}({\bf R}_{c_j})+ E_{j,\alpha} ({\bf R}_{c_j})\ .
\ee

Within the Born-Oppenheimer approximation, the derivatives of $|E_{j,\alpha} ({\bf R}_{c_j})\rangle$ with respect to ${\bf R}_{c_j}$ can be neglected.  In addition, assuming that the adiabatic electronic state is insensitive to the small displacements of the nuclear coordinate around a reference nuclear coordinate ${\bf R}_{c_j}^0$, we can approximate that
\be
|E_{j,\alpha} ({\bf R}_{c_j})\rangle \approx |E_{j,\alpha}({\bf R}_{c_j}^0)\rangle \equiv |E_{j,\alpha}\rangle \ , \label{eq:e_dk_rd}
\ee
where we have assumed that ${\bf R}_{c_j}^0$ can be defined commonly for all adiabatic states, namely, independent of $\alpha$, for the $j$th pigment. 

Taking the inner product of $\langle E_{j,\alpha}|$ with Eq. (\ref{eq:rd_hd}), employing Eqs. (\ref{eq:rd_psi}) and (\ref{eq:nonad_sch}), and applying the Born-Oppenheimer approximation, we obtain
\ben
&&\langle E_{j,\alpha}|\langle {\bf R}_{c_j}|\hat H_{j}|\Psi_{j}\rangle  \nonumber \\
&&= (T_{j,n}(\nabla_{c_j})+U_{j,\alpha} ({\bf R}_{c_j}) ) \Phi_{j,\alpha} ({\bf R}_{c_j}) \nonumber \\
&&=\langle {\bf R}_{c_j}| (\hat T_{j,n}+U_{j,\alpha} (\hat {\bf R}_{c_j}) )|\Phi_{j,\alpha}\rangle \ .
\een  
The above relation holds for any value of ${\bf R}_{c_j}$ (within the approximation of Eq. (\ref{eq:e_dk_rd})), and thus leads to the following identity:
\ben
\langle E_{j,\alpha}|\hat H_{j}|\Psi_{j}\rangle = (\hat T_{j,n}+U_{j,\alpha} (\hat {\bf R}_{c_j}) )|\Phi_{j,\alpha}\rangle \ .
\een

Applying the same procedure as described above for all other adiabatic electronic states and considering linear combination of 
them, we can express the general molecular quantum state of the $j$th chromophore as follows:
\be 
|\Psi_{j}\rangle=\sum_\alpha C_{j,\alpha} |\Phi_{j,\alpha}\rangle |E_{j,\alpha}\rangle \ ,\label{eq:psi_d_lc}
\ee 
where $C_{j,\alpha}$'s are complex coefficients satisfying the normalization condition $\sum_\alpha |C_{j,\alpha}|^2=1$.
Applying $\hat H_{j}$ to Eq. (\ref{eq:psi_d_lc}) and invoking the Born-Oppenheimer approximation for each adiabatic electronic state, we obtain
\be
\hat H_{j}|\Psi_j\rangle =\sum_\alpha C_{j,\alpha} (\hat T_{j,n}+U_{j,\alpha} (\hat {\bf R}_{c_j}) )|\Phi_{j,\alpha}\rangle |E_{j,\alpha}\rangle \ .
\ee
This expression is equivalent to the following representation of the Hamiltonian of the $j$th pigment molecule:
\be
\hat H_{j}=\sum_\alpha  (\hat T_{j,n}+U_{j,\alpha} (\hat {\bf R}_{c_j}) ) |E_{j,\alpha}\rangle\langle E_{j,\alpha}| \ .\label{eq:hd_bo}
\ee
The set of $|E_{j,\alpha}\rangle$'s forms a complete basis and constitutes the site basis of the Frenkel exciton states. 

In the direct product space of the electronic states constituting the $j$th and $k$th pigment molecules, the identity resolution for the electronic degrees of freedom can be expressed as 
\ben
\hat 1_{jk,e} &=&\sum_\alpha\sum_{\alpha'} |E_{j,\alpha} \rangle \langle E_{j,\alpha}| \otimes |E_{k,\alpha'}\rangle \langle E_{k,\alpha'}| \nonumber \\
&=&\sum_\alpha\sum_{\alpha'} |E_{j,\alpha}\rangle|E_{k,\alpha'}\rangle \langle E_{j,\alpha}| \langle E_{k,\alpha'}|\ .
\een
Each term constituting $\hat H_{jk}$ of Eq. (\ref{eq:hjk_def}) can be projected into this basis of electronic states. 
The first term representing nuclear-nuclear repulsion, $\hat V_{jk,nn}$, remains the same and diagonal in the electronic basis.   
The second term representing the attraction between electrons in the $j$th pigment and nuclei in the $k$th pigment can be expressed as 
\ben 
&&\hat V_{jk,ne}=\sum_\alpha \sum_{\alpha',\alpha''}  V_{jk,ne}^{\alpha'\alpha''} (\hat {\bf R}_{c_j}) \nonumber \\
&&\hspace{.5in} \times  |E_{j,\alpha}\rangle |E_{k,\alpha'}\rangle\langle E_{j,\alpha}| \langle E_{k,\alpha''}| \ ,
\een
where 
\ben
&&V_{jk,ne}^{\alpha'\alpha''} (\hat {\bf R}_{c_j}) \nonumber \\
&&=-\sum_{l=1}^{L_j}\sum_{i=1}^{N_k}\langle E_{k,\alpha'}|\frac{Z_{j,l} e^2}{|\hat {\bf R}_{j,l}-\hat {\bf r}_{_k,i}|}|E_{k,\alpha''}\rangle  \ . \label{eq:v_ne-def}
\een
Similarly, the third term representing the attraction between electrons of the $j$th pigment and the nuclei of the $k$th pigment can be expressed as
\ben 
&&\hat V_{jk,en}=\sum_{\alpha,\alpha'} \sum_{\alpha''}  V_{jk,en}^{\alpha\alpha'} (\hat {\bf R}_{c_k})  \nonumber \\
&&\hspace{.5in}\times |E_{j,\alpha}\rangle |E_{k,\alpha''}\rangle\langle E_{j,\alpha'}| \langle E_{k,\alpha''}| \ ,  
\een
where 
\ben
&&V_{jk,en}^{\alpha\alpha'} (\hat {\bf R}_{c_k}) \nonumber \\
&&=-\sum_{l=1}^{L_k}\sum_{i=1}^{N_j}\langle E_{j,\alpha}|\frac{Z_{k,l} e^2}{|\hat {\bf R}_{k,l}-\hat {\bf r}_{j,i}|}|E_{j,\alpha'}\rangle \ . \label{eq:v_en-def}
\een
Finally, the fourth term representing repulsion between electrons of the $j$th and $k$th pigment can be expressed as
\ben
&&\hat V_{jk,ee} = \sum_{\alpha,\alpha'}\sum_{\alpha'',\alpha'''} V_{jk,ee}^{\alpha\alpha',\alpha''\alpha'''}  \nonumber \\
&&\hspace{.5in} |E_{j,\alpha}\rangle |E_{k,\alpha''}\rangle\langle E_{j,\alpha'}| \langle E_{k,\alpha'''}| \ ,
\een
where
\ben
&&V_{jk,ee}^{\alpha\alpha',\alpha''\alpha'''}  \nonumber \\
&&=\sum_{i=1}^{N_j}\sum_{i'= 1}^{N_k}\langle E_{j,\alpha}|\langle E_{k,\alpha''}|\frac{e^2}{|\hat {\bf r}_{j,i}-\hat {\bf r}_{k,i'}|} |E_{j,\alpha'}\rangle|E_{k,\alpha'''}\rangle \ .\nonumber \\ \label{eq:v_ee-def}
\een

We here define the following electron density matrix function:
\be 
\rho_{j}^{\alpha'' \alpha'}({\bf r}_{j,1})=N_j\int \left (\prod_{i=2}^{N_j}d{\bf r}_{j,i}\right) \langle E_{j,\alpha'}|{\bf r}_{c_j}\rangle \langle {\bf r}_{c_j}|E_{j,\alpha''}\rangle \ ,
\ee
where ${\bf r}_{c_j}$ refers to the collection of all the electron coordinates of the $j$th pigment.
For $\alpha'=\alpha''$, this corresponds to the single electron density at ${\bf r}_{j,1}$.  On the other hand, for $\alpha'\neq \alpha''$, it becomes the transition density.  All the interaction terms between different pigment molecules shown above can then be expressed  in terms of the above electron density matrix.  First, Eq. (\ref{eq:v_ne-def}) can be shown to be 
\ben
&&V_{jk,ne}^{\alpha'\alpha''} (\hat {\bf R}_{c_j})\nonumber \\
&& =-\int d{\bf r}_{c_k} \sum_{l=1}^{L_j}\sum_{i=1}^{N_k}\langle E_{k,\alpha'}|{\bf r}_{c_k}\rangle\langle {\bf r}_{c_k}|\frac{Z_{j,l} e^2}{|\hat {\bf R}_{j,l}-\hat {\bf r}_{k,i}|}|E_{k,\alpha''}\rangle \nonumber \\
&&=- \sum_{l=1}^{L_j} \int d{\bf r}\ \rho^{\alpha''\alpha'}_{k}({\bf r})\frac{Z_{j,l} e^2}{|\hat {\bf R}_{j,l}- {\bf r}|} \ .
\een
Similarly, 
\ben
&&V_{jk,en}^{\alpha\alpha'} (\hat {\bf R}_{c_k})\nonumber \\
&& =-\int d{\bf r}_{c_j} \sum_{l=1}^{L_k}\sum_{i=1}^{N_j}\langle E_{j,\alpha}|{\bf r}_{c_j}\rangle\langle {\bf r}_{c_j}|\frac{Z_{k,l} e^2}{|\hat {\bf R}_{k,l}-\hat {\bf r}_{j,i}|}|E_{j,\alpha'}\rangle \nonumber \\
&&=-\sum_{l=1}^{L_k}\int d{\bf r}\ \rho_{j}^{\alpha'\alpha}({\bf r})\frac{Z_{k,l} e^2}{|\hat {\bf R}_{k,l}- {\bf r}|} \ .
\een
Finally, it is easy to show that 
\be
V_{jk,ee}^{\alpha\alpha',\alpha''\alpha'''} =\int d{\bf r} d{\bf r}' \rho_{j}^{\alpha'\alpha}({\bf r})\rho_{k}^{\alpha'''\alpha''}({\bf r}')\frac{e^2}{|{\bf r}-{\bf r}'|}  \ . \label{eq:vjk_ee}
\ee
This corresponds to the Coulomb interaction term between $\alpha\rightarrow \alpha'$ electronic transition of the $j$th pigment and $\alpha''\rightarrow \alpha'''$ electronic transition of the $k$th pigment.  

\subsubsection{Hamiltonian in the site excitation basis}
The site basis for the electronic states of the aggregates of pigments can be constructed by taking direct product of all the adiabatic electronic states.  
For the sake of simplicity, we here assume that the index $\alpha$ can be represented by nonnegative integers and there is no degeneracy in the ground and the first excited adiabatic electronic states of each pigment.  Thus, $\alpha=0$ represents the ground electronic state and $\alpha=1$ the first excited state of each pigment.
First, the ground electronic state of the aggregates can be expressed as  
\be
|g\rangle =\prod_{j=1}^{N_c}|E_{j,0}\rangle  \ . \label{eq:g-def}
\ee
The state where only the $j$th pigment is excited (site excitation state) is defined as 
\be
|s_j\rangle =\left (\prod_{k\neq j} |E_{k,0}\rangle\right)  |E_{j,1}\rangle \ , \label{eq:sj-def}
\ee
where the product is over all $k=1,\cdots,N_c$ except for $j$.   

The molecular Hamiltonian for the aggregate of pigments, $\hat H_c$, can be expressed in the basis spanned by $|g\rangle$ and $|s_j\rangle$ as defined above by employing the expressions obtained in the previous subsection.  First, the diagonal elements of $\hat H_c$ in the site excitation basis are given by
\ben
&&\langle g|\hat H_c|g\rangle=\sum_{j=1}^{N_c} \left (\hat T_{j,n}+U_{j,0}(\hat{\bf R}_{c_j})\right) \nonumber \\
&&\hspace{.5in}+\frac{1}{2}\sum_{j=1}^{N_c}\sum_{k\neq j}\left (V_{jk,nn}(\hat {\bf R}_{c_j},\hat {\bf R}_{c_{k}}) +V_{jk,ne}^{00}(\hat {\bf R}_{c_j})\right . \nonumber \\
&&\hspace{1in}\left . +V_{jk,en}^{00}(\hat {\bf R}_{c_{k}})+V_{jk,ee}^{00,00}\right)\ , \\
&&\langle s_j|\hat H_c|s_j\rangle=\langle g|\hat H_c|g\rangle +U_{j,1} (\hat {\bf R}_{c_j})-U_{j,0} (\hat {\bf R}_{c_j}) \nonumber \\
&&\hspace{.2in}+\frac{1}{2}\sum_{k\neq j}^{N_c}\left (V_{kj,ne}^{11}(\hat {\bf R}_{c_j}) - V_{kj,ne}^{00}(\hat {\bf R}_{c_j})\right . \nonumber \\
&&\hspace{.6in} +V_{jk,en}^{11}(\hat {\bf R}_{c_j}) - V_{jk,en}^{00}(\hat {\bf R}_{c_j}) \nonumber \\
&&\hspace{.6in}\left . +V_{kj,ee}^{00,11}-V_{kj,ee}^{00,00}+V_{jk,ee}^{11,00}-V_{jk,ee}^{00,00}\right)\ . 
\een
On the other hand, the off-diagonal elements are given by 
\ben
&&\langle s_j|\hat H_c|g\rangle=\frac{1}{2}\sum_{k\neq j}^{N_c}\left (V_{kj,ne}^{10}(\hat {\bf R}_{c_k}) +V_{jk,en}^{10}(\hat {\bf R}_{c_k}) \right )=\hat J_{jg}^{c,0}\nonumber \\
  \label{eq:j_jgc}\\
&&\langle g|\hat H_c|s_j\rangle=\frac{1}{2}\sum_{k\neq j}^{N_c}\left (V_{kj,ne}^{01}(\hat {\bf R}_{c_k}) +V_{jk,en}^{01}(\hat {\bf R}_{c_k}) \right )=\hat J_{gj}^{c,0}\nonumber \\
 \label{eq:j_gjc}\\
&&\langle s_j|\hat H_c|s_k\rangle =V_{jk,ee}^{10,01}=J_{jk}^{c,0} \ , \mbox{ for } j\neq k \ .  \label{eq:j_jk}
\een
In Eqs. (\ref{eq:j_jgc}) and (\ref{eq:j_gjc}), the fact that $V_{kj,ee}^{00,10}=V_{jk,ee}^{10,00}=V_{kj,ee}^{00,01}=V_{jk,ee}^{01,00}=0$ has been used, which can be proved from the fact that the integration of the transition density over the entire electronic coordinates becomes zero.

Let us define 
\ben
&&E_g^{c,0}=\sum_{j=1}^{N_c} U_{j,0}(\hat{\bf R}_{c_j}^0) \nonumber \\
&&\hspace{.2in}+\frac{1}{2}\sum_{j=1}^{N_c}\sum_{k\neq j}\left (V_{jk,nn}(\hat {\bf R}_{c_j}^0,\hat {\bf R}_{c_{k}}^0) +V_{jk,ne}^{00}(\hat {\bf R}_{c_j}^0)\right . \nonumber \\
&&\hspace{1in}\left . +V_{jk,en}^{00}(\hat {\bf R}_{c_k}^0)+V_{jk,ee}^{00,00}\right)\ , \label{eq:def-egc}
\een
\ben
&&E_j^{c,0}=E_g^c+ U_{j,1} (\hat {\bf R}_{c_j}^0)-U_{j,0} (\hat {\bf R}_{c_j}^0) \nonumber \\
&&\hspace{.2in}+\frac{1}{2}\sum_{k\neq j}^{N_c}\left (V_{kj,ne}^{11}(\hat {\bf R}_{c_k}^0) - V_{kj,ne}^{00}(\hat {\bf R}_{c_k}^0)\right . \nonumber \\
&&\hspace{.5in} +V_{jk,en}^{11}(\hat {\bf R}_{c_k}^0) - V_{jk,en}^{00}(\hat {\bf R}_{c_k}^0) \nonumber \\
&&\hspace{.5in}\left . +V_{kj,ee}^{00,11}-V_{kj,ee}^{00,00}+V_{jk,ee}^{11,00}-V_{jk,ee}^{00,00}\right) \ . \label{eq:def-ejc}
\een
Collecting all the terms involving nuclear degrees of freedom, we can also define 
\ben
&&\hat H_{c,b}^0=\sum_{j=1}^{N_c} \left (\hat T_{j,n}+U_{j,0}(\hat{\bf R}_{c_j}) - U_{j,0}(\hat{\bf R}_{c_j}^0)\right) \nonumber \\
&&\hspace{.2in}+\frac{1}{2}\sum_{j=1}^{N_c}\sum_{k\neq j}\left (V_{jk,nn}(\hat {\bf R}_{c_j},\hat {\bf R}_{c_{k}}) - V_{jk,nn}(\hat {\bf R}_{c_j}^0,\hat {\bf R}_{c_{k}}^0) \right . \nonumber \\
&&\hspace{.9in}+V_{jk,ne}^{00}(\hat {\bf R}_{c_j}) - V_{jk,ne}^{00}(\hat {\bf R}_{c_j}^0) \nonumber \\
&&\hspace{.9in}\left . +V_{jk,en}^{00}(\hat {\bf R}_{c_{k}}) - V_{jk,en}^{00}(\hat {\bf R}_{c_{k}}^0)\right)\ ,  \label{eq:hcb}\\
&&\hat B_{c,j}^0=U_{j,1} (\hat {\bf R}_{c_j})-U_{j,0} (\hat {\bf R}_{c_j}) - U_{j,1} (\hat {\bf R}_{c_j}^0)+U_{j,0} (\hat {\bf R}_{c_j}^0)\nonumber \\
&&\hspace{.2in}+\frac{1}{2}\sum_{k\neq j}^{N_c}\left (V_{kj,ne}^{11}(\hat {\bf R}_{c_k}) - V_{kj,ne}^{11}(\hat {\bf R}_{c_k}^0) \right .\nonumber \\
&&\hspace{.6in}-V_{kj,ne}^{00}(\hat {\bf R}_{c_k}) + V_{kj,ne}^{00}(\hat {\bf R}_{c_k}^0)\nonumber \\
&&\hspace{.6in} +V_{jk,en}^{11}(\hat {\bf R}_{c_k}) - V_{jk,en}^{11}(\hat {\bf R}_{c_k}^0) \nonumber \\
&&\hspace{.6in} \left . - V_{jk,en}^{00}(\hat {\bf R}_{c_k}) + V_{jk,en}^{00}(\hat {\bf R}_{c_k}^0) \right )\ . \label{eq:bcj}
\een
Collecting all these terms, we obtain the exciton-bath Hamiltonian defined by Eqs. (\ref{eq:h_c_ebh}).  It is important to note that the ground electronic state and the single exciton state in this section has been defined as direct product of those of independent pigment molecules.  In practice, these can be improved further by defining them as those accounting for the implicit effect of the other pigment molecules in the ground electronic states.   

\section{Modeling of Environmental effects}
In the context of the modeling of electronic processes in embedded systems, the most used multiscale approach is that combining the QM description of the chromophores with a classical description of the environment.
The success of such QM/classical formulation is mostly related to the accuracy with which the interactions between the QM and the classical parts are treated. 
Generally speaking, we can identify two different classes of QM/classical formulations, namely that describing the environment as a continuum, in terms of its macroscopic properties (QM/continuum), and that keeping an atomistic description through MM force fields (QM/MM). Both methods allow to account for the presence of the environment in the description of a molecular system at an affordable computational cost, increasing both the possibilities of molecular modeling and the manifold of treatable systems. 
In both versions of QM/classical models, an important common aspect is that the QM part can be modified in its electronic and nuclear degrees of freedom by the presence of the classical part. From a QM point of view, the Hamiltonian of the whole system can be written as:
\begin{equation}
\hat{H} = \hat{H}_{QM} + \hat{H}_{\textrm{env}} + \hat{H}_{\textrm{int}} 
\label{eq:HTOT}
\end{equation}
where $\hat{H}_{QM}$ is the Hamiltonian of the gas-phase (isolated) QM subsystem, $\hat{H}_{\textrm{env}}$ is the Hamiltonian of the rest of the system, which is purely classical, and finally $\hat{H}_{\textrm{int}}$ represents the interaction between the QM and the classical parts. In the QM/classical models the degrees of freedom of the sole environment are not of great relevance. In continuum solvation models, the solvent atomistic nature disappears together with the $\hat{H}_{\textrm{env}}$ term. In the case of QM/MM approach, $\hat{H}_{\textrm{env}}$ is maintained, but such term only adds a constant term to the total energy.

In continuum models, the QM subsystem is placed in a suitably shaped molecular cavity $C$ immersed in the dielectric medium representing the environment. The polarization response of the medium to the QM charge distribution is obtained by solving the Poisson equation of classical electrostatics \cite{tomasi-cr105}.
If we assume that the charge distribution ($\rho$) of the QM subsystem is entirely contained inside the cavity and that the dielectric outside the cavity is characterized by a scalar dielectric constant, we obtain
\begin{equation}
\begin{cases}
-\nabla^2 V(\br) = 4\pi \rho(\br) & \mbox{within the cavity } \\ 
-\epsilon \nabla^2V(\br)=0            & \mbox{outside the cavity}
\end{cases}
\label{eq:poisson2}
\end{equation}
A possible strategy to solve this equation is to write the electrostatic potential $V$ as a sum of the solute potential plus the contribution due to the reaction of the environment (i.e. the polarization of the dielectric).
Among the possible approaches to define the reaction potential, a very effective one is the apparent surface charges (ASC) approach, where an apparent surface density $\sigma(\bs)$ spread on the cavity surface $\Gamma$ is used to represent the reaction (polarization) of the dielectric \cite{tomasi-cr105}. 
From a computational point of view, the solution is achieved by partitioning the cavity surface into a set of finite elements and substituting  $\sigma(\bs)$ with a set of point charges ($q_t$), each corresponding to a surface element.
There have been different definitions of $q_t$ proposed, leading to different formulations of the so-called Polarizable Continuum Models (PCMs)\cite{mennucci-wires2}, or to Conductor-like Screening Model (COSMO)\cite{Klamt:2011ji}. 
Within this framework, the operator $\hat{H}_{\text{int}}$ in Eq. (\ref{eq:HTOT}) represents the electrostatic interaction between the QM charge density and the apparent charges on the surface of the cavity:
\begin{equation}
\label{eq:HPCM}
\hat{H}_{\textrm{int}}=
\hat{H}_\textrm{QM/PCM} = \sum_t q_t \hat{V}_{\text{QM}}(\br_t)\ ,
\end{equation}
where $q_t$ is the ASC at the position $\br_t$ in the Cartesian coordinate system, and $\hat{V}_{QM}(\br_t)$ is the electrostatic potential operator due to the QM charge density calculated at $\br_t$. The summation runs over all the $N_t$ surface elements. Since $\hat{H}_{\text{QM/PCM}}$ depends on the QM charge density which is modified by the environment through $\hat{H}_{\textrm{int}}$, a non-linear problem is obtained and its solution leads to a mutually polarized QM/continuum system.

Continuum models are often the ideal strategy to account for the bulk effects of the environment. On the other hand, the use of a structureless medium, does not allow to include the effects of the heterogeneity, as well as specific interactions between the two parts. 
In particular, when the heterogeneity of the environment acts at a small scale as in a protein matrix, where each position/orientation of the QM subsystem feels a different local environment or in the presence of specific interactions such as hydrogen bonds, an atomistic description is preferred.
In these cases,  QM/MM models represent a very effective strategy.
However, they also suffer from a drawback: the QM-environment interaction depends critically on the configuration of the environment, particularly those at short range. Therefore, to correctly account for the dynamical nature of the interactions, several configurations of the whole system need to be taken into consideration. This is not needed when a continuum approach is employed since it implicitly gives a configurationally sampled effect due to the use of macroscopic properties.
Indeed, the sampling issue may introduce an additional difficulty related to the fact that it often involves
classical MD simulation, which can incur errors
resulting from the inaccuracy of the force-field used  and the intrinsic approximation of classical MM simulations.

In the most common formulation, the MM part of QM/MM systems is described through a set of fixed point charges, generally placed at the atoms of the environment. The QM/classical interactions are therefore of electrostatic nature. The MM charges are chosen to best represent the molecular electrostatic properties. Eventually, higher multipole moments can be used to improve the electrostatic description. 

The $\hat{H}_{int}$ term in Eq. (\ref{eq:HTOT}) is given by the interaction between the electronic density of the QM subsystem $\rho_s(\br)$ and the MM charge distribution ($\hat{H}_{\textrm{QM/MM}}$).  Namely,
\begin{equation}
\label{eq:HeffQMMM}
\hat{H}_{\textrm{int}}=
\hat{H}_\textrm{QM/MM} = \sum_m q_m \hat{V}_{\text{QM}}(\br_m)\ ,
\end{equation}
where $q_m$'s are the fixed partial charges placed at position $\br_m$'s in the Cartesian coordinate system and $\hat{V}_{QM}(\br_m)$ is the electrostatic potential operator due to the QM charge density calculated at $\br_m$. The summation runs over all the number of  charges. 
This scheme is also known as ``electrostatic embedding" in the sense that the electronic structure of the QM part is modified by the presence of the charge distribution representing the environment and consequentially is polarized by it. 
However, the use of fixed point charges to describe the MM
system implies that, while the QM density is polarized by the MM one, the opposite is not true. This is a serious limitation, as the explicit response arising from the environment polarization can be crucial, particularly when charged or very polar systems are studied, or when electronic excitation processes are considered. In order to improve the QM/MM description, a possible strategy is to use a ``flexible" MM model which can be polarized back by the QM charge distribution.
Three main groups of methods include these mutual polarization effects \cite{Senn:2009gk,mennucci-pccp15}.
In the induced dipole (ID) approach, atomic polarizabilities are assigned to atoms that lead to induced dipoles in the presence of an electric field. The source of the electric field are the QM charge distribution, the MM point charges and the induced dipoles themselves. A self-consistent procedure is required to solve the QM/MM problem because the dipoles interact with each other and act back on the QM electron density. The parameters to be defined in addition to the MM charges are the atomic polarizabilities.
If Drude oscillators (DO) are used, instead, atoms are represented by a pair of point charges separated by a variable distance (namely a spring).  When interacting with an electric field, a dipole is generated. The parameters are the magnitude of the mobile charge and the force constant of the spring. Finally, when the fluctuating charges (FQ)  model is used, 
charges placed on the atoms are allowed to fluctuate so as to represent the charge flow within the molecule. Two sets of atomic parameters are needed (hardness and electronegativities) as the model is based on the electronegativity equalization principle, which states that, at equilibrium, the instantaneous electronegativity of each atom has the same value. 

In the ID approach, a set of atomic polarizabilities is assigned to MM atoms.  
The polarizability $\boldsymbol{\alpha}_i$ is a $3\times 3$ tensor. However, in most formulations of the model, only isotropic component $\alpha^{\textrm{iso}}$ is used.
Within this framework, the induced dipole moment $\bmu$ at the MM site $i$ can be written as
\begin{equation}
\label{eq:muind2}
\mu_{i}=\alpha^{\textrm{iso}}_{i} \left[ E_i^{\textrm{ext}} - \sum_{j \neq i}^{N_p} \mathbf{T}_{ij} \mu_j \right]
\end{equation}
where $E_i^{\textrm{ext}}$ is the electric field due to the QM subsystem and the set of MM atomic charges.  
The $\mathbf{T}_{ij}$ is known as the dipole-dipole interaction tensor and is defined as:
\begin{equation}
\label{eq:tensor}
\mathbf{T}_{ij} = \frac{f_e}{r_{ij}^3}\mathbf{I} - \frac{3f_t}{r_{ij}^5}
\left[ \begin{array}{ccc} 
r_x^2 & r_x r_y & r_x r_z \\ 
r_y r_x & r_y^2 & r_y r_z \\ 
r_z r_x & r_z r_y & r_z^2 
\end{array} \right]
\end{equation}
where $\mathbf{I}$ is the 3 x 3 unit tensor and $r_x$, $r_y$ and $r_z$ are Cartesian components of the vector connecting the two atoms $i$ and $j$. The $f_e$ and $f_t$ factors are distance-dependent screening functions that depend on the specific dipole interaction models.

In recent years, atomistic (MM) and continuum approaches have been coupled to give fully polarizable QM/MM/continuum approaches \cite{Steindal:2011cj,Lipparini:2011hu,boulanger-jctc8,Caprasecca:2012dk}.  In these methods,
on top of the polarizable discrete model, one adds a further external polarizable continuum layer, coupled to the former. 
In this way the continuum description of the environment is combined together with the polarizable QM/MM, thus obtaining a three level model which allows to exploit the advantages of each method.  The polarizable MM model is used to describe short-range directional interactions, whereas the continuum one accounts for long-range (or bulk) effects.

When hybrid QM/classical models are used to describe ultrafast processes such as electronic excitations in embedded chromophores, a new aspect regarding the characteristic response time of the environmental degrees of freedom has to be taken into account. When a vertical excitation occurs, in fact, the electronic (or dynamic) component of the response, which is assumed to be sufficiently fast, immediately follows any change in the chromophore charge distribution.  On the contrary, the slower (or inertial) response that arises from nuclear and molecular motions remains frozen in its initial state. The possible delay of the slower component results in a nonequilibrium regime, which eventually relaxes into a new equilibrium where all the degrees of freedom of the environment reach the equilibrium with the excited state chromophore.
Especially for highly polar environments, \textit{equilibrium} and \textit{nonequilibrium} regimes represent very different configurations and their energy difference is generally known as ``reorganization energy."

The \textit{nonequilibrium} regime can be properly taken into account with both the QM/continuum and QM/MM models. In the first case, if we adopt the ASC framework, the fast (or dynamic) response will be represented in terms of apparent charges obtained  from the the optical component ($\epsilon_{\infty}$) of the dielectric permittivity instead of the static one. 
By contrast, the slow response is obtained as the difference of the equilibrium charges and the dynamic ones.
In the QM/MM framework, instead, the slow component is automatically taken into account if the MM charge distribution does not change during the excitation process (charges are fixed in position and in value). For what concerns the fast component, this can be properly taken into account only if a polarizable embedding is considered.

\section{Lineshape and response functions}
The model considered here is more general than what is considered in the main text because we assume here that the EBH can have additional time dependence due to fluctuating energies and couplings.
Thus, the total Hamiltonian including interaction with the radiation has the following form:
\be
\hat H_T(t)=E_g(t)|g\rangle\langle g|+\hat H_{ex}(t)+\hat H_{int}(t)\ , \label{eq:hamil_tot}
\ee
where $\hat H_{int}(t)$ is the matter-radiation interaction Hamiltonian and  
\be
\hat H_{ex}(t)=\hat H_e(t)+\sum_{j=1}^{N_c}\sum_{k=1}^{N_c} \hat B_{jk}|s_j\rangle\langle s_k|+\hat H_b \ , 
\ee
with 
\be
\hat H_e(t)=\sum_{j=1}^{N_c} E_j(t)|s_j\rangle\langle s_j|+\sum_{j=1}^{N_c}\sum_{k\neq j}J_{jk}(t)|s_j\rangle\langle s_k| \ .
\ee
The time dependences in $E_g(t)$, $E_j(t)$, and $J_{jk}(t)$ represent the influence of all other sources that are not represented by the bath Hamiltonian.  The details of these time dependent terms depend on specific experimental conditions under which a specific spectroscopic measurement is made.

The eigenstates of $\hat H_e(t)$ are denoted as $|\varphi_j(t)\rangle$'s, and we introduce $M_{kj}(t)=\langle s_k|\varphi_j(t)\rangle$.  Thus, 
\be
|s_k\rangle=\sum_j M_{kj}^*(t)|\varphi_j(t)\rangle \ .
\ee
The transition dipole vector for the excitation from $|g\rangle$ to $|s_k\rangle$, $\mbox{\boldmath$\mu$}_k$, is assumed to be time dependent in general.  
Then, the polarization operator for the transitions to the single exciton space at time $t$ is given by 
\ben
\hat {\bf P}(t)&=&\sum_{k} \mbox{\boldmath$\mu$}_k(t) \left (|s_k\rangle \langle g|+|g\rangle\langle s_k|\right)  \nonumber \\
&=&\sum_j \left ({\bf D}_j(t) |\varphi_j(t)\rangle\langle g|+|g\rangle\langle \varphi_j(t)| {\bf D}_j(t)  \right)  \ , \label{eq:p_def}
\een
where ${\bf D}_j(t)=\sum_k \mbox{\boldmath$\mu$}_k (t) M_{kj}^*(t)$.  

The time evolution operator for the total Hamiltonian Eq. (\ref{eq:hamil_tot}) is denoted as 
\be 
\hat {\mathcal U}_T(t,t_0)=\exp_{(+)}\left\{-\frac{i}{\hbar}\int_{t_0}^t dt' \hat H_T(t')\right\}\ , 
\ee
where $(+)$ represents chronological time ordering.

Thus, given that the total density operator at time $t_0$ is prepared to be $\hat \rho(t_0)$, at time $t$,  it evolves into
\be
\hat \rho(t)=\hat {\mathcal U}_T(t,t_0)\hat \rho(t_0) \hat{\mathcal U}_T^\dagger(t,t_0) \ . \label{eq:rho_t_gen}
\ee

For the discussion that follows, it is convenient to introduce the total Hamiltonian (without matter-radiation interaction) in the ground electronic state and the manifold of single exciton states as follows:
\ben
&&\hat H_g(t)=E_g(t)|g\rangle\langle g|+\hat H_b\ , \\
&&\hat H(t)=E_g(t)|g\rangle\langle g|+\hat H_{ex}(t) \ .
\een
We also denote the collection of all the time dependent parameters as 
\be
\Gamma (t)\equiv (E_g(t), E_j(t)\mbox{'s}, J_{jk}(t)\mbox{'s}, \mbox{\boldmath$\mu$}_k(t)\mbox{'s})
\ee

\subsection{Absorption} 
For $t\leq t_0$,  the system is in the ground electronic state and the bath is in thermal equilibrium.  Therefore, 
\be
\hat \rho (t_0)=|g\rangle \langle g| \rho_b \ , \label{eq:initial-abs}
\ee 
where 
$\hat \rho_b=e^{-\beta \hat H_b}/Tr_b\{e^{-\beta \hat H_b}\}$ with $\beta=1/k_BT$.   For a monochromatic radiation with frequency $\omega$ and polarization {\boldmath{${\eta}$}}, the matter-radiation interaction Hamiltonian (within the semiclassical approximation for the radiation and the rotating wave approximation) can be expressed as
\be
\hat H_{int}(t)=Ae^{-i\omega t} |D(t)\rangle \langle g|+A^*e^{i\omega t}|g\rangle\langle D(t)| \ ,
\ee
where $A$ is the amplitude of the radiation including the wave vector, and
\be 
|D(t)\rangle=\sum_j \mbox{\boldmath{${\eta}$}} \cdot  {\bf D}_j (t)|\varphi_j(t)\rangle \ .
\ee  

Expanding Eq. (\ref{eq:rho_t_gen}) up to the second order of $\hat H_{int}(t)$ and introducing the following identity operator in the single exciton space, 
\be 
\hat  I_e=\sum_j|s_j\rangle\langle s_j| \ ,
\ee 
the probability to find the population of exciton at time $t$, when approximated up to the second order of matter-radiation interaction, becomes
\ben 
&&P_e(t)=\frac{1}{\hbar^2}\int_{t_0}^t dt'\int_{t_0}^t dt''\ Tr\left \{\hat I_e \hat {\mathcal U}(t,t')\hat H_{int}(t') \right . \nonumber \\
&&\hspace{.3in}\times \left . \hat {\mathcal U} (t',t_0)\hat \rho (t_0) \hat {\mathcal U}^\dagger (t'',t_0)\hat H_{int}(t'')\hat {\mathcal U}^\dagger  (t,t'')\right\} \nonumber \\
&&=\frac{|A|^2}{\hbar^2}\int_{t_0}^t dt' \int_{t_0}^t dt'' e^{i\omega (t''-t')}Tr\left\{ \hat {\mathcal U}_g(t_0,t'')|g\rangle\langle D(t'')| \right. \nonumber \\
&&\hspace{.3in}\times \left . \hat {\mathcal U}_{ex}(t'',t') |D(t')\rangle\langle g|\hat {\mathcal U}_g(t',t_0)\hat \rho(t_0)\right\} \ . \label{eq:pet-second}
\een
where
\ben
&&\hat {\mathcal U}(t,t_0)=\exp_{(+)}\left\{-\frac{i}{\hbar}\int_{t_0}^t d t'  \hat H(t') \right\} \ ,\\ 
&&\hat {\mathcal U}_g(t,t_0)=e^{-\frac{i}{\hbar} \int_{t_0}^t d\tau E_g(\tau)}|g\rangle\langle g|e^{-\frac{i}{\hbar}\hat H_b(t-t_0)} \ ,\\
&&\hat {\mathcal U}_{ex}(t,t_0)=\exp_{(+)}\left\{-\frac{i}{\hbar}\int_{t_0}^t d t' \hat H_{ex}(t') \right\} \ .
\een
Taking time derivative of Eq. (\ref{eq:pet-second}), one obtains
\ben
&&\frac{d}{dt}P_e(t)=2 \frac{|A|^2}{\hbar^2} {\rm Re} \int_{t_0}^t dt' e^{i\omega (t-t')} Tr\left\{ \hat {\mathcal U}_g(t_0,t) \right. \nonumber \\
&&\times \left . |g\rangle\langle D(t)|\hat {\mathcal U}_{ex}(t,t') |D(t')\rangle\langle g|\hat {\mathcal U}_g(t',t_0)\hat \rho(t_0)\right\} \ .
\een

Let us introduce the following ``ideal" lineshape by dividing the above time derivative with appropriate normalization factor: 
\ben
&&I_{id} [\omega,\Gamma(t)]=\frac{\hbar^2}{2\pi |A|^2}\frac{d}{dt} P_e(t) \nonumber \\
&&=\frac{1}{\pi}{\rm Re} \int_{t_0}^{t} dt'\ e^{i\omega (t-t')} e^{\frac{i}{\hbar} \int_{t'}^t d\tau' E_g(\tau')} \nonumber \\ 
&&\times Tr\left\{\hat {\mathcal U}_{ex}(t,t') |D(t')\rangle \langle D(t)|\hat \rho_b e^{\frac{i}{\hbar} H_b(t-t')}\right\} \ . \label{eq:i_id_gen}
\een
In general, the absorption line shape can be defined as 
\be
I(\omega)=\lim_{t\rightarrow t_s}\langle I_{id}[\omega;\Gamma (t)]\rangle_{\Gamma (t)}\ .
\ee

For the simple case where there is no time dependent fluctuation of $E_g(t)$ and parameters constituting $\hat H_{ex}(t)$,  Eq. (\ref{eq:i_id_gen}) reduces to 
\ben
&&I_{id} (\omega)=\frac{1}{\pi}{\rm Re} \int_{0}^{t-t_0} d\tau \ e^{i\omega \tau} e^{\frac{i}{\hbar} \tau E_g} \nonumber \\ 
&&\times Tr\left\{e^{-\frac{i}{\hbar}\tau \hat H_{ex}} |D\rangle \langle D|\hat \rho_b e^{\frac{i}{\hbar}\tau  H_b}\right\} \ . \label{eq:i_id_gen-nt}
\een
Taking the limit of $t-t_0\rightarrow \infty$ and making average over the disorder, orientation, and polarization leads to Eq. (\ref{eq:abs-lhc}).

\subsection{Emission}
Assume that the matter is in single exciton space and is in equilibrium at time $t_0$.  Then, the initial density operator can be expressed as 
\be
\hat \rho (t_0) =\frac{e^{-\beta \hat H_{ex}(t_0)}}{Tr\{e^{-\beta \hat H_{ex}(t_0)} \}} \ .
\ee
For the above initial density operator, following a procedure similar to obtaining Eq. (\ref{eq:pet-second}), the population of the ground state at time $t$, when approximated up to the second order of matter-radiation interaction, can be shown to be
\ben 
&&P_g(t)=\frac{1}{\hbar^2}\int_{t_0}^t dt'\int_{t_0}^t dt''\ Tr\left \{|g\rangle\langle g| \hat {\mathcal U}(t,t')\hat H_{int}(t') \right . \nonumber \\
&&\hspace{.3in}\times \left . \hat {\mathcal U} (t',t_0)\hat \rho (t_0) \hat {\mathcal U}^\dagger (t'',t_0)\hat H_{int}(t'')\hat {\mathcal U}^\dagger  (t,t'')\right\} \nonumber \\
&&=\frac{|A|^2}{\hbar^2}\int_{t_0}^t dt' \int_{t_0}^t dt'' e^{i\omega (t'-t'')}Tr\left\{ \hat {\mathcal U}_{ex}(t_0,t'')|D(t'')\rangle \langle g| \right. \nonumber \\
&&\hspace{.3in}\times \left . \hat {\mathcal U}_g(t'',t') |g\rangle\langle D(t')|\hat {\mathcal U}_{ex}(t',t_0)\hat \rho(t_0)\right\} \ .
\een
Then, 
\ben
&&E_{id} [\omega,\Gamma(t)]=\frac{\hbar^2}{2\pi |A|^2}\frac{d}{dt} P_g(t)\nonumber \\
&&=\frac{1}{\pi}{\rm Re} \int_{t_0}^{t} dt'\ e^{-i\omega (t-t')} e^{-\frac{i}{\hbar} \int_{t'}^t d\tau' E_g(\tau')} \nonumber \\ 
&&\times Tr\left\{\hat {\mathcal U}_{ex}(t_0,t) e^{-\frac{i}{\hbar}(t-t')\hat H_b} |D(t)\rangle \langle D(t')|\hat {\mathcal U}_{ex}(t',t_0)\hat \rho(t_0)\right\} \nonumber \\ \label{eq:e_id_gen}
\een
In general, the emission line shape can be defined as 
\be
E(\omega)=\lim_{t\rightarrow t_s}\langle E_{id}[\omega;\Gamma (t)]\rangle_{\Gamma (t)}\ .
\ee

For the simple case where there is no time dependent fluctuation of $E_g(t)$ and parameters constituting $\hat H_{ex}(t)$, Eq. (\ref{eq:e_id_gen}) reduces to 
\ben
&&E_{id} [\omega,\Gamma(t)]=\frac{1}{\pi}{\rm Re} \int_{0}^{t-t_0} d\tau\ e^{-i\omega \tau} e^{-\frac{i}{\hbar} \tau E_g} \nonumber \\ 
&&\times Tr\left\{e^{-\frac{i}{\hbar}\tau \hat H_b} |D\rangle \langle D|\hat \rho_{ex} e^{\frac{i}{\hbar}\tau \hat H_{ex}}\right\} \ , \label{eq:e_id_gen-nt}
\een
where $\hat \rho_{ex}=e^{-\beta\hat H_{ex}}/Tr\{e^{-\beta \hat H_{ex}}\}$. 
Taking the limit of $t-t_0\rightarrow \infty$ and making average over the disorder, orientation, and polarization of the above expression leads to Eq. (\ref{eq:ems-lhc}).
\subsection{Four wave mixing spectroscopy}

For three incoming pulses, the  matter-radiation interaction Hamiltonian is  given by 
\ben
&&\hat H_{int}(t)=\sum_{\nu=1}^3\sum_j {\bf E}_\nu(t-t_\nu)\cdot  {\bf D}_j(t) |\varphi_j(t)\rangle \langle g| e^{i{\bf k}_\nu\cdot {\bf r}-i\omega_\nu t} \nonumber \\
&&\hspace{0.5in}+ {\rm H.c.}  \nonumber \\
&&=\sum_{\nu=1}^3 E_\nu(t-t_\nu) |D_\nu (t)\rangle \langle g| e^{i{\bf k}_\nu\cdot {\bf r}-i\omega_\nu t} + {\rm H.c.} \ .\label{eq:h_int_def}
\een    
In the above expression, ${\bf E}_\nu (t-t_\nu)=E_\nu (t-t_\nu) \mbox{\boldmath{${\eta}$}}_\nu $, with $E_\nu (t-t_\nu)$ and $\mbox{\boldmath{${\eta}$}}_\nu$ respectively defining the amplitude and unit polarization vector of the $\nu$th pulse.  It is assumed that $t_3 \geq t_2\geq t_1$. 
In the last line of Eq. (\ref{eq:h_int_def}), $|D_\nu(t)\rangle$ is the sum of all the exciton states weighted by the components of the transition dipoles along the direction $\mbox{\boldmath{${\eta}$}}_\nu$ and has the following expression:
\be 
|D_\nu(t)\rangle=\sum_j \mbox{\boldmath{${\eta}$}}_\nu \cdot  {\bf D}_j(t) |\varphi_j(t)\rangle \ .
\ee  
In representing the final expressions, it is useful to introduce the following polarization operators: 
\ben
&&\hat P_{g\nu}(t)=|g\rangle\langle D_\nu(t)| \ ,\\
&&\hat P_{\nu g}(t)=|D_\nu(t)\rangle\langle g|=\hat P_{g\nu}^\dagger(t) \ .
\een

Expanding $\hat {\mathcal U}_T(t,t_0)$ and $\hat {\mathcal U}_T^\dagger(t,t_0)$ with respect to $\hat H_{int}(t)$, and collecting all the terms of the third order, we find the following third order components of the density operator: 
\be
\hat \rho^{(3)}(t_m)=\hat \rho_{I}(t_m)+\hat \rho_{I}^\dagger(t_m)+\hat \rho_{II}(t_m)+\hat \rho_{II}^\dagger (t_m) \comma
\ee
where
\ben
\hat \rho_{I}(t_m)&=&-\frac{i}{\hbar^3}\int_{0}^{t_m} dt\int_{0}^{t} dt' \int_{0}^{t_m} dt''\ \hat {\mathcal U}(t_m,t) \nonumber \\
&&\times \hat H_{int}(t)\hat {\mathcal U}(t,t')\hat H_{int}(t')\hat {\mathcal U}(t',t_0)\hat \rho(t_0)\nonumber \\
&&\times \hat {\mathcal U}^\dagger (t'',t_0)\hat H_{int}(t'')\hat {\mathcal U}^\dagger(t_m,t'') \ ,  \label{eq:rho_I}
\\
\hat \rho_{II}(t_m)&=&-\frac{i}{\hbar^3}\int_{0}^{t_m} dt \int_{0}^{t} dt' \int_{0}^{t'} dt'' \ \hat {\mathcal U}(t_m,t_0) \hat \rho (t_0) \nonumber \\
&&\times \hat {\mathcal U}^\dagger (t'',t_0)  \hat H_{int}(t'') \hat {\mathcal U}^\dagger (t',t'') \hat H_{int}(t') \nonumber \\
&&\times \hat {\mathcal U}^\dagger (t,t')  \hat H_{int}(t)\hat {\mathcal U}^\dagger (t_m,t) \ .  \label{eq:rho_II}
\een

In Eq. (\ref{eq:rho_I}),
the integration over $t''$ can be split into three regions,
$0 < t''<t'$, $t'<t''<t$, and $t <t''<t_m$.    Relabeling the dummy time integration variables in each region such that $t\geq t' \geq t''$,  the three terms can be rewritten so as to have the same time integration boundaries as $\rho_{II}(t_m)$.  The resulting third order components can therefore be expressed as 
\ben
\hat \rho^{(3)}(t_m)&=&-\frac{i}{\hbar^3}\int_{0}^{t_m} dt \int_{0}^t dt' \int_{0}^{t'} dt''\sum_{j=1}^4 \hat{\mathcal T}_j (t_m,t,t',t'') \nonumber \\
&&+{\rm H.c.}\ , \label{eq:rho_3}
\een 
where 
\ben
&&\hat{\mathcal T}_1(t_m,t,t',t'')\equiv  \hat {\mathcal U}(t_m,t')\hat H_{int}(t')\hat {\mathcal U}(t',t'')\hat H_{int}(t'')\nonumber \\
&&\hspace{.1in}\times  \hat {\mathcal U}(t'',t_0)\hat \rho(t_0){\mathcal U}^\dagger (t,t_0)\hat H_{int}(t) {\mathcal U}^\dagger (t_m,t) \label{eq:t1}\ ,  \\
&&\hat {\mathcal T}_2(t_m,t,t',t'')\equiv \hat {\mathcal U}(t_m,t)\hat H_{int}(t)\hat {\mathcal U}(t,t'')\hat H_{int}(t'')\nonumber \\
&&\hspace{.1in} \times \hat {\mathcal U}(t'',t_0) \hat \rho(t_0)\hat {\mathcal U}^\dagger(t',t_0)\hat H_{int}(t') \hat {\mathcal U}^\dagger (t_m,t')\ , \label{eq:t2}\\
&&\hat {\mathcal T}_3(t_m,t,t',t'')\equiv \hat {\mathcal U}(t_m,t)\hat H_{int}(t)\hat {\mathcal U}(t,t')\hat H_{int}(t')\nonumber \\
&&\hspace{.1in}\times \hat {\mathcal U}(t',t_0)\hat \rho(t_0)\hat {\mathcal U}^\dagger (t'',t_0)\hat H_{int}(t'') \hat {\mathcal U}(t_m,t'')\ , \label{eq:t3}\\
&&\hat {\mathcal T}_4(t_m,t,t',t'')\equiv  \hat {\mathcal U}(t_m,t_0)\hat \rho(t_0)\hat {\mathcal U}^\dagger (t'',t_0)\hat H_{int}(t'')\nonumber \\
&&\hspace{.1in} \times \hat {\mathcal U}^\dagger (t',t'') \hat H_{int}(t')\hat {\mathcal U}^\dagger (t,t')\hat H_{int}(t)\hat {\mathcal U}^\dagger (t_m,t)\ .  \label{eq:t4}
\een

The corresponding third order contribution to the polarization can be calculated by taking the trace of the scalar product between $\hat {\bf P}(t_m)$, Eq. (\ref{eq:p_def}), and $\hat \rho^{(3)}(t_m)$, Eq. (\ref{eq:rho_3}).  The resulting expression for the third order polarization at time $t_m$  can be shown to be
\ben
&&\bar{\bf P}^{(3)}(t_m)\equiv Tr\{\hat {\bf P}(t_m)\hat \rho^{(3)}(t_m)\} =\frac{2}{\hbar^3}{\rm Im}\sum_{j=1}^4 \int_{0}^{t_m}dt\nonumber \\
&&\times \int_{0}^{t} dt' \int_{0}^{t'}dt'' Tr\left\{\hat {\bf P}(t_m) \hat {\mathcal T}_j (t_m,t,t',t'')\right\} \ ,  \label{eq:p4t}
\een
where $``{\rm Im}"$ implies imaginary part of the complex function.

Let us assume that we are interested in the polarization along the direction of $\mbox{\boldmath{${\eta}$}}_m$ at time $t_m$, and also define  
\ben
&&|D_m(t_m)\rangle=\sum_j  \mbox{\boldmath{${\eta}$}}_m \cdot {\bf D}_j(t) |\varphi_j(t)\rangle \ ,\\ 
&&\hat P_{mg}(t_m)=|D_m(t_m)\rangle\langle g| \ .
\een
Then, taking scalar product of $\mbox{\boldmath{${\eta}$}}_m$ with the integrand of Eq. (\ref{eq:p4t}) and considering only those terms where interactions with $E_1$, $E_2$, and $E_3$  occur in the chronological order at $t''$, $t'$, and $t$, respectively, we obtain the following general expressions: 
\ben
&&\int_{0}^{t_m} dt \int_{0}^t dt' \int_{0}^{t'} dt''\ \mbox{\boldmath{${\eta}$}}_m\cdot Tr\left\{\hat {\bf P} \hat {\mathcal T}_1(t_m,t,t',t'')\right\}\nonumber \\
&&= e^{-i({\bf k}_3+{\bf k}_2 -{\bf k}_1)\cdot {\bf r}}e^{i(\omega_3t_3+\omega_2t_2-\omega_1 t_1)} \nonumber \\
&&\hspace{.1in}\times \int_{0}^{t_m} dt \int_{0}^t dt' \int_{0}^{t'} dt'' \chi^{(1)}(t_m,t,t',t'')\nonumber \\
&&\hspace{.3in}\times E^*_3 (t-t_3)e^{i\omega_3(t-t_3)}E^*_2 (t'-t_2)e^{i\omega_2(t'-t_2)} \nonumber \\
&&\hspace{.3in}\times E_1 (t''-t_1)e^{-i\omega(t''-t_1)} \,  \label{eq:pt1}  
\een
\ben
&&\int_{0}^{t_m} dt \int_{0}^t dt' \int_{0}^{t'} dt''\ \mbox{\boldmath{${\eta}$}}_m\cdot Tr\left\{ \hat {\bf P} \hat {\mathcal T}_2(t_m,t,t',t'')\right\}\nonumber \\
&&=e^{-i({\bf k}_3 +{\bf k}_2 -{\bf k}_1) \cdot{\bf r}}e^{i(\omega_3 t_3+\omega_2 t_2-\omega_1 t_1)}\nonumber \\
&&\hspace{.1in}\times \int_{0}^{t_m} dt \int_{0}^t dt' \int_{0}^{t'} dt'' \chi^{(2)}(t_m,t,t',t'') \nonumber \\
&&\hspace{.3in}\times E^*_3 (t-t_3)e^{i\omega_3(t-t_3)} E^*_2 (t'-t_2) e^{i\omega_2(t'-t_2)}  \nonumber \\
&&\hspace{.3in}\times E_1 (t''-t_1)e^{-i\omega_1 (t''-t_1)} \ , \label{eq:pt2}
\een
\ben
&&\int_{0}^{t_m} dt \int_{0}^t dt' \int_{0}^{t'} dt''\ \mbox{\boldmath{${\eta}$}}_m\cdot Tr\left\{\hat {\bf P} \hat {\mathcal T}_3(t_m,t,t',t'')\right\}\nonumber \\
&&= e^{-i({\bf k}_3-{\bf k}_2+{\bf k}_1)\cdot {\bf r}} e^{i(\omega_3 t_3-\omega_2t_2+\omega_1t_1)}\nonumber \\
&&\hspace{.1in}\times   \int_{0}^{t_m} dt \int_{0}^t dt' \int_{0}^{t'} dt''  \chi^{(3)}(t_m,t,t',t'')  \nonumber \\
&&\hspace{.3in} \times E^*_3(t-t_3) e^{i\omega_3(t-t_3)}E_2 (t'-t_2) e^{-i\omega_2(t'-t_2)} \nonumber \\
&&\hspace{.3in} \times E^*_1(t''-t_1)e^{i\omega_1(t''-t_1)} \ , \label{eq:pt3} 
\een
\ben
&&\int_{0}^{t_m} dt \int_{0}^t dt' \int_{0}^{t'} dt''\ \mbox{\boldmath{${\eta}$}}_m\cdot Tr\left\{ \hat {\bf P} \hat {\mathcal T}_4(t_m,t,t',t'')\right\}\nonumber \\
&&=e^{-i({\bf k}_3-{\bf k}_2+{\bf k}_1)\cdot {\bf r}} e^{i(\omega_3t_3-\omega_2t_2+\omega_1 t_1)} \nonumber \\
&&\hspace{.1in} \times   \int_{0}^{t_m} dt \int_{0}^t dt' \int_{0}^{t'} dt''  \chi^{(4)}(t_m,t,t',t'')\nonumber \\
&&\hspace{.3in}\times  E_3^*(t-t_3)e^{i\omega_3(t-t_3)}E_2(t'-t_2) e^{-i\omega_2 (t'-t_2)}\nonumber \\
&&\hspace{.3in}\times  E_1^* (t''-t_1) e^{i\omega_1(t''-t_1)}  \  . \label{eq:pt4}
\een
where  $\chi^{(k)}$, $k=1-4$, are response functions.  Under the assumption that excited state absorption or double exciton formation can be neglected, these can be expressed as
\ben 
&&\chi^{(1)}(t_m,t,t',t'')=Tr\left\{\hat {\mathcal U}_g(t_m,t') \hat P_{g2}(t')\ \hat {\mathcal U}_{ex}(t',t'')\  \right . \nonumber \\
&&\hspace{.6in}\times \hat P_{1g}(t'')\hat {\mathcal U}_{g}(t'',t_0)\hat \rho (t_0) \hat {\mathcal U}^\dagger_g (t,t_0)\nonumber \\
&&\hspace{.6in}\times \left .\hat P_{g3}(t)\  \hat {\mathcal U}^\dagger_{ex} (t_m,t) \hat P_{mg}(t_m)\right\} \ ,
\een
\ben
&&\chi^{(2)}(t_m,t,t',t'')=Tr\left\{\hat {\mathcal U}_g (t_m,t) \hat P_{g3}(t)\ \hat {\mathcal U}_{ex}(t,t'')\ \right . \nonumber \\
&&\hspace{.6in} \times   \hat P_{1g}(t'')\hat {\mathcal U}_g(t'',t_0) \hat \rho (t_0) \hat {\mathcal U}_g^\dagger (t',t_0) \nonumber \\
&&\hspace{.6in}\times \left .\hat P_{g2}(t') \hat {\mathcal U}_{ex}^\dagger (t_m,t') \hat P_{mg}(t_m) \right\} \ , 
\een
\ben
&&\chi^{(3)}(t_m,t,t',t'')=Tr\left\{\hat {\mathcal U}_g (t_m,t) \hat P_{g3}(t)\ \hat {\mathcal U}_{ex}(t,t') \ \ \right . \nonumber \\
&& \hspace{.6in}\times \hat P_{2g}(t')\hat {\mathcal U}_{g}(t',t_0)\hat \rho (t_0) \ \hat {\mathcal U}_g^\dagger(t'',t_0) \nonumber \\
&&\hspace{.6in}\left. \times  \hat P_{g1}(t'') \  \hat {\mathcal U}_{ex}^\dagger (t_m,t'') \hat P_{mg}(t_m) \right\}  \ , 
\een
\ben
&&\chi^{(4)}(t_m,t,t',t'')=Tr\left\{\hat {\mathcal U}_g(t_m,t_0) \hat \rho (t_0) \hat {\mathcal U}_g^\dagger (t'',t_0)\   \right . \nonumber \\
&&\hspace{.6in}\times  \hat P_{g1}(t'')\hat {\mathcal U}_{ex}^\dagger (t',t'')\hat P_{2g}(t')\hat {\mathcal U}_{g}^\dagger (t,t') \nonumber \\
&&\hspace{.6in}\left. \times \hat P_{g3}(t) \hat {\mathcal U}_{ex}^\dagger (t_m,t) \hat P_{mg}(t_m)  \right\} \ .
\een 
The above general expressions for the response functions are equivalent to those in previous works \cite{mukamel,jonas-arpc54,cho-jpcb109,cho-cr108,schlau-cohen-cp386} within the assumption that only single exciton states contribute.  Fourier transforms of these with respect to $t'-t''$ and $t_m - t$  can be related to the spectra of 2DES if proper averaging over the ensemble of disorder is made. 

Now, for the initial state given by Eq. (\ref{eq:initial-abs}), the response functions can be simplified as follows:
\ben 
&&\chi^{(1)}(t_m,t,t',t'')=e^{\frac{i}{\hbar}\int_{t''}^{t'} d\tau E_g(\tau)} e^{-\frac{i}{\hbar}\int_{t}^{t_m} d\tau E_g(\tau) }  \nonumber \\
&& \hspace{.1in}\times Tr\left\{e^{-\frac{i}{\hbar}(t_m-t')\hat H_b} |D_m(t_m)\rangle \langle D_2(t')| \hat {\mathcal U}_{ex}(t',t'') \right . \nonumber \\
&&\hspace{.1in}\times \left . \hat \rho_b |D_1(t'')\rangle\langle D_3(t)| e^{\frac{i}{\hbar}(t-t'')\hat H_b}\hat {\mathcal U}^\dagger_{ex} (t_m,t) \right\} \ ,
\een
\ben
&&\chi^{(2)}(t_m,t,t',t'')=e^{-\frac{i}{\hbar}\int_{t}^{t_m} d\tau E_g(\tau)} e^{\frac{i}{\hbar}\int_{t''}^{t'} d\tau E_g(\tau) } \nonumber \\
&&\hspace{.1in}\times Tr\left\{e^{-\frac{i}{\hbar}(t_m-t)\hat H_b} |D_m(t_m)\rangle \langle D_3(t)|\hat {\mathcal U}_{ex}(t,t'') \right . \nonumber \\
&& \hspace{.1in}\times \left . \hat \rho_b |D_1(t'')\rangle\langle D_2(t')|  e^{\frac{i}{\hbar}(t'-t'')\hat H_b}\hat {\mathcal U}_{ex}^\dagger (t_m,t') \right\} \ , 
\een
\ben
&&\chi^{(3)}(t_m,t,t',t'')=e^{-\frac{i}{\hbar}\int_{t}^{t_m} d\tau E_g(\tau)} e^{-\frac{i}{\hbar}\int_{t''}^{t'} d\tau E_g(\tau) } \nonumber \\
&&\hspace{.1in}\times Tr\left\{ e^{-\frac{i}{\hbar}(t_m-t)\hat H_b} |D_m(t_m)\rangle \langle D_3(t)| \hat {\mathcal U}_{ex}(t,t') \right . \nonumber \\
&& \hspace{.1in}\left . \times  \hat \rho_b |D_2(t')\rangle\langle D_1(t'')| e^{\frac{i}{\hbar}(t'-t'')\hat H_b} \hat {\mathcal U}_{ex}^\dagger (t_m,t'') \right\}  \ , 
\een
\ben
&&\chi^{(4)}(t_m,t,t',t'')=e^{-\frac{i}{\hbar}\int_{t}^{t_m} d\tau E_g(\tau)} e^{-\frac{i}{\hbar}\int_{t''}^{t'} d\tau E_g(\tau) } \nonumber \\
&&\hspace{.1in}\times Tr\left\{e^{-\frac{i}{\hbar}(t_m-t'')\hat H_b}|D_m(t_m)\rangle \langle D_1(t'')|\hat \rho_b  \hat {\mathcal U}_{ex}^\dagger (t',t'') \right . \nonumber \\
&&\hspace{.1in}\left . \times |D_2(t')\rangle\langle D_3(t)| e^{\frac{i}{\hbar}(t-t')\hat H_b}\hat {\mathcal U}_{ex}^\dagger (t_m,t)  \right\} \ .
\een 

For the case where the Hamiltonians are time independent, the above expressions become Eqs. (\ref{eq:chi1-nt})-(\ref{eq:chi4-nt}).

%

\end{document}